\documentclass[11pt]{article}
\pdfoutput=1
\usepackage[nosort]{cite}

\usepackage{xcolor}

\usepackage{epsfig}
\usepackage{amsfonts}
\usepackage{amscd}
\usepackage{latexsym}
\usepackage{amsmath,amssymb}
\usepackage{amsthm}
\usepackage{verbatim}
\usepackage{setspace}
\usepackage{color}
\usepackage{fancyhdr}
\usepackage{tikz}
\usepackage{tikz-cd}
\usepackage{slashed}
\usepackage{soul}
\usepackage{multirow}

\usetikzlibrary{calc}
\usetikzlibrary{topaths}
\usetikzlibrary{decorations}
\usetikzlibrary{decorations.pathmorphing}

\usetikzlibrary{external}
\usetikzlibrary{zx-calculus}

\usepackage{draft}
\usepackage{hyperref}
\usepackage{graphicx,color,subfig}
\usepackage{cite}
\usepackage{mciteplus}
\usepackage{skak}
\usepackage{centernot}

\usepackage{enumitem}
\usepackage{crossreftools}

\numberwithin{equation}{section}

\renewcommand{\b}[1]{\boldsymbol{#1}}

\newcommand{\cnot}{\operatorname{CNOT}}
\newcommand{\cz}{\operatorname{CZ}}
\newcommand{\had}{\operatorname{H}}

\def\({\left(}
\def\){\right)}

\def\>{\rangle}
\def\<{\langle}

\NewExpandableDocumentCommand{\zxSolidBox}{O{}mmm}{
\xdef\Loopu{}
\foreach \c in {0,...,\numexpr #3-1\relax} { \xdef\Loopu{\Loopu(\zxGetNameRelativeNode{0}{\c})} }
\xdef\Loopd{}
\foreach \c in {0,...,\numexpr #3-1\relax} { \xdef\Loopd{\Loopd(\zxGetNameRelativeNode{\numexpr #2-1\relax}{\c})} }
\xdef\Loopl{}
\foreach \r in {0,...,\numexpr #2-1\relax} { \xdef\Loopl{\Loopl(\zxGetNameRelativeNode{\r}{0})} }
\xdef\Loopr{}
\foreach \r in {0,...,\numexpr #2-1\relax} { \xdef\Loopr{\Loopr(\zxGetNameRelativeNode{\r}{\numexpr #3-1\relax})} }
  \zxCont[fill=none,rounded corners=0pt,solid,opacity=1,zx thickness wires style,draw=blue,rectangle,
  fit margins={all=0.5mm},
  #1]{#2}{#3}[][\Loopu \Loopd \Loopl \Loopr]{#4}
}
\NewExpandableDocumentCommand{\zxDashedBox}{O{}mmm}{
\xdef\Loopu{}
\foreach \c in {0,...,\numexpr #3-1\relax} { \xdef\Loopu{\Loopu(\zxGetNameRelativeNode{0}{\c})} }
\xdef\Loopd{}
\foreach \c in {0,...,\numexpr #3-1\relax} { \xdef\Loopd{\Loopd(\zxGetNameRelativeNode{\numexpr #2-1\relax}{\c})} }
\xdef\Loopl{}
\foreach \r in {0,...,\numexpr #2-1\relax} { \xdef\Loopl{\Loopl(\zxGetNameRelativeNode{\r}{0})} }
\xdef\Loopr{}
\foreach \r in {0,...,\numexpr #2-1\relax} { \xdef\Loopr{\Loopr(\zxGetNameRelativeNode{\r}{\numexpr #3-1\relax})} }
  \zxCont[fill=none,rounded corners=0pt,solid,opacity=1,zx thickness wires style,draw=white,rectangle,append after command={(\tikzlastnode.north west) edge[zx thickness wires style,blue,densely dashed] (\tikzlastnode.north east) (\tikzlastnode.south west) edge[zx thickness wires style,blue,densely dashed] (\tikzlastnode.south east) (\tikzlastnode.north west) edge[zx thickness wires style,blue] (\tikzlastnode.south west) (\tikzlastnode.north east) edge[zx thickness wires style,blue] (\tikzlastnode.south east)},
  fit margins={left=0.5mm,right=0.5mm,top=0,bottom=0},
  #1]{#2}{#3}[][\Loopu \Loopd \Loopl \Loopr]{#4}
}

\zxConfigureExternalSystemCallAuto
\zxExternalSuffix{}

\begin{document}

\begin{titlepage}

\hfill YITP-SB-2024-11\\
\hfill \\

\title{Tensor networks for non-invertible  symmetries  in 3+1d and beyond}

\author{Pranay Gorantla$^{1}$, Shu-Heng Shao$^{2,3}$, Nathanan Tantivasadakarn$^{4}$}

\address{${}^{1}$Kadanoff Center for Theoretical Physics \& Enrico Fermi Institute, University of Chicago}
\address{${}^{2}$Yang Institute for Theoretical Physics, Stony Brook University}
\address{${}^{3}$Center for Theoretical Physics, Massachusetts Institute of Technology}
\address{${}^{4}$Walter Burke Institute for Theoretical Physics and Department of Physics, California Institute of Technology}

\vspace{2.0cm}

\begin{abstract}\noindent

Tensor networks provide a natural language for non-invertible symmetries in general Hamiltonian lattice models. 
We use ZX-diagrams, which are tensor network presentations of quantum circuits, to define a non-invertible operator implementing the Wegner duality in 3+1d lattice $\mathbb{Z}_2$ gauge theory. 
The non-invertible algebra, which mixes with lattice translations, can be efficiently computed using  ZX-calculus. 
We further deform the $\mathbb{Z}_2$ gauge theory while preserving the duality and find a model with nine exactly degenerate ground states on a torus, consistent with the Lieb-Schultz-Mattis-type constraint imposed by the symmetry. 
Finally, we provide a ZX-diagram presentation of the non-invertible duality operators (including non-invertible parity/reflection symmetries) of generalized Ising models based on graphs, encompassing the 1+1d Ising model, the three-spin Ising model, the Ashkin-Teller model, and the 2+1d plaquette Ising model. 
The mixing (or lack thereof) with spatial symmetries is understood from a unifying perspective based on graph theory.

\end{abstract}

\end{titlepage}

\eject

\tableofcontents

\section{Introduction}

The symmetry principle in theoretical physics has been generalized in many different directions in recent years. 
In particular, it has become increasingly clear that symmetries in quantum system need not be invertible. 
These \textit{non-invertible symmetries} are not described by group theory, and go beyond the paradigm set by Wigner's theorem. 
Nonetheless, they exist ubiquitously in many familiar quantum field theories and lattice models, leading to new conservation laws and selection rules. 
See \cite{McGreevy:2022oyu,Cordova:2022ruw,Brennan:2023mmt,Bhardwaj:2023kri,Schafer-Nameki:2023jdn, Luo:2023ive,Shao:2023gho,Carqueville:2023jhb} for reviews.

The simplest example of non-invertible symmetries is the Kramers-Wannier (KW) symmetry of the 1+1d Ising lattice model \cite{Grimm:1992ni,Oshikawa:1996dj,Ho:2014vla,Hauru:2015abi,Aasen:2016dop}. 
In particular, it has been realized recently that in the quantum Hamiltonian Ising lattice model, the operator algebra of this non-invertible symmetry mixes with the lattice translation \cite{Seiberg:2023cdc, Seiberg:2024gek}.\footnote{On the other hand, a general fusion category  can be realized with no mixing with the lattice translation on an anyonic chain \cite{Feiguin:2006ydp,Aasen:2020jwb}, which generally does not have a tensor product Hilbert space.} 
This symmetry implies a Lieb-Schultz-Mattis (LSM) type constraint \cite{Levin:2019ifu,Seiberg:2024gek}, forbidding a trivially gapped phase. 
See, for examples, \cite{Inamura:2021szw,Tan:2022vaz,Eck:2023gic,Mitra:2023xdo,Sinha:2023hum,Fechisin:2023dkj,Yan:2024eqz,Okada:2024qmk,Seifnashri:2024dsd,Bhardwaj:2024wlr,Chatterjee:2024ych,Bhardwaj:2024kvy,Khan:2024lyf,Jia:2024bng,Lu:2024ytl,Li:2024fhy,Ando:2024hun,ODea:2024tkt} for recent discussions of non-invertible symmetries in 1+1d lattice systems, and \cite{Delcamp:2023kew,Inamura:2023qzl,Moradi:2023dan,Cao:2023doz, Cao:2023rrb,ParayilMana:2024txy,Spieler:2024fby,Choi:2024rjm,Hsin:2024aqb,Cao:2024qjj} for 2+1d examples.

The most natural generalization of the Kramers-Wannier duality in 1+1d is the  Wegner duality in 3+1d lattice $\bZ_2$ gauge theory \cite{Wegner:1971app}.\footnote{Note that in 2+1d, the Ising lattice model is not self-dual under gauging; rather, it is mapped to the lattice $\mathbb{Z}_2$ gauge theory. The corresponding non-invertible symmetry in $\mathbb{Z}_2$ gauge theory coupled to Ising matter  is constructed in \cite{Choi:2024rjm} and generalized in Section \ref{sec:prod}. This is related to the  fact that gauging a $\mathbb{Z}_2^{(q)}$ symmetry in $(d+1)$d returns a dual $\mathbb{Z}_2^{(d-q-1)}$ symmetry \cite{Gaiotto:2014kfa,Tachikawa:2017gyf}. (The superscript $(q)$ denotes the form degree of a $q$-form global symmetry.) Hence, the dual symmetry can be isomorphic to the original one only if $q=0, d=1$ and $q=1, d=3$.} 
While the Kramers-Wannier duality exchanges the broken  and unbroken phases of an ordinary $\bZ_2^{(0)}$ symmetry, the Wegner duality exchanges those of a $\bZ_2^{(1)}$ one-form global symmetry.  
At the self-dual point, the Wegner duality leads to a non-invertible symmetry.
The topological defect associated with this symmetry  was constructed in the Euclidean lattice gauge theory in \cite{Koide:2021zxj}, and its continuum counterpart was found in \cite{Choi:2021kmx,Kaidi:2021xfk}. 

In this work, we derive the corresponding non-invertible duality \textit{operator} in the quantum Hamiltonian lattice $\bZ_2$ gauge theory in 3+1d on a tensor product Hilbert space, generalizing the KW operator of the 1+1d transverse-field Ising model.

To this end, we use the  \textit{tensor network} formalism \cite{Orus:2018dya,Cirac:2020obd} to present these non-invertible operators in general dimensions. 
On the lattice, tensor networks are a natural tool to describe non-unitary operators.
In particular, non-invertible symmetries in one spatial dimension have been recently studied on the lattice using a class of tensor networks  called matrix product operators \cite{Verstraete:2004gdw,Zwolak:2004nwu,Bultinck:2015bot,Vanhove:2018wlb,Lootens:2021tet,Garre-Rubio:2022uum,Molnar:2022nmh,Ruiz-de-Alarcon:2022mos,Lootens:2022avn,Lootens:2023wnl,Seiberg:2024gek}.
In this work, we use a particular formulation of tensor networks called \textit{ZX-calculus} \cite{Coecke:2008lcg,Duncan:2009ocf} (see \cite{vandeWetering:2020giq} for a review).
ZX-calculus is a graphical language to reason about linear maps between qubits. It is a concrete realization of a more abstract program of diagrammatic reasoning in quantum physics \cite{Selinger_2010,Coecke_Kissinger_2017}.
It has roots in quantum foundations, but has recently gained popularity in practical applications in quantum information ranging from optimizing quantum circuits \cite{Kissinger:2019alv,de_Beaudrap_2020,deBeaudrap:2020iwu} to concepts in quantum error correction and fault-tolerant quantum computation \cite{de_Beaudrap_2020_2,Kissinger:2022cyj,Khesin:2023qah,Bombin:2023dyc,Bauer:2023awl,Bombin:2023slv,townsend2023floquetifying}.

The central object of ZX-calculus is a \emph{ZX-diagram}. It is a string diagram/tensor network representation of a quantum circuit, or more generally, a linear map between qubits. One of the most important features of a ZX-diagram is that it can be deformed arbitrarily, while maintaining the connectivity, without changing the linear map it represents. In other words, \emph{only topology of the diagram matters}. Furthermore, a ZX-diagram can be modified to another topologically-inequivalent diagram using a simple (and finite) set of local moves known as \emph{rewrite rules}. Much of the power of ZX-calculus derives from the \emph{completeness} of these rewrite rules, i.e., any reasoning about linear maps between qubits can be done entirely graphically using a finite sequence of rewrite rules. This is in contrast to the most general formulation of tensor networks, where a general product of tensors increases the bond dimension, and does not have an obvious algorithm to simplify the tensors (especially in the case of high connectivity).

\begin{figure}
\centering\tikzsetfigurename{intro-D}
$\mathsf D_{\text{l} \leftarrow \text{s}} \sim ~~$\begin{ZX}[circuit]
& \zxElt{\vdots} \dar[] & \zxN{}\\[-8pt]
& \zxZ{} \dar[H] \ar[r] & \zxN{}\\
\rar & \zxZ{} \dar[H] & \zxN{}\\
& \zxZ{} \dar[H] \ar[r] & \zxN{}\\
\rar & \zxZ{} \dar[H] & \zxN{}\\
& \zxZ{} \dar[H] \ar[r] & \zxN{}\\
\rar & \zxZ{} \dar[H] & \zxN{}\\
& \zxElt{\smash\vdots} & \zxN{}
\end{ZX} ~~~~~~~~~~~~~~~ $\mathsf D_{\text{p}\leftarrow\text{l}} \sim ~~$ \raisebox{-0.5\height}{\includegraphics[scale=0.17]{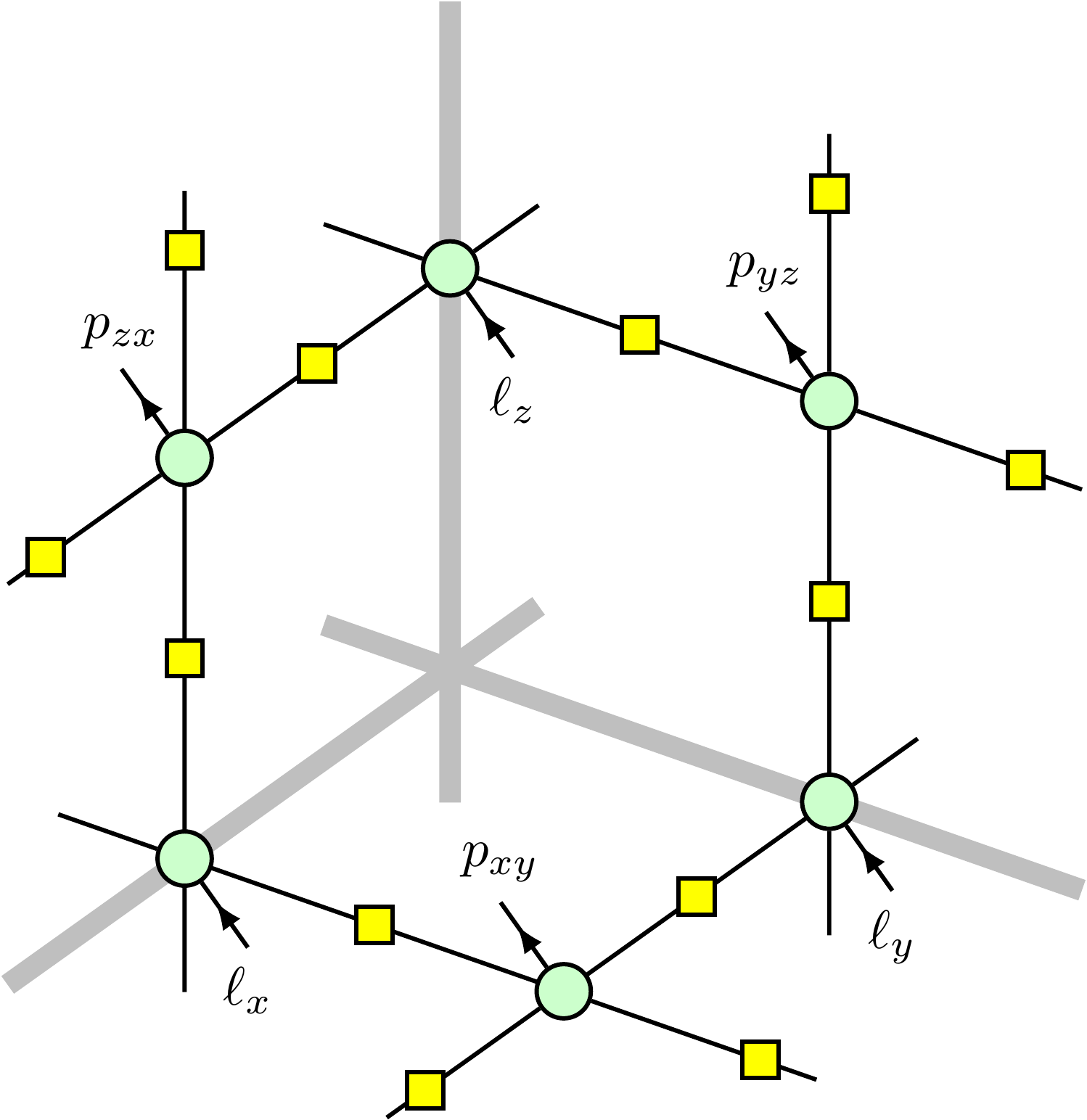}}
\\~\\
(a) ~~~~~~~~~~~~~~~~~~~~~~~~~~~~~~~~~~~~~~~~~~~(b)~~~~~~~~~~~~~~
\caption{The tensor network representations, or the ZX-diagrams, of (a) the Kramers-Wannier duality map in 1+1d and (b) the Wegner duality map in 3+1d. In (a), time flows from left to right, so the input legs are on the sites of the periodic chain and the output legs are on the links. In (b), the flow of time is indicated by the arrows, so the input legs are on the links (thick gray lines) of the periodic cubic lattice and the output legs are on the plaquettes. (We show only the part of the diagram in a unit cell; the full diagram is obtained by repeating and gluing this diagram periodically in the three spatial directions.)}\label{fig:intro-zx-D}
\end{figure}

The Kramers-Wannier duality operator in 1+1d and the Wegner duality operator in 3+1d admit a simple presentation as ZX-diagrams: first, we construct the duality \emph{maps} represented by the ZX-diagrams shown in Figure \ref{fig:intro-zx-D}, and then, we compose them with \emph{half-translation maps} (not shown in the figure). In both cases, the duality map takes the original lattice to the dual lattice, and the half-translation map takes it back to the original lattice. In the end, we get an \emph{operator} on the original Hilbert space.

Using ZX-calculus, we derive the non-invertible operator algebras for the 1+1d KW duality symmetry and that for the 3+1d Wegner duality symmetry in a uniform manner. 
In particular, we find that the algebra of the Wegner duality operator $\mathsf{D}$ involves a lattice translation in the $(1,1,1)$ direction, generalizing the 1+1d observation in \cite{Levin:2019ifu,Seiberg:2024gek}. See Table \ref{tbl:D^2}. 
The non-invertible operator $\mathsf{D}$ exchanges the product state with the 3+1d toric code ground state. 
We also define a condensation operator  $\mathsf{C}$ on the lattice, which creates a 3+1d toric code ground state from a product state.  

We further discuss a deformation of the lattice $\bZ_2$ gauge theory while preserving the non-invertible Wegner duality symmetry. 
This is a direct generalization of the deformation of the 1+1d Ising model in \cite{OBrien:2017wmx}. 
We argue for an LSM-type constraint for the non-invertible symmetry, which implies that the deformed model cannot be trivially gapped.

Finally, our tensor network presentation of the non-invertible symmetry can be readily extended to generalized transverse-field Ising models (TFIMs) based on arbitrary bipartite graphs.
We present the condition under which such a model has a non-invertible duality symmetry, and derive a condition for when the non-invertible operator algebra mixes with the spatial symmetry.
We then discuss several examples of non-invertible symmetries in this unifying presentation. 
In some examples, we find a non-invertible parity (or reflection) symmetry. 

The rest of this paper is organized as follows. Section \ref{sec:zx-review} reviews some of the essential elements of the ZX-calculus. Section \ref{sec:1d-review} demonstrates the utility of tensor network representation and ZX-calculus in constructing and analyzing the 1+1d Kramers-Wannier duality operator. Section \ref{sec:3d-review} gives a brief review of the 3+1d lattice $\mathbb Z_2$ gauge theory, gauging the $\mathbb Z_2$ 1-form symmetry, and Wegner duality. Section \ref{sec:3d-noninv} gives the explicit tensor network representation of the 3+1d Wegner duality operator and the derivation of its operator algebra using ZX-calculus. Section \ref{sec:graph} presents the tensor network representation of the non-invertible duality operator and its algebra in the generalized TFIM based on an arbitrary bipartite graph. Section \ref{sec:discuss} concludes with a discussion of the results and potential future directions. Appendix \ref{app:zx-3d} gives the details of several derivations involving the Wegner duality operator and the condensation operator using ZX-calculus. And finally, Appendix \ref{app:higher-quan-sym} gives the tensor network representation of the higher quantum symmetry operators associated with higher-gauging the $\mathbb Z_2$ 1-form symmetry on the lattice.

\renewcommand{\arraystretch}{1.7}
\begin{table}
\centering
\begin{tabular}{|c|c|c|}
\hline
 & 1+1d & 3+1d
\tabularnewline
\hline
Continuum & $\mathcal D^2 = 1+\eta$ & $\mathcal D(M)^2 = \frac1{|H^0(M,\mathbb Z_2)|} \sum_{\Sigma\in H_2(M,\mathbb Z_2)} \eta(\Sigma) =: {\cal C}(M)$
\tabularnewline
\hline
Lattice & $\mathsf D^2 = (1+\eta) T$ & $\mathsf D^2 = \frac1{2^{V}}\, T_{1,1,1} \sum_{\widehat \Sigma} \eta(\widehat \Sigma) = :\mathsf C T_{1,1,1}$
\tabularnewline
\hline
\end{tabular}
\caption{The operator algebras of the Kramers-Wannier  duality operator in 1+1d and the Wegner duality operator in 3+1d with themselves in the continuum and on the lattice. 
Here, $\eta$ is the ordinary $\bZ_2$ symmetry operator  in 1+1d , while $\eta(\Sigma)$ is a 1-form symmetry operator on a 2-cycle $\Sigma$ in 3+1d. 
Finally, $M$ is a spatial 3-manifold, $T$ is the generator of spatial translation in 1+1d, and $T_{1,1,1}$ is the generator of spatial translation in the $(1,1,1)$ direction in 3+1d.\label{tbl:D^2}}
\end{table}
\renewcommand{\arraystretch}{1}

\section{Review of ZX-calculus}\label{sec:zx-review}
A ZX-diagram is a graphical representation of a linear map of qubits.\footnote{This includes states (or kets), which are linear maps with domain $\mathbb C$, and their adjoints (or bras), which are linear maps with codomain $\mathbb C$.} It is built out of the following generators.
\tikzsetfigurename{ZX-def}
\begin{enumerate}
\item Z-spider: $\zx{\leftManyDots{n} \zxZ{\alpha} \rightManyDots{m}} := |0\>^{\otimes m} \<0|^{\otimes n} + e^{i\alpha} |1\>^{\otimes m} \<1|^{\otimes n}$.
\item X-spider: $\zx{\leftManyDots{n} \zxX{\alpha} \rightManyDots{m}} := |+\>^{\otimes m} \<+|^{\otimes n} + e^{i\alpha} |-\>^{\otimes m} \<-|^{\otimes n}$.
\item Hadamard, $\had$: $\zx[circuit]{\zxN{}\rar &\zxH{} \rar &\zxN{}} := \frac1{\sqrt2}(|0\>\<0| + |0\>\<1| + |1\>\<0| - |1\>\<1|)$.
\item Identity: $\zx[circuit]{\zxN{}\rar&[16pt]\zxN{}} := |0\>\<0| + |1\>\<1|$.
\item SWAP: $\zx[circuit]{\zxN{}\ar[rd,S] &\zxN{}\\ \zxN{}\ar[ru,S] &\zxN{}} := |00\>\<00| + |01\>\<10| + |10\>\<01| + |11\>\<11|$.
\item A Bell pair (ket): $\zx[circuit]{\zxN{}\ar[d,C]\\ \zxN{}} := |00\> + |11\>$.
\item A Bell pair (bra): $\zx[circuit]{\zxN{}\ar[d,C-]\\ \zxN{}} := \<00| + \<11|$.
\end{enumerate}
Here, $n,m\in \mathbb N = \{0,1,2,\ldots\}$ and $\alpha \sim \alpha + 2\pi$. It is customary to omit the phase $\alpha$ inside the Z/X-spiders if $\alpha=0$. The following are elementary examples of ZX-diagrams of some familiar states and operators.
\tikzsetfigurename{ZX-example}
\begin{enumerate}
\item Eigenstates of Pauli $Z$: $\zx[circuit]{\zxX{}\rar&\zxN{}} = \sqrt2 |0\>$ and $\zx[circuit]{\zxX{\pi}\rar&\zxN{}} = \sqrt2 |1\>$.
\item Eigenstates of Pauli $X$: $\zx[circuit]{\zxZ{}\rar&\zxN{}} = \sqrt2 |+\>$ and $\zx[circuit]{\zxZ{\pi}\rar&\zxN{}} = \sqrt2 |-\>$.
\item A Bell pair (ket): $\zx[circuit]{\zxN{}\ar[d,C,Z]\\ \zxN{}} = \zx[circuit]{\zxN{}\ar[d,C,X]\\ \zxN{}} = \zx[circuit]{\zxN{}\ar[d,C]\\ \zxN{}} = |00\> + |11\>$.
\item GHZ state: $\zx{&\\ \zxZ{} \ar[ru,('] \ar[r] \ar[rd,(] & \\ &} = |000\> + |111\>$.\footnote{Here, a helpful mnemonic is that the \textbf{G}HZ state is constructed from the \textbf{G}reen Z-spider.}
\item Pauli $Z$: $\zx[circuit]{\zxN{}\rar& \zxZ{\pi} \rar&\zxN{}} = |0\>\<0| - |1\>\<1|$.
\item Pauli $X$: $\zx[circuit]{\zxN{}\rar& \zxX{\pi} \rar&\zxN{}} = |+\>\<+| - |-\>\<-|$.
\item Controlled-NOT gate, $\cnot_{i,j}$: $\sqrt{2}~\zx[circuit]{\zxN-{i} \rar & \zxZ{} \rar \dar & \zxN-{i} \\ \zxN-{j} \rar & \zxX{} \rar & \zxN-{j}} = |0\>\<0|_i \otimes I_j + |1\>\<1|_i \otimes X_j$.
\item Controlled-$Z$ gate, $\cz_{i,j}$: $\sqrt{2}~\zx[circuit]{\zxN-{i} \rar & \zxZ{} \rar \dar[H] & \zxN-{i} \\ \zxN-{j} \rar & \zxZ{} \rar & \zxN-{j}} = \frac12(1+Z_i + Z_j - Z_i Z_j)$.
\end{enumerate}

Note that the ZX-diagrams of $\cnot$ and $\cz$ gates involve more than one generator. Indeed, one can build more general maps by combining ZX-diagrams as follows: let $D_1$ and $D_2$ be two diagrams with associated matrices denoted as $[D_1]$ and $[D_2]$, respectively.
\tikzsetfigurename{ZX-compose}
\begin{enumerate}
\item The diagram associated with the \emph{tensor product} $[D_1]\otimes[D_2]$ is obtained by placing them together. For example,
\ie
\begin{ZX}[circuit]
\zxX{}\rar&\zxN-{i}\\[-4pt]
\zxX{\pi}\rar&\zxN-{j}
\end{ZX}
= (\sqrt2)^2 |0\>_i\otimes|1\>_j~.
\fe
\item The diagram associated with the \emph{product} or \emph{composition} $[D_2] [D_1]$, provided it is well-defined, is obtained by joining the outputs of $D_1$ to the inputs of $D_2$.\footnote{Unless otherwise specified, we place the inputs (outputs) on the left (right) so that ``time'' flows from left to right. Where this is not possible---for example, in higher dimensional systems---we use ingoing (outgoing) arrows to represent the inputs (outputs).} For example,
\ie
\zx[circuit]{\zxN{} \rar & \zxZ{\pi} \rar & \zxX{\pi} \rar & \zxN{}} = XZ = -iY~.
\fe
\end{enumerate}
In fact, any map of qubits can be represented as a ZX-diagram built out of the above generators using the above composition rules. This property is known as \emph{universality}.

Given two ZX-diagrams, one natural question is if they represent the same map. There is a set of rewrite rules, known as the ZX-calculus, that allows one to transform a given diagram to obtain another diagram representing the same map. Some of them are listed below (the full set can be found in \cite{vandeWetering:2020giq} for instance).
\tikzsetfigurename{ZX-rule}
\begin{enumerate}
\item[({\crtcrossreflabel{\text{SF}}[sf]})] Spider fusion: $\zx{\leftManyDots{n} \zxZ{\alpha} \dar \rightManyDots{m}\\[8pt] \leftManyDots{k} \zxZ{\beta} \rightManyDots{l}} = \zx{\leftManyDots{n+k} \zxZ{\alpha+\beta} \rightManyDots{m+l}}$ and similarly for the X-spiders.
\item[({\crtcrossreflabel{\text{I}}[i]})] Identity removal: $\zx[circuit]{\zxN{}\ar[r,Z]&[16pt]\zxN{}} = \zx[circuit]{\zxN{}\ar[r,X]&[16pt]\zxN{}} = \zx[circuit]{\zxN{}\rar&[16pt]\zxN{}}$.
\item[({\crtcrossreflabel{\text{HC}}[hc]})] Hadamard cancellation: $\zx[circuit]{\zxN{} \rar[H] & \zxN{} \rar[H] & \zxN{}} = \zx[circuit]{\zxN{}\rar&[16pt]\zxN{}}$.
\item[({\crtcrossreflabel{\text{SC}}[sc]})] State copy: $\zx{\zxZ{}\rar & \zxX{\alpha} \rightManyDots{n}} =(\frac{1}{\sqrt2})^{n-1}~\zx[circuit]{\zxZ{} \rar&\zxN{}\\ \zxZ{}\rar \ar[u,3 vdotsr={$n$}]&\zxN{}}$ and similarly with Z/X-spiders exchanged.
\item[({\crtcrossreflabel{\text{Pi}}[pi]})] $\pi$ commutation: $\zx{\zxN{}\rar&\zxZ{\pi}\rar & \zxX{\alpha} \rightManyDots{n}} = e^{i\alpha}~\zx{\zxN{}&\zxN{} &\zxZ{\pi} \rar&\zxN{}\\[-4pt] \zxN{} \rar& \zxX{-\alpha} \ar[ru,<.] \ar[rd,<.] & &\\[-4pt] \zxN{} &\zxN{} &\zxZ{\pi}\rar \ar[uu,3 vdotsr={$n$}]&\zxN{}}$ and similarly with Z/X-spiders exchanged.
\item[({\crtcrossreflabel{\text{CC}}[cc]})] Color change: $\zx[circuit]{&\zxN{}&\zxN{}&\zxN{}\\[-6pt] &\zxN{} & \zxX{\alpha} \ar[lu,<.,H={pos=0.7}] \ar[ld,<.,H={pos=0.7}] \ar[ru,<.,H={pos=0.7}] \ar[rd,<.,H={pos=0.7}] & &\\[-6pt] &\zxN{}\ar[uu,3 vdots={$n$}] &\zxN{} &\zxN{}\ar[uu,3 vdotsr={$m$}]} = \zx{\leftManyDots{n} \zxZ{\alpha} \rightManyDots{m}}$ and similarly with Z/X-spiders exchanged.
\item[({\crtcrossreflabel{\text{B}}[b]})] Bialgebra: $\zx[circuit]{\zxN{}\rar& \zxX{} \ar[dr] \rar & \zxZ{} \ar[dl] \rar & \zxN{}\\ \zxN{} \rar& \zxX{} \rar & \zxZ{} \rar &\zxN{}} = \frac1{\sqrt2}~\zx[circuit]{\zxN{}&\zxN{}& &\zxN{}\\[-6pt] \zxN{} & \zxZ{} \ar[lu,<.] \ar[ld,<.] \rar & \zxX{} \ar[ru,<.] \ar[rd,<.] & &\\[-6pt] \zxN{} &\zxN{} & &\zxN{}}$.
\item[({\crtcrossreflabel{\text{Hf}}[hf]})] Hopf: $\zx[circuit]{\zxN{} \rar & \zxZ{} \ar[r,(.] \ar[r,('] & \zxX{} \rar & \zxN{}} = \frac12~\zx[circuit]{\zxN{} \rar & \zxZ{} & \zxX{} \rar & \zxN{}}$.
\item[({\crtcrossreflabel{\text{S}}[s]})] Scalar: $\zx[circuit]{\zxZ{}} = 2$, $\zx[circuit]{\zxZ{\pi}} = 0$, $\zx[circuit]{\zxZ{\alpha}} = 1+e^{i\alpha}$, $\zx[circuit]{\zxX{\alpha} \rar & \zxZ{}} = \sqrt{2}$, $\zx[circuit]{\zxX{\alpha} \rar & \zxZ{\pi}} = \sqrt{2}e^{i\alpha}$, $\zx[circuit]{\zxX{} \rar \ar[r,(.] \ar[r,('] & \zxZ{}} = \frac1{\sqrt{2}}$, and similarly with Z/X spiders exchanged.
\end{enumerate}
Some more useful rules derived from the above rules are listed below:
\tikzsetfigurename{ZX-derived-rule}
\begin{enumerate}
\item[({\crtcrossreflabel{\text{CC'}}[cc']})] Color change (version 2): $\zx[circuit]{&\zxN{}&\zxN{}&\zxN{}\\[-6pt] &\zxN{} & \zxX{\alpha} \ar[lu,<.] \ar[ld,<.] \ar[ru,<.,H={pos=0.7}] \ar[rd,<.,H={pos=0.7}] & &\\[-6pt] &\zxN{}\ar[uu,3 vdots={$n$}] &\zxN{} &\zxN{}\ar[uu,3 vdotsr={$m$}]} = \zx[circuit]{&\zxN{}&\zxN{}&\zxN{}\\[-6pt] &\zxN{} & \zxZ{\alpha} \ar[lu,<.,H={pos=0.7}] \ar[ld,<.,H={pos=0.7}] \ar[ru,<.] \ar[rd,<.] & &\\[-6pt] &\zxN{}\ar[uu,3 vdots={$n$}] &\zxN{} &\zxN{}\ar[uu,3 vdotsr={$m$}]}$ and similarly with Z/X-spiders exchanged, which follow from a combination of \ref{hc} and \ref{cc}.
\item[({\crtcrossreflabel{\text{GB}}[gb]})] Generalized bialgebra: $\begin{ZX}[circuit]
\zxN{} \rar & \zxX{} \rar \ar[rd] &\zxZ{} \rar & \zxN{}\\
\zxN{} \rar & \zxX{}\ar[u,3 vdots={$n$}] \rar \ar[ru] &\zxZ{} \rar \ar[u,3 vdotsr={$m$}] & \zxN{}
\end{ZX}
= \left(\frac{1}{\sqrt2}\right)^{(n-1)(m-1)}~\begin{ZX}[]
\leftManyDots{n} \zxZ{} \rar &[\zxWCol] \zxX{} \rightManyDots{m}
\end{ZX}$, which follows from \ref{sc} when $n=0$ or $m=0$, from \ref{i} when $n=1$ or $m=1$, and from a repeated application of \ref{b} and \ref{sf} when $n\ge2$ and $m\ge2$.
\end{enumerate}
We remark that many of the above rules may be familiar to readers acquainted with Hopf algebras. In particular, the above rules can be viewed as an extension of the group Hopf algebra $\mathbb C[\mathbb Z_2]$, where the phaseless 3-legged X-spider with two input legs and one output leg plays the role of multiplication (i.e., addition in $\mathbb Z_2$), and the phaseless 3-legged Z-spider with one input leg and two output legs plays the role of comultiplication. The associativity and coassociativity of the Hopf algebra essentially enables spider fusion \ref{sf}.

Another implicitly assumed rule when working with ZX-diagrams is that ``only topology of the diagram matters,'' i.e., we can move any part of a diagram (except for inputs/outputs) arbitrarily while maintaining the connectivity without changing the map it represents. For example, the ZX-diagram for the Controlled NOT gate (up to factors of $\sqrt2$) can be written as
\tikzsetfigurename{ZX-CNOT}
\ie
\zx[circuit]{\zxN-{i} \rar &[-4pt] \zxZ{} \rar \ar[dr,sIs] \zxSlice{} &[12pt] \zxN{} \rar &[-4pt] \zxN-{i} \\ \zxN-{j} \rar & \zxN{} \rar & \zxX{} \rar & \zxN-{j}}
= \zx[circuit]{\zxN-{i} \rar & \zxZ{} \rar \dar & \zxN-{i} \\ \zxN-{j} \rar & \zxX{} \rar & \zxN-{j}}
= \zx[circuit]{\zxN-{i} \rar &[-4pt] \zxN{} \rar \zxSlice{} &[12pt] \zxZ{} \rar \ar[dl,sIs] &[-4pt] \zxN-{i} \\ \zxN-{j} \rar & \zxX{} \rar & \zxN{} \rar & \zxN-{j}}~,
\fe
where we draw the red dashed line as a visual aid to divide the diagrams on the left and right into two subdiagrams. Each subdiagram is a tensor product of some generators and the two subdiagrams are composed by gluing their inputs/outputs along the red dashed line. The equality of the two diagrams on the left and right is a manifestation of the slogan ``only topology matters.'' This is why there is no ambiguity in the definition of the diagram in the middle.

These rules are so powerful that one can prove entirely graphically whatever one can prove using matrices, i.e., if $D_1$ and $D_2$ represent the diagrams of two matrices, denoted as $[D_1]$ and $[D_2]$, then $[D_1] = [D_2]$ if and only if there is a finite sequence of rules that transform $D_1$ to $D_2$. The ``if'' direction is known as \emph{soundness} and the ``only if'' direction is known as \emph{completeness}.\footnote{One has to add more rules to the list above so that the ZX-calculus is complete. We do not go into the details here and point interested readers to \cite{vandeWetering:2020giq}.}

In many cases, a diagram contains several copies of a smaller diagram. In such cases, it is convenient to use the annotated !-box (read as ``bang box''). For example,
\ie\label{bangbox}
\tikzsetnextfilename{bangbox1}
\begin{ZX}[circuit]
\zxZ{} \ar[rd] & &\\[-4pt]
\zxN{} & \zxZ{\alpha_i} \rar \zxSolidBox{1}{2}{above:${\scriptstyle i=1,2,3}$} &\\[-4pt]
\zxX{} \ar[ru] & &
\end{ZX}
:= \tikzsetnextfilename{bangbox2}
\begin{ZX}[circuit]
\zxZ[yshift=-8pt]{} \ar[rd] \rar \ar[rdd] & \zxZ{\alpha_1} \rar &\\[-8pt]
\zxN{} & \zxZ{\alpha_2} \rar &\\[-8pt]
\zxX[yshift=8pt]{} \ar[ru] \rar \ar[ruu] & \zxZ{\alpha_3} \rar &
\end{ZX}~.
\fe
Here, the part of the diagram inside the annotated !-box (the blue box) is repeated three times, and each copy, labelled by $i=1,2,3$, is connected to the part of the diagram outside the annotated !-box in the same way.

It is also convenient to define a ``periodic'' version of the annotated !-box. For example,
\ie
\tikzsetnextfilename{per-bangbox1}
\begin{ZX}[circuit,baseline=-3pt]
\zxN{} \zxDashedBox{4}{3}{above: ${\scriptstyle i=1,2,3}$} & & \zxN{}\\
\ar[r] & \zxZ{\alpha_i} \uar[H] \dar[H] & \zxN{}\\
& \zxZ{} \dar \ar[ru,S] &\\[-8pt]
\zxN{} & \zxN{} & \zxN{}
\end{ZX}
:= \tikzsetnextfilename{per-bangbox2}
\begin{ZX}[circuit]
& \zxN{} \dar[H] & \zxN{} \ar[l,C'] & \zxN{}\\
\rar & \zxZ{\alpha_1} \dar[H] & \zxN{} & \zxN{}\\
& \zxZ{} \dar[H] \rar & \zxN{} \ar[ru,S] & \zxN{}\\
\rar & \zxZ{\alpha_2} \dar[H] & \zxN{} & \zxN{}\\
& \zxZ{} \dar[H] \rar & \zxN{} \ar[ru,S] & \zxN{}\\
\rar & \zxZ{\alpha_3} \dar[H] & \zxN{} & \zxN{}\\
& \zxZ{} \dar \rar & \zxN{} \ar[ru,S] & \zxN{}\\[-8pt]
& \zxN{} \ar[r,C.] & \zxN{} \ar[uuuuuuu] & \zxN{}
\end{ZX}~.
\fe
It is very similar to the annotated !-box except that the wires that end on the dashed blue lines are glued together in a periodic way. Note that when there are no wires that end on the dashed lines, there is no difference between the two boxes, so we will use the periodic version only when there are wires that end on the dashed lines. Moreover, in what follows, it is always understood that $i$ labels the sites of a periodic 1d chain with $L$ sites, so we omit the annotation ``$i=1,\ldots,L$'' above the box.

Lastly, for completeness, we mention the following. Given a ZX-diagram $D$ representing a map $[D]$, we can obtain the ZX-diagram of
\begin{enumerate}
\item its \emph{conjugate} $[D]^*$ by flipping the signs of all the phases in the spiders in $D$,
\item its \emph{transpose} $[D]^\intercal$ by flipping the diagram about the vertical axis (or reversing the flow of time), and 
\item its \emph{adjoint} $[D]^\dagger$ by doing both.
\end{enumerate}

\section{Non-invertible Kramers-Wannier operator in 1+1d}\label{sec:1d-review}

In this section, we review the non-invertible Kramer-Wannier operator of the  critical Ising lattice model in 1+1d using the language of ZX-calculus. 

Consider a periodic chain of $L$ sites, labelled by $i=1,\ldots,L$, with a qubit at each site, i.e., $\mathcal H_i \cong \mathbb C^2$. The total Hilbert space is the tensor product $\mathcal H = \bigotimes_i \mathcal H_i$. The Pauli operators acting on the site $i$ are labelled as $Z_i$ and $X_i$. 
An eigenstate of the $Z_i$'s is denoted as $|\{m_i\}\> := \bigotimes_i |m_i\>$ with $ m_i \in \{0,1\}$. 
The corresponding ZX-diagram is
\ie
|\{m_i\}\> = \tikzsetnextfilename{Z-state}\frac1{2^{L/2}}~\zx{\zxElt{\vdots} & \\ \zxX{m_{i-1}\pi} \rar& \\ \zxX{m_i\pi} \rar& \\ \zxX{m_{i+1}\pi} \rar&\\ \zxElt{\smash\vdots} &} = \tikzsetnextfilename{Z-state-per}\frac1{2^{L/2}}~\zx[baseline=-3pt]{\zxX{m_i\pi} \rar \zxSolidBox{1}{2}{} &\zxN-{i}}~,
\fe
Similarly, an eigenstate of the $X_i$'s is denoted as $|\{(-)^{m_i}\}\> := \bigotimes_i |(-)^{m_i}\>$ with $m_i \in \{0,1\}$. 
The corresponding ZX-diagram is
\ie
|\{(-)^{m_i}\}\> = \tikzsetnextfilename{X-state}\frac1{2^{L/2}}~\zx{\zxElt{\vdots} & \\ \zxZ{m_{i-1}\pi} \rar& \\ \zxZ{m_i\pi} \rar& \\ \zxZ{m_{i+1}\pi} \rar&\\ \zxElt{\smash\vdots} &} = \tikzsetnextfilename{X-state-per}\frac1{2^{L/2}}~\zx[baseline=-3pt]{\zxZ{m_i\pi} \rar \zxSolidBox{1}{2}{} &\zxN-{i}}~.
\fe

While most of our discussions below concern with only the operators on the Hilbert space $\cal H$, it is useful to have in mind a specific example of a Hamiltonian. 
We start with the Hamiltonian of the 1+1d transverse-field Ising model
\ie\label{1d-H}
H = -J \sum_{i=1}^L Z_i Z_{i+1} - h \sum_{i=1}^L X_i~,
\fe
but our discussion applies to more general Hamiltonians with the same symmetries of interest. (See \eqref{1d-deformedHlambda}  for an example.) 
The Ising Hamiltonian is translationally invariant, i.e., $H$ commutes with the generator of translation
\ie
T = \tikzsetnextfilename{T-def}
\begin{ZX}[]
&[-4pt] \zxElt{\vdots}\ar[rd,<',start anchor=south] &[-4pt] \zxN{}\\[4pt]
\zxN+{}\ar[rrd,S] & & \zxN+{}\\[8pt]
\zxN+{}\ar[rrd,S] & & \zxN+{}\\[8pt]
\zxN+{}\ar[rd,'>,end anchor=north] & & \zxN+{}\\[4pt]
& \zxElt{\smash\vdots} & \zxN{}
\end{ZX}
= \tikzsetnextfilename{T-def-per}
\begin{ZX}[circuit,baseline=-3pt]
\zxN{}\zxDashedBox{3}{3}{}&\ar[rd,<']&\\[\zxSRow]
\zxN-{i}\ar[rd,'>]&&\zxN-{i}\\[\zxSRow]
&&
\end{ZX}~.
\fe
It has a global $\mathbb Z_2$ symmetry generated by the operator
\ie\label{1d-eta-def}
\eta = \prod_i X_i = \tikzsetnextfilename{eta-def}
\begin{ZX}[]
& \zxElt{\vdots} & \zxN{}\\[-4pt]
\rar & \zxX{\pi} \rar & \zxN{}\\
\rar & \zxX{\pi} \rar & \zxN{}\\
\rar & \zxX{\pi} \rar & \zxN{}\\[-4pt]
& \zxElt{\smash\vdots} & \zxN{}
\end{ZX}
= \tikzsetnextfilename{eta-def-per}
\begin{ZX}[circuit,baseline=-3pt]
\zxN-{i} \rar \zxSolidBox{1}{3}{} & \zxX{\pi} \rar & \zxN-{i}
\end{ZX}~.
\fe

\subsection{Non-invertible Kramers-Wannier duality operator}
When $J=h$, known as the critical Ising model, it also has a non-invertible Kramers-Wannier  duality symmetry, generated by the operator $\mathsf D$ which exchanges the Ising and transverse field terms in the Hamiltonian, i.e.,
\ie\label{1d-D-action}
\mathsf D X_i = Z_i Z_{i+1} \mathsf D~,\qquad \mathsf D Z_i Z_{i+1} = X_{i+1} \mathsf D~.
\fe
 There are several equivalent representations of the duality operator. Below we list a few:
 \begin{enumerate}
 \item $\mathsf D = \frac{1}{\sqrt{2}} e^{-2\pi i L/8} (1+\eta) e^{\frac{i\pi}4 X_1} e^{\frac{i\pi}4 Z_1 Z_2} e^{\frac{i\pi}4 X_2} e^{\frac{i\pi}4 Z_2 Z_3} e^{\frac{i\pi}4 X_3} \cdots e^{\frac{i\pi}4 X_{L-1}} e^{\frac{i\pi}4 Z_{L-1} Z_L} e^{\frac{i\pi}4 X_L}$ \cite{Chen:2023qst,Seiberg:2023cdc},
 \item $\mathsf D = \frac12 (1+\eta) \had_1 \cz_{1,2} \had_2 \cz_{2,3} \had_3 \cdots \had_{L-1} \cz_{L-1,L} \had_L (1+\eta)$ \cite{Seiberg:2024gek},
 \item $\mathsf D= \Tr (\mathbb{U}^1 \mathbb{U}^2 \cdots \mathbb{U}^L)$,~where $\mathbb{U}^i =\left(\begin{smallmatrix}\ket{0}\bra{+}_i & \ket{0}\bra{-}_i\\\ket{1}\bra{-}_i & \ket{1}\bra{+}_i  \end{smallmatrix}\right) $ \cite{Tantivasadakarn:2021vel,Seiberg:2024gek}, and
 \item $\mathsf D = \sum_{\{m_i\}}|\{(-)^{m_{i-1}+m_i}\}\>\<\{m_i\}| = \frac1{2^{L/2}}\sum_{\{m_i\},\{\hat m_i\}} (-1)^{\sum_i \hat m_i (m_{i-1}+m_i)} |\{\hat m_i\}\>\<\{m_i\}|$ \cite{Aasen:2016dop,Li:2023ani}.
 \end{enumerate}
From the fourth expression, we derive the ZX-diagrammatic presentation of the duality operator:
\ie\label{1d-D-def}
\mathsf D 
&= \tikzsetfigurename{D-def}
2^{L/2}~\begin{ZX}[circuit]
& \zxElt{\vdots} \dar[] & \zxN{}\\[-8pt]
& \zxZ{} \dar[H] \ar[rd,S] & \zxN{}\\
\rar & \zxZ{} \dar[H] & \zxN{}\\
& \zxZ{} \dar[H] \ar[rd,S] & \zxN{}\\
\rar & \zxZ{} \dar[H] & \zxN{}\\
& \zxZ{} \dar[H] \ar[rd,S] & \zxN{}\\
\rar & \zxZ{} \dar[H] & \zxN{}\\
& \zxElt{\smash\vdots} & \zxN{}
\end{ZX}
\overset{\ref{cc'}}{=} 2^{L/2}~\begin{ZX}[circuit]
& \zxElt{\vdots} \dar[] & \zxN{}\\[-8pt]
& \zxX{} \dar[] \ar[rd,S,H] & \zxN{}\\
\rar & \zxZ{} \dar[] & \zxN{}\\
& \zxX{} \dar[] \ar[rd,S,H] & \zxN{}\\
\rar & \zxZ{} \dar[] & \zxN{}\\
& \zxX{} \dar[] \ar[rd,S,H] & \zxN{}\\
\rar & \zxZ{} \dar[] & \zxN{}\\
& \zxElt{\smash\vdots} & \zxN{}
\end{ZX}
\overset{\ref{cc'}}{=} 2^{L/2}~\begin{ZX}[circuit]
& \zxElt{\vdots} \dar[] & \zxN{}\\[-8pt]
& \zxZ{} \dar[] \ar[rd,S] & \zxN{}\\
\rar[H] & \zxX{} \dar[] & \zxN{}\\
& \zxZ{} \dar[] \ar[rd,S] & \zxN{}\\
\rar[H] & \zxX{} \dar[] & \zxN{}\\
& \zxZ{} \dar[] \ar[rd,S] & \zxN{}\\
\rar[H] & \zxX{} \dar[] & \zxN{}\\
& \zxElt{\smash\vdots} & \zxN{}
\end{ZX}
\\
&= \tikzsetfigurename{D-def-per}
2^{L/2}~\begin{ZX}[circuit,baseline=-3pt]
\zxN{} \zxDashedBox{4}{3}{} & \zxN{} & \zxN{}\\[-8pt]
& \zxZ{} \uar \ar[rd,S] &\\
\zxN-{i}\ar[r] & \zxZ{} \uar[H] \dar[H] & \zxN-{i}\\
\zxN{} & & \zxN{}
\end{ZX}
\overset{\ref{cc'}}{=} \tikzsetnextfilename{D-def-per2}
2^{L/2}~\begin{ZX}[circuit,baseline=-3pt]
\zxN{} \zxDashedBox{4}{3}{} & \zxN{} & \zxN{}\\[-8pt]
& \zxX{} \uar \ar[rd,S,H] &\\
\zxN-{i}\ar[r] & \zxZ{} \uar \dar & \zxN-{i}\\
\zxN{} & & \zxN{}
\end{ZX}
\overset{\ref{cc'}}{=} \tikzsetnextfilename{D-def-per3}
2^{L/2}~\begin{ZX}[circuit,baseline=-3pt]
\zxN{} \zxDashedBox{4}{3}{} & \zxN{} & \zxN{}\\[-8pt]
& \zxZ{} \uar \ar[rd,S] &\\
\zxN-{i}\ar[r,H] & \zxX{} \uar \dar & \zxN-{i}\\
\zxN{} & & \zxN{}
\end{ZX}~,
\fe
where we have used the color change \ref{cc'} rewrite rule to provide equivalent presentations. 
The left or right (but not the middle) blue box in the last line gives the tensor $\mathbb{U}^i$ in the Matrix Product Operator (MPO) \cite{Verstraete:2004gdw,Zwolak:2004nwu} presentation equivalent to the third expression above.

To verify that these are indeed the ZX-diagrams of the matrix in the fourth expression, let us compute the matrix elements using ZX-calculus:
\tikzsetfigurename{D-matrixelem}
\ie
\mathsf D |\{m_i\}\> &= \frac{2^{L/2}}{2^{L/2}}~\begin{ZX}[circuit,baseline=-3pt]
\zxN{} \zxDashedBox{4}{3}{} & \zxN{} & \zxN{}\\[-8pt]
& \zxX{} \uar \ar[rd,S,H] &\\
\zxX{m_i \pi}\ar[r] & \zxZ{} \uar \dar & \zxN-{i}\\
\zxN{} & & \zxN{}
\end{ZX}
\overset{\ref{pi},\ref{sc}}{=} \frac1{2^{L/2}}~\begin{ZX}[circuit,baseline=-3pt]
\zxN{} \zxDashedBox{5}{2}{} & \zxN{}\\[-4pt]
\zxX{} \uar \ar[rd,S,H] &\\[-4pt]
\zxX{m_i \pi} \uar  & \zxN-{i}\\[-4pt]
\zxX{m_i \pi} \dar  & \zxN{}\\[-4pt]
\zxN{} & \zxN{} 
\end{ZX}
\overset{\ref{sf}}{=} \frac1{2^{L/2}}~\begin{ZX}[circuit,baseline=-3pt]
\zxX{(m_{i-1}+m_i) \pi} \rar[H] \zxSolidBox{1}{2}{} & \zxN-{i}
\end{ZX}
\\
&\overset{\ref{cc}}{=} \frac1{2^{L/2}}~\begin{ZX}[circuit,baseline=-3pt]
\zxZ{(m_{i-1}+m_i) \pi} \rar \zxSolidBox{1}{2}{} & \zxN-{i}
\end{ZX}
= |\{(-)^{m_{i-1}+m_i}\}\>~.
\fe
One can also modify the ZX-diagrams in \eqref{1d-D-def} to make the equivalence to the other expressions manifest.\footnote{In deriving the first expression using ZX-calculus, it is useful to note the Euler decomposition of the Hadamard gate:
\ie
\tikzsetfigurename{H-Euler}
\zx{\rar & \zxH{} \rar &} = e^{-i\pi/4}~\zx{\rar & \zxFracZ{\pi}{2} \rar & \zxFracX{\pi}{2} \rar & \zxFracZ{\pi}{2} \rar &} = e^{-i\pi/4}~\zx{\rar & \zxFracX{\pi}{2} \rar & \zxFracZ{\pi}{2} \rar & \zxFracX{\pi}{2} \rar &}~.
\fe}

It is sometimes useful to define an auxiliary Hilbert space $\mathcal H_\text{l}$ on the links, in the same way as the Hilbert space on the sites, which we now denote as $\mathcal H_\text{s}$ for clarity, and define the following maps between $\mathcal H_\text{s}$ and $\mathcal H_\text{l}$.
\ie
&\mathsf D_{\text{l}\leftarrow \text{s}} = \tikzsetnextfilename{Dl-s-def}
2^{L/2}~\begin{ZX}[circuit,baseline=-3pt]
\zxN{} \zxDashedBox{4}{3}{} & \zxN{} & \zxN{}\\[-8pt]
& \zxZ{} \uar \ar[r] & \zxN-{i-\frac12}\\
\zxN-{i}\ar[r] & \zxZ{} \uar[H] \dar[H] & \zxN{}\\
\zxN{} & & \zxN{}
\end{ZX}~,\quad
T_{\text{l}\leftarrow \text{s}} = \tikzsetnextfilename{Tl-s-def}
\begin{ZX}[circuit,baseline=-3pt]
\zxDashedBox{4}{3}{}&[-4pt]\ar[rd,<']&[-4pt]\\[-8pt]
&&\zxN-{i-\frac12}\\
\zxN-{i}\ar[rd,'>]&&\\
\zxN{}&&
\end{ZX}~,
\\
&\mathsf D_{\text{s}\leftarrow \text{l}} = \tikzsetnextfilename{Ds-l-def}
2^{L/2}~\begin{ZX}[circuit,baseline=-3pt]
\zxN{} \zxDashedBox{4}{3}{} & \zxN{} & \zxN{}\\[-8pt]
\zxN-{i-\frac12}\ar[r] & \zxZ{} \uar &\\
& \zxZ{} \dar[H] \uar[H] \ar[r] & \zxN-{i}\\
& & \zxN{}
\end{ZX}~,\quad
T_{\text{s}\leftarrow \text{l}} = \tikzsetnextfilename{Ts-l-def}
\begin{ZX}[circuit,baseline=-3pt]
\zxSolidBox{4}{2}{}&\\[-8pt]
\zxN-{i-\frac12}\ar[rd,S]&\\
&\zxN-{i}\\
\zxN{}&
\end{ZX}~,
\fe
where the sites are labelled by the integer $i$ (modulo $L$), whereas the links are labelled by the half-integer $i-\frac12$ (modulo $L$).
These duality maps were studied in \cite{Aasen:2016dop,Lootens:2021tet,Tantivasadakarn:2021vel,Tantivasadakarn:2022hgp,Lootens:2022avn,Li:2023ani}---for instance,
\ie
\mathsf D_{\text{l}\leftarrow \text{s}} \tikzsetnextfilename{Dl-s-measure}
\overset{\ref{sf}}{=} 2^{L/2}~\begin{ZX}[circuit,baseline=-3pt]
\zxN{} \zxDashedBox{4}{5}{} &[-16pt]\zxN{} & \zxN{} & \zxN{} &[-16pt]\\[-8pt]
&\zxZ{} \rar & \zxZ{} \uar \ar[r] & \zxN{} & \zxN-{i-\frac12}\\
\zxN-{i}& \zxN{}\ar[r] & \zxZ{} \uar[H] \dar[H] \rar & \zxZ{} &\\
&\zxN{} & & \zxN{} &
\end{ZX}
= \left( \bigotimes_i \<{+}|_i \right) \left( \prod_i \cz_{i,i-\frac12}\cz_{i,i+\frac12} \right) \left( \bigotimes_i |{+}\>_{i+\frac12} \right)~,
\fe
which is the expression for the Kramers-Wannier duality map in terms of the cluster entangler \cite{Tantivasadakarn:2021vel,Tantivasadakarn:2022hgp}. Furthermore, changing the inputs on sites $i$ to outputs in $\mathsf D_{\text{l}\leftarrow \text{s}}$ gives the ZX-diagram of the cluster state associated with the nontrivial $\mathbb Z_2 \times \mathbb Z_2$ SPT phase.

In terms of these maps, we can write
\ie\label{1d-half-trans-D}
T = \tikzsetnextfilename{Ts-l-Tl-s}
\begin{ZX}[circuit,baseline=-3pt]
\zxN{}\zxDashedBox{3}{3}{}&\ar[rd,<']&\\[\zxSRow]
\zxN-{i}\ar[rd,'>]&&\zxN-{i}\\[\zxSRow]
&&
\end{ZX}
= T_{\text{s}\leftarrow\text{l}}T_{\text{l}\leftarrow \text{s}}~,\qquad
\mathsf D &= \tikzsetnextfilename{Ts-l-Dl-s}
2^{L/2}~\begin{ZX}[circuit,baseline=-3pt]
\zxN{} \zxDashedBox{4}{3}{} & \zxN{} & \zxN{}\\[-8pt]
& \zxZ{} \uar \ar[rd,S] &\\
\zxN-{i}\ar[r] & \zxZ{} \uar[H] \dar[H] & \zxN-{i}\\
\zxN{} & & \zxN{}
\end{ZX}
= T_{\text{s}\leftarrow \text{l}} \mathsf D_{\text{l}\leftarrow \text{s}}
\\
&= \tikzsetnextfilename{Ds-l-Tl-s}
2^{L/2}~\begin{ZX}[circuit,baseline=-3pt]
\zxN{} \zxDashedBox{4}{4}{} &[-4pt] \ar[rd,<'] &[-4pt] \zxN{} & \zxN{}\\[-8pt]
& & \zxZ{} \uar &\\
\zxN-{i}\ar[rd,'>] & & \zxZ{} \dar[H] \uar[H] \ar[r] & \zxN-{i}\\
\zxN{} & & & \zxN{}
\end{ZX}
= \mathsf D_{\text{s}\leftarrow \text{l}}T_{\text{l}\leftarrow \text{s}}~.
\fe
are operators on $\mathcal H_\text{s}$.

\subsection{Action on operators}

Let us compute the action of $\mathsf D$ on the  $\mathbb Z_2$-symmetric, local operators. Such operators are generated by products of $X_i$ and $Z_i Z_{i+1}$, so it suffices to understand the action on these operators. First, we have
\tikzsetfigurename{DX=ZZD}
\ie
\mathsf D X_i &= 2^{L/2}~\begin{ZX}[circuit]
&& \zxElt{\vdots} & \zxN{}\\[-8pt]
&& \zxZ{} \uar \ar[rd,S] & \zxN{}\\
\zxN-{i}\rar & \zxX{\pi} \rar & \zxZ{} \uar[H] & \zxN-{i}\\
&& \zxZ{} \uar[H] \ar[rd,S] & \zxN{}\\
\zxN-{i+1} \rar & \zxN{} \rar & \zxZ{} \uar[H] & \zxN-{i+1}\\
&& \zxElt{\smash\vdots} \uar[H] & \zxN{}
\end{ZX}
\overset{\ref{pi}}{=} 2^{L/2}~\begin{ZX}[circuit]
& \zxElt{\vdots} & \zxN{}\\[-8pt]
& \zxZ{} \uar \ar[rdd,S] & \zxN{}\\
& \zxX{\pi} \uar[H] & \zxN{}\\[-8pt]
\zxN-{i}\rar & \zxZ{} \uar & \zxN-{i}\\[-8pt]
& \zxX{\pi} \uar & \zxN{}\\
& \zxZ{} \uar[H] \ar[rd,S] & \zxN{}\\
\zxN-{i+1} \rar & \zxZ{} \uar[H] & \zxN-{i+1}\\
& \zxElt{\smash\vdots} \uar[H] & \zxN{}
\end{ZX}
\\
&\overset{\ref{cc'}}{=} 2^{L/2}~\begin{ZX}[circuit]
& \zxElt{\vdots} & \zxN{}\\[-8pt]
& \zxZ{} \uar \ar[rdd,S] & \zxN{}\\[-8pt]
& \zxZ{\pi} \uar & \zxN{}\\
\zxN-{i}\rar & \zxZ{} \uar[H] & \zxN-{i}\\
& \zxZ{\pi} \uar[H] & \zxN{}\\[-8pt]
& \zxZ{} \uar \ar[rd,S] & \zxN{}\\
\zxN-{i+1} \rar & \zxZ{} \uar[H] & \zxN-{i+1}\\
& \zxElt{\smash\vdots} \uar[H] & \zxN{}
\end{ZX}
\overset{\ref{sf}}{=} 2^{L/2}~\begin{ZX}[circuit]
& \zxElt{\vdots} & \zxN{}\\[-8pt]
& \zxZ{} \uar \ar[rd,S] & \zxN{} &\\
\zxN-{i}\rar & \zxZ{} \uar[H] & \zxZ{\pi} \rar & \zxN-{i}\\
& \zxZ{} \uar[H] \ar[rd,S] & \zxN{} &\\
\zxN-{i+1} \rar & \zxZ{} \uar[H] & \zxZ{\pi} \rar & \zxN-{i+1}\\
& \zxElt{\smash\vdots} \uar[H] & \zxN{} &
\end{ZX}
= Z_i Z_{i+1} \mathsf D~.
\fe
Similarly, one can show that $\mathsf D Z_i Z_{i+1} = X_{i+1} \mathsf D$. 
Therefore, $\mathsf D$ indeed generates the KW duality transformation.

More generally, for $m_i \in \{0,1\}$, we have
\tikzsetfigurename{DprodX=prodZD}
\ie
\mathsf D \prod_i X_i^{m_i} &=  2^{L/2}~\begin{ZX}[circuit,baseline=-3pt]
\zxDashedBox{4}{4}{} & \zxN{} & \zxN{} &[\zxHCol] \zxN{}\\[-8pt]
& & \zxX{} \uar \ar[rd,H,S] &\\
\zxN-{i}\rar & \zxX{m_i\pi}\ar[r] & \zxZ{} \uar \dar & \zxN-{i}\\
\zxN{} & & & \zxN{} 
\end{ZX}
\overset{\ref{pi}}{=} 2^{L/2}~\begin{ZX}[circuit,baseline=-3pt]
\zxDashedBox{6}{3}{} & \zxN{} &[\zxHCol] \zxN{}\\[-4pt]
& \zxX{} \uar \ar[rdd,H,S] &\\[-4pt]
& \zxX{m_i\pi} \uar & \zxN{}\\[-4pt]
\zxN-{i}\rar & \zxZ{} \uar \dar & \zxN-{i}\\[-4pt]
& \zxX{m_i\pi} \dar & \zxN{}\\[-4pt]
\zxN{} & & \zxN{} 
\end{ZX}
\\
&\overset{\ref{sf}}{=} 2^{L/2}~\begin{ZX}[circuit,baseline=-3pt]
\zxN{} \zxDashedBox{4}{3}{} & \zxN{} &[\zxHCol] \zxN{}\\[-8pt]
& \zxX{(m_i+m_{i-1})\pi} \uar \ar[rd,H,S] &\\
\zxN-{i}\ar[r] & \zxZ{} \uar \dar & \zxN-{i}\\
\zxN{} & & \zxN{}
\end{ZX}
\overset{\ref{sf},\ref{cc'}}{=} 2^{L/2}~\begin{ZX}[circuit,baseline=-3pt]
\zxN{} \zxDashedBox{4}{4}{} & \zxN{} &[\zxHCol] \zxN{} & \\[-8pt]
& \zxX{} \uar \ar[rd,H,S] & & \\
\zxN-{i}\ar[r] & \zxZ{} \uar \dar & \zxZ{(m_i + m_{i-1})\pi} \rar & \zxN-{i}\\
\zxN{} & & \zxN{} &
\end{ZX}
\\
&= \prod_i Z_i^{m_i + m_{i-1}} \mathsf D = \prod_i (Z_i Z_{i+1})^{m_i} \mathsf D~,
\fe
and similarly, $\mathsf D \prod_i (Z_i Z_{i+1})^{m_i} = \prod_i X_{i+1}^{m_i} \mathsf D$. 
Consider  a special case when $m_1=m_2=\cdots =m_{k-1}=1$ for some $k>2$, and all the other $m_i$'s are 0. We find that $\mathsf D$ maps a pair of order operators ($Z$) to the disorder operator  ($X$):
\ie
\mathsf D Z_1 Z_k = X_2 X_3 \cdots X_k \mathsf D~.
\fe
Consider an even more special case where $m_i = 1$ for all $i$, we get
\ie\label{1d-D-eta}
\mathsf D \eta = \eta \mathsf D =  \mathsf D~.
\fe

\subsection{Operator algebra}\label{sec:1d-op-alg}
In this subsection, we compute the fusion rules of $\eta$, $\mathsf D$, and $T$ using ZX-calculus.  In particular, we will rederive 
\ie\label{1+1dD2}
\mathsf D^2 = (1+\eta)T~.
\fe
 of  \cite{Seiberg:2023cdc,Seiberg:2024gek}
 from ZX-calculus. 

The fusion rules involving $\eta$ and $T$ are:
\ie
&\eta^2 = \tikzsetfigurename{etasquare}
\begin{ZX}[circuit,baseline=-3pt]
\zxN-{i} \rar \zxSolidBox{1}{4}{} & \zxX{\pi} \rar & \zxX{\pi} \rar & \zxN-{i}
\end{ZX}
\overset{\ref{sf}}{=} \begin{ZX}[circuit,baseline=-3pt]
\zxN-{i} \rar \zxSolidBox{1}{3}{} & \zxX{} \rar & \zxN-{i}
\end{ZX}
\overset{\ref{i}}{=} \begin{ZX}[circuit,baseline=-3pt]
\zxN-{i} \rar \zxSolidBox{1}{2}{} &[\zxWCol] \zxN-{i}
\end{ZX}
= 1~,
\\
&T\eta = \tikzsetfigurename{T-eta}
\begin{ZX}[circuit,baseline=-3pt]
\zxN{}\zxDashedBox{3}{4}{}&&\ar[rd,<']&\\[\zxSRow]
\zxN-{i}\rar & \zxX{\pi} \ar[rd,'>]&&\zxN-{i}\\[\zxSRow]
&&&
\end{ZX}
= \begin{ZX}[circuit,baseline=-3pt]
\zxN{}\zxDashedBox{3}{4}{}&\ar[rd,<']&&\\[\zxSRow]
\zxN-{i}\ar[rd,'>]&& \zxX{\pi} \rar &\zxN-{i}\\[\zxSRow]
&&&
\end{ZX}
=\eta T~,
\\
&T^L = \tikzsetnextfilename{TtotheL}
\begin{ZX}[circuit,baseline=-3pt]
\zxDashedBox{3}{5}{}&\ar[rd,<']&&\ar[rd,<']&\\[\zxSRow]
\zxN-{i}\ar[rd,'>]&&\zxN-{\cdots}\ar[rd,'>]&&\zxN-{i}\\[\zxSRow]
\zxN{}&&\zxN{}&&
\end{ZX}
=1~.
\fe
To see the last line of this equation explicitly, consider $L=3$ for concreteness:
\ie
T = \tikzsetnextfilename{T-L=3}
\begin{ZX}[circuit]
\zxN-{1} \ar[rd,S] &[4pt] \zxN-{1}\\
\zxN-{2} \ar[rd,S] & \zxN-{2}\\
\zxN-{3} \ar[ruu,S] & \zxN-{3}
\end{ZX}
\implies T^3 = \tikzsetfigurename{TtotheL-L=3}
\begin{ZX}[circuit]
\zxN-{1} \ar[rd,S,green!70!black] &[4pt] \zxN{} \ar[rd,S,red] &[4pt] \zxN{} \ar[rd,S,blue] &[4pt] \zxN-{1}\\
\zxN-{2} \ar[rd,S,blue] & \zxN{} \ar[rd,S,green!70!black] & \zxN{} \ar[rd,S,red] & \zxN-{2}\\
\zxN-{3} \ar[ruu,S,red] & \zxN{} \ar[ruu,S,blue] & \zxN{} \ar[ruu,S,green!70!black] & \zxN-{3}
\end{ZX}
= \begin{ZX}[circuit]
\zxN-{1} \ar[r,green!70!black] &[4pt] \zxN-{1} \\
\zxN-{2} \ar[r,blue] & \zxN-{2} \\
\zxN-{3} \ar[r,red] & \zxN-{3}
\end{ZX}
=1~,
\fe
where we colored the wires for clarity.

Let us now compute the fusion rules involving $\mathsf D$ using ZX-calculus. We already showed that $\mathsf D \eta = \eta \mathsf D = \mathsf D$. Moreover, it is easy to see that $\mathsf D$ commutes with $T$. We are then left with the fusion of $\mathsf D$ with itself:
\tikzsetfigurename{Dsquare=DDT}
\ie
\mathsf D^2 = \mathsf D_{\text{s}\leftarrow \text{l}}T_{\text{l}\leftarrow \text{s}} T_{\text{s}\leftarrow \text{l}} \mathsf D_{\text{l}\leftarrow \text{s}}
= 2^L~\begin{ZX}[circuit,baseline=-3pt]
\zxN{} \zxDashedBox{4}{5}{} & \zxN{} &[-4pt] \ar[rd,<'] &[-4pt] \zxN{} &\zxN{} \\[-8pt]
& \zxZ{} \uar \ar[rdd,'>] & & \zxZ{} \uar &\\
\zxN-{i}\ar[r] & \zxZ{} \dar[H] \uar[H] & & \zxZ{} \dar[H] \uar[H] \ar[r] & \zxN-{i}\\
\zxN{} & & & & \zxN{} 
\end{ZX}
= 2^L~\begin{ZX}[circuit,baseline=-3pt]
\zxN{} \zxDashedBox{4}{5}{} & \zxN{} & \zxN{} &[-4pt]\zxN{} \ar[rdd,<'] &[-4pt]\\[-8pt]
& \zxZ{} \uar \ar[r] & \zxZ{} \uar & &\\
\zxN-{i}\ar[r] & \zxZ{} \uar[H] \dar[H] & \zxZ{} \uar[H] \dar[H] \ar[rd,'>] & & \zxN-{i}\\
\zxN{} & & & & \zxN{} 
\end{ZX}
= T \mathsf D_{\text{s}\leftarrow \text{l}} \mathsf D_{\text{l}\leftarrow \text{s}}~.
\fe
(As an aside, note that $T_{\text{l}\leftarrow \text{s}} T_{\text{s}\leftarrow \text{l}}$ is the generator of translation on $\mathcal H_\text{l}$.) Similarly, we have
\ie
\mathsf D^2 = \mathsf D_{\text{s}\leftarrow \text{l}} \mathsf D_{\text{l}\leftarrow \text{s}} T~.
\fe
We now show that $\mathsf D_{\text{s}\leftarrow \text{l}} \mathsf D_{\text{l}\leftarrow \text{s}} = 1+\eta$ using ZX-calculus:
\tikzsetfigurename{Dsquare=C}
\ie
\mathsf D_{\text{s}\leftarrow \text{l}}\mathsf D_{\text{l}\leftarrow \text{s}} &= 2^L~\begin{ZX}[circuit,baseline=-3pt]
\zxN{} \zxDashedBox{4}{4}{} & \zxN{} & \zxN{} &\zxN{} \\[-8pt]
& \zxZ{} \uar \ar[r] & \zxZ{} \uar &\\
\zxN-{i}\ar[r] & \zxZ{} \uar[H] \dar[H] & \zxZ{} \uar[H] \dar[H] \ar[r] & \zxN-{i}\\
\zxN{} & & & \zxN{} 
\end{ZX}
\overset{\ref{cc'}}{=} 2^L~\begin{ZX}[circuit,baseline=-3pt]
\zxN{} \zxDashedBox{4}{4}{} & \zxN{} & \zxN{} &\zxN{} \\[-8pt]
& \zxX{} \uar \ar[r] & \zxX{} \uar &\\
\zxN-{i}\ar[r] & \zxZ{} \uar \dar & \zxZ{} \uar \dar \ar[r] & \zxN-{i}\\
\zxN{} & & & \zxN{} 
\end{ZX}
\overset{\ref{sf}}{=} 2^L~\begin{ZX}[circuit,baseline=-3pt]
\zxN{} \zxDashedBox{5}{5}{} &[-4pt] &[-6pt] &[-6pt] &[-4pt] \\[-4pt]
& & \zxX{} \uar &  &\\[-8pt]
\zxN-{i}\ar[r] & \zxZ{} \ar[ru] \ar[rd] & & \zxZ{} \ar[lu] \ar[ld] \ar[r] & \zxN-{i}\\[-8pt]
& & \zxX{} \dar & &\\[-4pt]
\zxN{} & \zxN{} & & \zxN{} &
\end{ZX}
\\
&\overset{\ref{b}}{=} \frac{2^L}{2^{L/2}}~\begin{ZX}[circuit,baseline=-3pt]
\zxN{} \zxDashedBox{4}{4}{} & &[-6pt] &[-6pt] \\[-4pt]
& \zxZ{} \uar \ar[dd] & & \\[-4pt]
\zxN-{i}\ar[rr] & & \zxX{} \ar[lu] \ar[r] & \zxN-{i}\\
\zxN{} & \zxN{} & & \zxN{}
\end{ZX}
\overset{\ref{sf}}{=} 2^{L/2}~\begin{ZX}[circuit,baseline=-9pt]
& \zxZ{} \ar[d] & \\[-4pt]
\zxN-{i} \ar[r]\zxSolidBox{1}{3}{}  & \zxX{} \ar[r] & \zxN-{i}
\end{ZX} =: \mathsf C~,
\fe
where $\mathsf C$ is the condensation operator. It is convenient to define a more general operator
\ie\label{1d-Cn-def}
\mathsf C_n := \tikzsetnextfilename{Cn-def-per}
2^{L/2}~\begin{ZX}[circuit,baseline=-9pt]
& \zxZ{n\pi} \ar[d] & \\[-4pt]
\zxN-{i} \ar[r]\zxSolidBox{1}{3}{}  & \zxX{} \ar[r] & \zxN-{i}
\end{ZX}~,
\fe
where $n=0$ gives the condensation operator $\mathsf C= \mathsf C_0$.

We want to show that $\mathsf C_n = 1+(-1)^n\eta$. We have,
\tikzsetfigurename{C=1+eta-CNOT}
\ie
\mathsf C_n = 2^{L/2}~\begin{ZX}[circuit]
\zxZ[xshift=8pt]{n\pi} \ar[dr] \ar[ddr] &[-4pt] &\\[-8pt]
\zxN-{1} \ar[r] & \zxX{} \ar[d,3 vdots] \rar & \zxN-{1}\\
\zxN-{L}\ar[r] & \zxX{} \rar & \zxN-{L}\\
\end{ZX}
\overset{\ref{sf}}{=} 2^{L/2}~\begin{ZX}[circuit]
\zxZ{} \rar & \zxZ{} \ar[d] \rar & \zxZ{} \ar[dd] \rar & \zxZ{n\pi}\\
\zxN-{1} \ar[r] & \zxX{} \ar[dr,3 vdots] \ar[ur,3 dots] \rar & \zxN{} \rar & \zxN-{1}\\
\zxN-{L} \ar[r] & \zxN{} \rar & \zxX{} \rar & \zxN-{L}\\
\end{ZX}
= \frac{2^{L/2}}{2^{L/2}}\cdot (\sqrt{2})^2~\<(-)^n|_\mathsf{a} \left(\prod_{i=1}^L \cnot_{\mathsf{a},i}\right) |+\>_\mathsf{a}~,
\fe
where $\mathsf{a}$ represents the auxiliary wire above $1$. Now, we have
\ie
\prod_{i=1}^L \cnot_{\mathsf{a},i} &= \prod_{i=1}^L (|0\>\<0|_\mathsf{a} \otimes I + |1\>\<1|_\mathsf{a} \otimes X_i)
= |0\>\<0|_\mathsf{a} \otimes I + |1\>\<1|_\mathsf{a} \otimes \eta~.
\fe
It follows that
\ie
\mathsf C_n = 2\<(-)^n|_\mathsf{a} \left(|0\>\<0|_\mathsf{a} \otimes I + |1\>\<1|_\mathsf{a} \otimes \eta\right) |+\>_\mathsf{a} = 1+(-1)^n \eta~.
\fe
Therefore, we proved using ZX-calculus that $\mathsf C_n = 1+(-1)^n \eta$, and hence $\mathsf D^2 = (1+\eta)T$. 
More concretely, for $L=3$, we have
\tikzsetfigurename{Dsquare=CT}
\ie
\mathsf D^2 &=
2^{3/2}~\begin{ZX}[circuit]
\zxZ[xshift=8pt]{} \ar[dr] \ar[ddr] \ar[dddr] &[-4pt] & &[4pt]\\[-8pt]
\zxN-{1}\ar[r] & \zxX{} \rar &\zxN{} \ar[rd,S] & \zxN-{1}\\
\zxN-{2}\ar[r] & \zxX{} \rar &\zxN{} \ar[rd,S] & \zxN-{2}\\
\zxN-{3}\ar[r] & \zxX{} \rar &\zxN{} \ar[ruu,S] & \zxN-{3}\\
\end{ZX}
=T(1+\eta)
\\
&= 2^{3/2}~\begin{ZX}[circuit]
&[4pt] \zxZ[xshift=8pt]{} \ar[dr] \ar[ddr] \ar[dddr] &[-4pt] &\\[-8pt]
\zxN-{1} \ar[rd,S] & \zxN{} \ar[r] & \zxX{} \rar & \zxN-{1}\\
\zxN-{2} \ar[rd,S] & \zxN{}] \ar[r] & \zxX{} \rar & \zxN-{2}\\
\zxN-{3} \ar[ruu,S] & \zxN{}\ar[r] & \zxX{} \rar & \zxN-{3}\\
\end{ZX}
=(1+\eta)T~.
\fe

It follows from \eqref{1d-D-eta} that $\mathsf D \mathsf C = \mathsf C \mathsf D = 2\mathsf D$. Let us verify this using ZX-calculus:
\tikzsetfigurename{DC}\ie
\mathsf D_{\text{l}\leftarrow\text{s}} \mathsf C &= 2^L~\begin{ZX}[circuit,baseline=-3pt]
&[-4pt] \zxN{} \zxDashedBox{4}{4}{} & & &[\zxHCol] \zxN{}\\[-8pt]
\zxZ{}\ar[rrd] & & & \zxX{} \uar \ar[r,H] & \zxN-{i-\frac12}\\
&\zxN-{i}\rar & \zxX{} \ar[r] & \zxZ{} \uar \dar &\\
& & \zxN{} & \zxN{} & \zxN{}
\end{ZX}
\overset{\ref{b}}{=} 2^L\cdot 2^{L/2}~\begin{ZX}[circuit,baseline=-3pt]
&[-4pt] \zxN{} \zxDashedBox{5}{4}{} & & &[\zxHCol] \zxN{}\\[-8pt]
\zxZ{}\ar[rrd] & & & \zxX{} \uar \ar[r,H] & \zxN-{i-\frac12}\\[-4pt]
& & \zxZ{} \ar[r] \ar[rd] & \zxX{} \uar & \zxN{}\\
&\zxN-{i}\rar & \zxZ{} \ar[r] \ar[ru] & \zxX{} \dar &\\
& & \zxN{} & \zxN{} & \zxN{}
\end{ZX}
\\
&\overset{\ref{sf}}{=} 2^{3L/2}~\begin{ZX}[circuit,baseline=-3pt]
&[-4pt] \zxN{} \zxDashedBox{5}{3}{} & &[\zxHCol] \zxN{}\\[-8pt]
\zxZ{}\ar[rrdd,(.] \ar[rr] & & \zxX{} \uar \ar[r,H] & \zxN-{i-\frac12}\\
&\zxN-{i}\rar & \zxZ{} \uar \dar &\\
& & \zxX{} \dar & \zxN{}\\[-8pt]
& & \zxN{} & \zxN{}
\end{ZX}
\overset{\ref{sf}}{=} 2^{3L/2}~\begin{ZX}[circuit,baseline=-3pt]
&[-4pt] \zxN{} \zxDashedBox{4}{3}{} & &[\zxHCol] \zxN{}\\[-8pt]
\zxZ{} \ar[rr,(.] \ar[rr,('] & & \zxX{} \uar \ar[r,H] & \zxN-{i-\frac12}\\
&\zxN-{i}\rar & \zxZ{} \uar \dar &\\
& & \zxN{} & \zxN{}
\end{ZX}
\\
&\overset{\ref{sf}}{=} 2^{3L/2}~\begin{ZX}[circuit,baseline=-3pt]
&[-4pt] \zxN{} \zxDashedBox{4}{3}{} & &[\zxHCol] \zxN{}\\[-8pt]
\zxZ{} \rar & \zxZ{} \ar[r,(.] \ar[r,('] & \zxX{} \uar \ar[r,H] & \zxN-{i-\frac12}\\
&\zxN-{i}\rar & \zxZ{} \uar \dar &\\
& & \zxN{} & \zxN{}
\end{ZX}
\overset{\ref{hf}}{=} \frac{2^{3L/2}}{2^L}~\begin{ZX}[circuit,baseline=-3pt]
&[-4pt] \zxN{} \zxDashedBox{4}{3}{} & &[\zxHCol] \zxN{}\\[-8pt]
\zxZ{} \rar & \zxZ{} & \zxX{} \uar \ar[r,H] & \zxN-{i-\frac12}\\
&\zxN-{i}\rar & \zxZ{} \uar \dar &\\
& & \zxN{} & \zxN{}
\end{ZX}
\\
&\overset{\ref{sf}}{=} 2^{L/2}~\begin{ZX}[circuit,baseline=-3pt]
&[-4pt] \zxN{} \zxDashedBox{4}{3}{} & &[\zxHCol] \zxN{}\\[-8pt]
\zxZ{} & & \zxX{} \uar \ar[r,H] & \zxN-{i-\frac12}\\
&\zxN-{i}\rar & \zxZ{} \uar \dar &\\
& & \zxN{} & \zxN{}
\end{ZX}
\overset{\ref{s}}{=} 2\cdot 2^{L/2}~\begin{ZX}[circuit,baseline=-3pt]
\zxN{} \zxDashedBox{4}{3}{} & &[\zxHCol] \zxN{}\\[-8pt]
& \zxX{} \uar \ar[r,H] & \zxN-{i-\frac12}\\
\zxN-{i}\rar & \zxZ{} \uar \dar &\\
& \zxN{} & \zxN{}
\end{ZX}
= 2\mathsf D_{\text{l}\leftarrow\text{s}}~.
\fe
Multiplying $T_{\text{s}\leftarrow\text{l}}$ on the left on both sides gives the desired fusion rule. Showing that $\mathsf C \mathsf D = 2\mathsf D$ is similar.

For the sake of completeness, let us also compute $\mathsf C^2 = 2\mathsf C$ using ZX-calculus:
\tikzsetfigurename{Csquare}
\ie
\mathsf C^2 &= 2^L~\begin{ZX}[circuit,baseline=-9pt]
& \zxZ{} \ar[d] & \zxZ{} \ar[d] & \\[-4pt]
\zxN-{i} \ar[r]\zxSolidBox{1}{4}{} & \zxX{} \ar[r] & \zxX{} \ar[r] & \zxN-{i}
\end{ZX}
\overset{\ref{sf}}{=} 2^L~\begin{ZX}[circuit,baseline=-9pt]
\zxZ{} \rar &[-4pt] \zxZ{} \ar[rd] &[-8pt] &[-8pt] \zxZ{} \ar[ld] &[-4pt] \zxZ{} \lar \\[-4pt]
\zxN{} \zxSolidBox{2}{5}{} & & \zxX{} & &\\[-4pt]
\zxN-{i} \ar[rr] & & \zxX{} \ar[u] \ar[rr] & & \zxN-{i}
\end{ZX}
\overset{\ref{gb}}{=} \frac{2^L}{2^{(L-1)/2}}~\begin{ZX}[circuit,baseline=-9pt]
\zxZ{} \ar[rr] &[-4pt] &[-8pt] \zxX{} \ar[d] &[-8pt] &[-4pt] \zxZ{} \ar[ll] \\[-4pt]
\zxN{} & & \zxZ{} & &\\[-4pt]
\zxN-{i} \ar[rr] \zxSolidBox{1}{5}{} & & \zxX{} \ar[u] \ar[rr] & & \zxN-{i}
\end{ZX}
\\
&\overset{\ref{sc},\ref{sf}}{=} \frac{2^{(L+1)/2}}{\sqrt2}~\begin{ZX}[circuit,baseline=-9pt]
\zxZ{} & \zxZ{} &\\[-4pt]
\zxN-{i} \ar[r] \zxSolidBox{1}{3}{} & \zxX{} \ar[u] \ar[r] & \zxN-{i}
\end{ZX}
\overset{\ref{s}}{=} 2\cdot 2^{L/2}~\begin{ZX}[circuit,baseline=-9pt]
& \zxZ{} &\\[-4pt]
\zxN-{i} \ar[r] \zxSolidBox{1}{3}{} & \zxX{} \ar[u] \ar[r] & \zxN-{i}
\end{ZX}
= 2\mathsf C~.
\fe

\tikzexternaldisable

\subsection{Action on states}\label{sec:1d-groundstates}

The phase diagram of the  Ising model \eqref{1d-H} is well-understood \cite{Pfeuty:1970qrn}: the $\mathbb Z_2$ global symmetry is spontaneously broken (ferromagnetic phase) for $J>h$, whereas it is preserved (paramagnetic phase) for $J<h$. At the critical point $J=h$, there is a second order phase transition captured by the Ising CFT at low energies.

Let us comment on the ground states and the superselection sectors in these phases. 
In the limit $J \gg h$ (or when $h=0$), the Ising term dominates. 
In finite volume, there are two low-lying states whose energy splitting is exponentially small in $L$. 
In infinite volume, the Hilbert space splits into two superselection sectors:
\ie\label{01}
|0\cdots0\>~,\qquad |1\cdots1\>~,
\fe
On the other hand, in the limit $J\ll h$ (or when $J=0$), the transverse field term dominates, so there is a unique ground state,
\ie\label{plus}
|{+}\cdots{+}\>~.
\fe

In fact, these three product states $|0\cdots 0\>,|1\cdots1\>,|{+}\cdots{+}\>$  are the exact ground states of a special case ($\lambda=1$) of the lattice model introduced in \cite{OBrien:2017wmx}:
\ie\label{1d-deformedHlambda}
H_\lambda = -J \left[\sum_i (Z_i Z_{i+1} + X_i) - \frac{\lambda}2 \sum_i (X_{i-1} Z_i Z_{i+1} + Z_i Z_{i+1} X_{i+2})\right]~.
\fe
For any $\lambda$, this Hamiltonian preserves both the $\bZ_2$ and the KW duality symmetries. 
In the thermodynamic limit, the model is in the $c=1/2$ Ising CFT phase for $0 \le \lambda \lesssim0.856$,  reaches the $c=7/10$ tricritical Ising CFT point at $\lambda \sim 0.856$, and becomes gapped with three superselection sectors for $\lambda \gtrsim 0.856$. 
These phases are  consistent with the LSM-type constraint from the non-invertible symmetry in \cite{Seiberg:2024gek}.
In particular, this model was shown to have exact three-fold degeneracy even at finite $L$ when $\lambda = 1$.

Irrespective of the Hamiltonian, one can ask how the symmetry operators $\eta$ and $\mathsf D$ act on these states \cite{Seiberg:2024gek}:
\ie
&\eta |0\cdots0\> = |1\cdots1\>~,\qquad \eta |1\cdots1\> = |0\cdots0\>~,\quad &&\eta |{+}\cdots{+}\> = |{+}\cdots{+}\>~,
\\
&\mathsf D |0\cdots0\> = \mathsf D |1\cdots1\> = |{+}\cdots{+}\>~,\quad &&\mathsf D |{+}\cdots{+}\> = |0\cdots0\> + |1\cdots1\>~.
\fe
This basis corresponds to the three superselection sectors of the Hamiltonian \eqref{1d-deformedHlambda} in the thermodynamic limit. Alternatively, one can work in a basis that diagonalizes the symmetry operators:
\ie
&\frac1{\sqrt2} |{+}\cdots{+}\> \pm {1\over \sqrt{2}}|\text{GHZ}^+\>~,\quad &&\mathsf D = \pm \sqrt2~,\eta = 1~,
\\
&|\text{GHZ}^-\>~,\quad &&\mathsf D = 0~,\eta = -1~,
\fe
where
\ie
|\text{GHZ}^\pm\> :=  {1\over\sqrt{2}} (|0\cdots0\> \pm |1\cdots1\>)~.
\fe
The duality operator exchanges $|\text{GHZ}^+\>$ with the product state $|{+}\cdots{+}\>$:
\ie\label{Dexchange2d}
\mathsf{D} |{+}\cdots{+}\> =\sqrt{2} |\text{GHZ}^+\>~,\qquad
\mathsf{D} |\text{GHZ}^+\> = \sqrt{2} |{+}\cdots{+}\>~.
\fe

\section{Review of 3+1d lattice $\mathbb Z_2$ gauge theory}\label{sec:3d-review}

We denote the sites, links, plaquettes, and cubes of a 3d cubic lattice by $s,\ell, p , c$, respectively.  
Let $L_x, L_y, L_z$ be the number of sites in each direction and assume periodic boundary conditions. 
We place a qubit associated with a local Hilbert space $\mathcal H_\ell \cong \mathbb{C}^2$ on every link. It is the $\mathbb{Z}_2$ 1-form gauge field. We denote the Pauli operators acting on link $\ell$ as $X_\ell, Z_\ell$. 
The total Hilbert space is $  \mathcal H  = \bigotimes_\ell \mathcal H_\ell  $.

Next, we impose the Gauss law for the $\mathbb{Z}_2$ 1-form gauge field strictly as an operator equation at every site:
\ie\label{gauss}
G_s = \prod_{\ell \ni s} X_\ell=1~.
\fe
We denote the Hilbert space subject to this constraint as $\widetilde{\mathcal H}$, which  is not a tensor product of local factors.
We will later relax this operator equation and impose it energetically.

\begin{figure}
\centering
\hfill \raisebox{-0.5\height}{\includegraphics[scale=0.22]{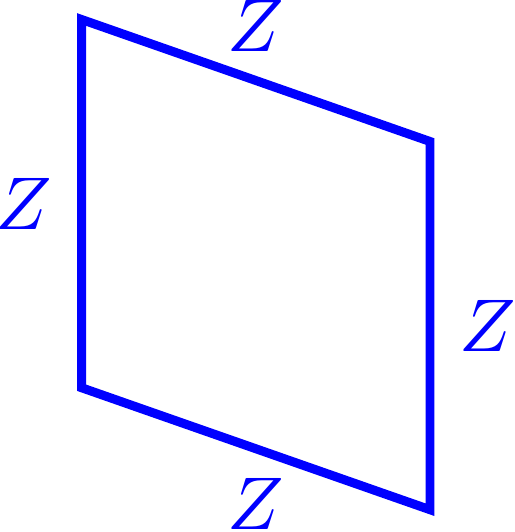}} \hfill \raisebox{-0.5\height}{\includegraphics[scale=0.22]{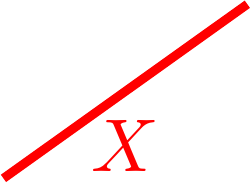}} \hfill $G_s = $~\raisebox{-0.5\height}{\includegraphics[scale=0.22]{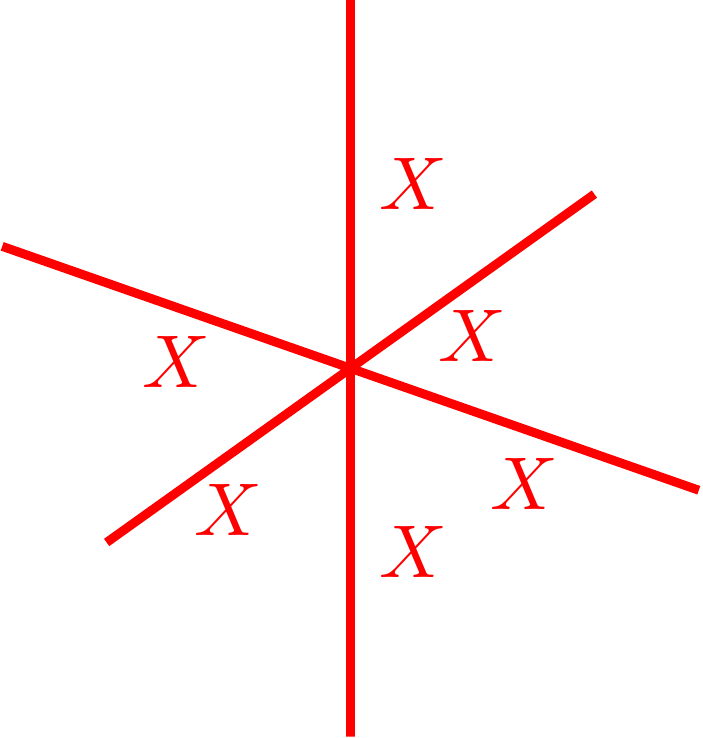}} \hfill ~
\\~\\
\hfill (a)~ \hfill (b)~~~~~~ \hfill (c) \hfill ~~~~~~~~~
\caption{The local terms and the Gauss law of the Hamiltonian \eqref{tH} of the lattice $\mathbb Z_2$ gauge theory. (a) The magnetic flux term on a plaquette (first term of \eqref{tH}), (b) the electric field term on a link (second term of \eqref{tH}), and (c) the Gauss law operator $G_s$ at a site.\label{fig:Hterms}}
\end{figure}

We start with the standard Hamiltonian for the lattice gauge theory:
\ie\label{tH}
\widetilde H =   -  J  \sum_p \prod_{\ell \in p}  Z_\ell    - h  \sum_\ell X_\ell~, 
\fe
which acts on the constrained Hilbert space $\widetilde{\mathcal H}$. 
The first term is the analog of the magnetic flux term, while the second term is the analog of the electric field term  in the continuum. (See Figure \ref{fig:Hterms} for an illustration of these terms.)
We will later study deformations of this Hamiltonian while preserving all the symmetries of interest.

\subsection{Topological versus non-topological 1-form global symmetries}

The central global symmetry that will be important to us is a lattice   global symmetry generated by 
\ie\label{eta-def}
\eta(\widehat \Sigma ) = \prod_{\ell \in \widehat \Sigma }X_\ell~,
\fe
where $\widehat \Sigma$ is a 2-dimensional surface on the dual lattice.  
The product is over all the links orthogonal to $\widehat \Sigma$, or equivalently, over all the dual plaquettes on $\widehat \Sigma$.  
In contrast to an ordinary global symmetry, this symmetry operator acts only on part of the space and commutes with the Hamiltonian $\widetilde H$ for any $\widehat \Sigma$. 
Since the symmetry operator is a product of the electric field $X_\ell$  on a codimension-1 surface in the 3-dimensional space, we refer to it as an electric  1-form global symmetry denoted by $\mathbb{Z}_2^{(1)}$. 
 (The superscript stands for the codimension of the symmetry operator in space.)

When the Gauss law is imposed as a strict operator equation as in \eqref{gauss}, this 1-form global symmetry is \textit{topological} in the sense that  
\ie
\eta( \widehat \Sigma ) = \eta( \widehat \Sigma' )~,\quad\text{if}\quad \widehat \Sigma \sim  \widehat \Sigma'~,
\fe
where $\widehat \Sigma \sim  \widehat \Sigma'$ means that they are in the same homology class on the dual lattice.  
That is, the symmetry operator depends only on the homological class of the 2-cycle $ \widehat \Sigma$.  Therefore, the $\mathbb Z_2$ 1-form symmetry is generated by the operators associated with the three fundamental non-contractible surfaces.

Similar to the continuum, this topological 1-form symmetry cannot be violated by any local operator deformation to the Hamiltonian. 
For instance, a term like $ Z_\ell$ violates the Gauss law \eqref{gauss} and is excluded from the Hamiltonian as a valid deformation.

Alternatively, we can relax the Gauss law constraint \eqref{gauss}, and impose it  energetically in the Hamiltonian as:
\ie\label{H}
 H =   -  J  \sum_p \prod_{\ell \in p}  Z_\ell    - h  \sum_\ell X_\ell - g \sum_s \prod_{\ell \ni s} X_\ell~.
\fe
This Hamiltonian acts on the tensor product Hilbert space $\mathcal H$.   
When $g\to \infty$, it reduces to the previous case where the Gauss law is enforced strictly. 
When $h=0$, this is the 3+1d toric code.\footnote{Moreover, the model with $h=0$ has an additional (non-topological) magnetic 2-form symmetry. This is analogous to the (topological) 1-form and 2-form symmetries in the modified Villain (integer BF) Euclidean lattice model for the 3+1d $\mathbb Z_2$ gauge theory \cite{Gorantla:2021svj}.}

When the Gauss law is relaxed such as in $ H$, the 1-form symmetry is no longer topological: $\eta(\widehat \Sigma)$ is a distinct operator acting on $\mathcal H$ for each dual surface $\widehat \Sigma$. 
Now that the Gauss law is relaxed, the non-topological 1-form symmetry can be violated by local operators, such as  $Z_\ell$. 
The difference between topological and non-topological higher-form symmetries on the lattice have been previously discussed in \cite{Seiberg:2019vrp,Qi:2020jrf,Choi:2024rjm}.\footnote{Topological/non-topological 1-form symmetries are  respectively referred to as   relativistic/non-relativistic 1-form symmetries  in \cite{Seiberg:2019vrp}, and as non-faithful/faithful in \cite{Qi:2020jrf}. Here we adopt the terminology topological/non-topological symmetries as in \cite{Choi:2024rjm} because it directly reflects the property of the  operators. }

This is a general feature: a higher-form global symmetry in the continuum, or a topological higher-form symmetry on the lattice cannot be violated by local operators. On the other hand, a non-topological one on the lattice can be violated by local operator deformations. 

\subsection{Defects for the 1-form symmetry}

Having discussed the conserved operator $\eta(\widehat \Sigma)$, we move on to the defect for the 1-form symmetry on the lattice.  
We work with the more general system where the Gauss law is only enforced energetically. 
The 1-form symmetry defect is represented as a modification of the Hamiltonian along a curve $\widehat \gamma$ on the dual lattice. 
We refer to the Hamiltonian with a defect as the defect Hamiltonian, which is given by
\ie
H_\eta(\widehat \gamma)
=  - J \sum_p (-1)^{\< \widehat \gamma,p \> }  \prod_{\ell\in p}Z_\ell 
- h \sum_\ell X_\ell
 - g \sum_s \prod_{\ell \ni s}X_\ell~,
\fe
where $\< \widehat \gamma,p\>$ is the intersection number of $\widehat \gamma$ and $p$.
The subscript and the superscript stand for the type and the location of the defect, respectively. 
The defect is topological in the sense that we can move its location by a product of local unitaries. 
Specifically, let $\widehat \gamma'$ be another curve that is homological to $\widehat \gamma$, and let $\widehat \Sigma$ be  the surface bounded by $\widehat \gamma - \widehat \gamma'$. Then
\ie
\left( \prod_{\ell \in \widehat \Sigma }  X_\ell \right) 
 H_\eta(\widehat \gamma) 
\left( \prod_{\ell \in \widehat \Sigma }  X_\ell \right) ^{-1} 
= H_\eta(\widehat \gamma')~.
\fe
where the product is over all dual faces, or equivalently, links $\ell$, lying on the dual surface $\widehat \Sigma$. 
Note that even though the 1-form symmetry operator $\eta(\widehat \Sigma)$ is not topological, the associated defect is.

\subsection{Gauging the 1-form symmetry and the Wegner duality}\label{sec:gauging}

The 1-form symmetry \eqref{eta-def} is on-site \cite{Wen:2018zux} and is free of 't Hooft anomaly. 
It means that it is compatible with a trivially gapped phase, and it can be gauged.
Here, we gauge the 1-form global symmetry, following the discussion in Appendix B of \cite{Choi:2021kmx} and Sec. IIIB of \cite{Chen:2020msl}. This generalizes the prescription for gauging a 0-form symmetry outlined in \cite{Levin:2012yb,Seiberg:2024gek}.

For concreteness, we focus on the Hamiltonian in \eqref{tH}, 
but most of the conclusions hold true for more general  Hamiltonian invariant under the 1-form symmetry. 
The local term of such a general Hamiltonian is a product of $X_\ell$ and $\prod_{\ell\in p} Z_\ell$.

\begin{figure}
\centering
\hfill $G_\ell = $~\raisebox{-0.5\height}{\includegraphics[scale=0.22]{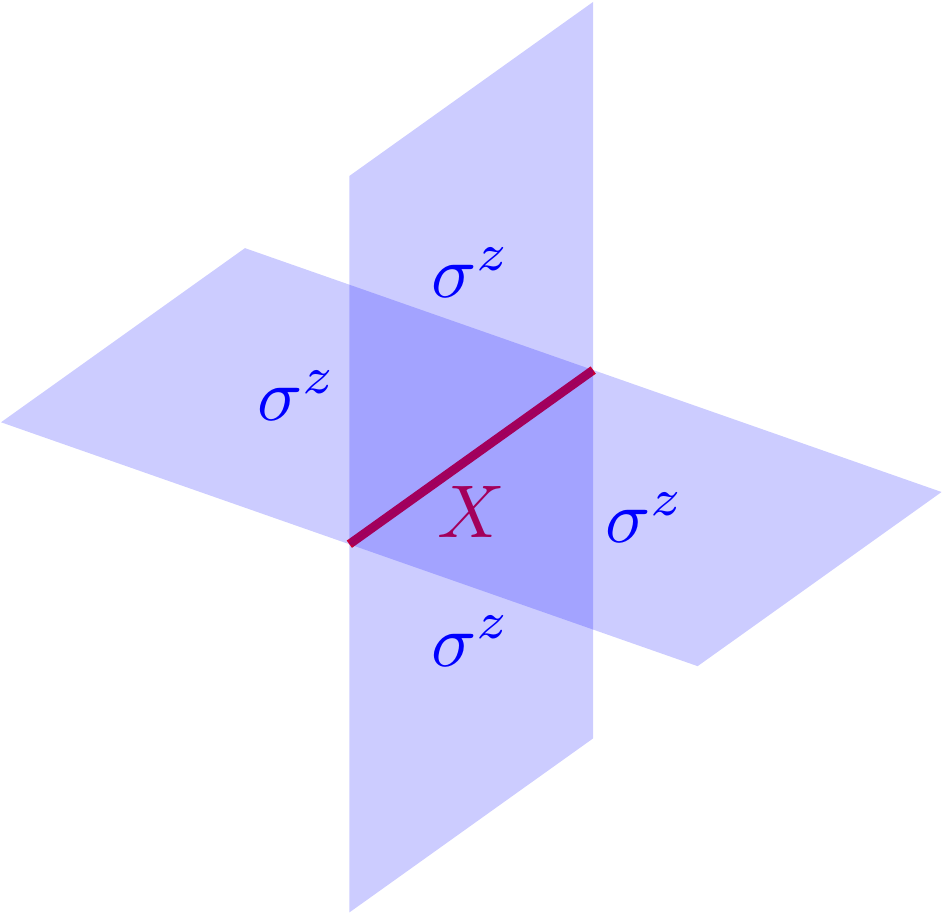}} \hfill \raisebox{-0.5\height}{\includegraphics[scale=0.22]{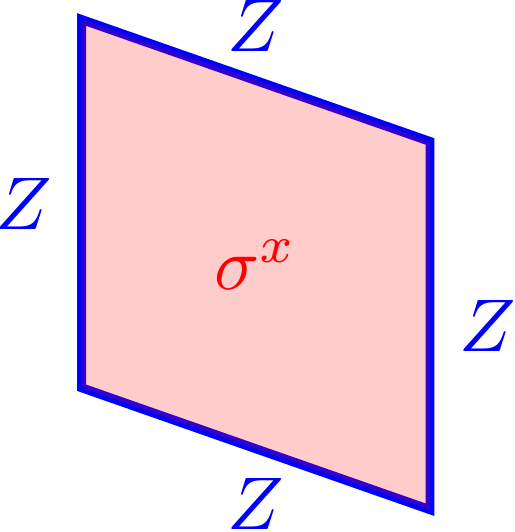}} \hfill \raisebox{-0.5\height}{\includegraphics[scale=0.22]{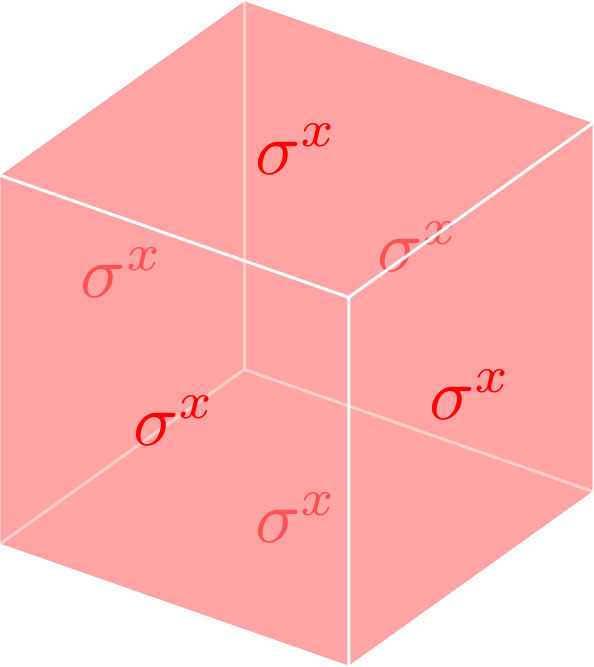}} \hfill ~
\\~\\
\hfill ~~~~~~~(a)~~~~~~~~~~~~~ \hfill ~~~(b) \hfill ~~~~(c) \hfill \hfill
\caption{The Gauss law and the local terms of the gauged Hamiltonian \eqref{gaugedH}. (a) The Gauss law operator $G_\ell$ for the 2-form gauge field on a link, (b) the minimal coupling term on a plaquette (first term of \eqref{gaugedH}), and (c) the flux term of the 2-form gauge field on a cube (third term of \eqref{gaugedH}). The second term of \eqref{gaugedH} is the same as Figure \ref{fig:Hterms}(b).\label{fig:gaugedHterms}}
\end{figure}

We gauge the 1-form global symmetry $\eta$ by introducing a qubit with Pauli operators $\sigma^{a}_p$ ($a=x,y,z$) on every plaquette $p$. 
In the continuum, they represent the 2-form $\mathbb{Z}_2$ gauge fields for the 1-form symmetry. 
The gauging consists of three steps:
\begin{enumerate}
\item Gauss law: Impose the Gauss law for the 2-form gauge field at every link:
\ie
G_\ell = X_\ell \prod_{p \ni \ell} \sigma^z_p  =1~.
\fe
With this new Gauss law, the original one $G_s= \prod_{\ell\ni s}X_\ell=1$ is trivially satisfied, and we no longer have to impose the latter by hand.
\item Minimal coupling:  Couple the gauge fields minimally to the Hamiltonian by replacing the plaquette term  $\prod_{\ell\in p} Z_\ell$ by  $\sigma^x_p \prod_{\ell\in p}Z_\ell$. 
\item Flux term:  We add a gauge-invariant term $  \prod_{p\in c}\sigma^x_p$ 
 of the 2-form gauge fields for every cube to the Hamiltonian.  This term is analogous to the magnetic field square term in the continuum. 
\end{enumerate}
The gauged Hamiltonian is then 
\ie\label{gaugedH}
\widehat H  = - J \sum_p \sigma^x_p \prod_{\ell\in p} Z_\ell - h \sum_\ell X_\ell  -  \widehat g  \sum_c\prod_{p\in c}\sigma^x_p ~,
\fe
subject to the  Gauss law constraint for the 2-form gauge fields:
\ie
G_\ell = X_\ell \prod_{p\ni \ell}\sigma^z_p=1~.
\fe
See Figure \ref{fig:gaugedHterms} for an illustration of these terms.

\begin{figure}
\centering
\hfill \raisebox{-0.5\height}{\includegraphics[scale=0.22]{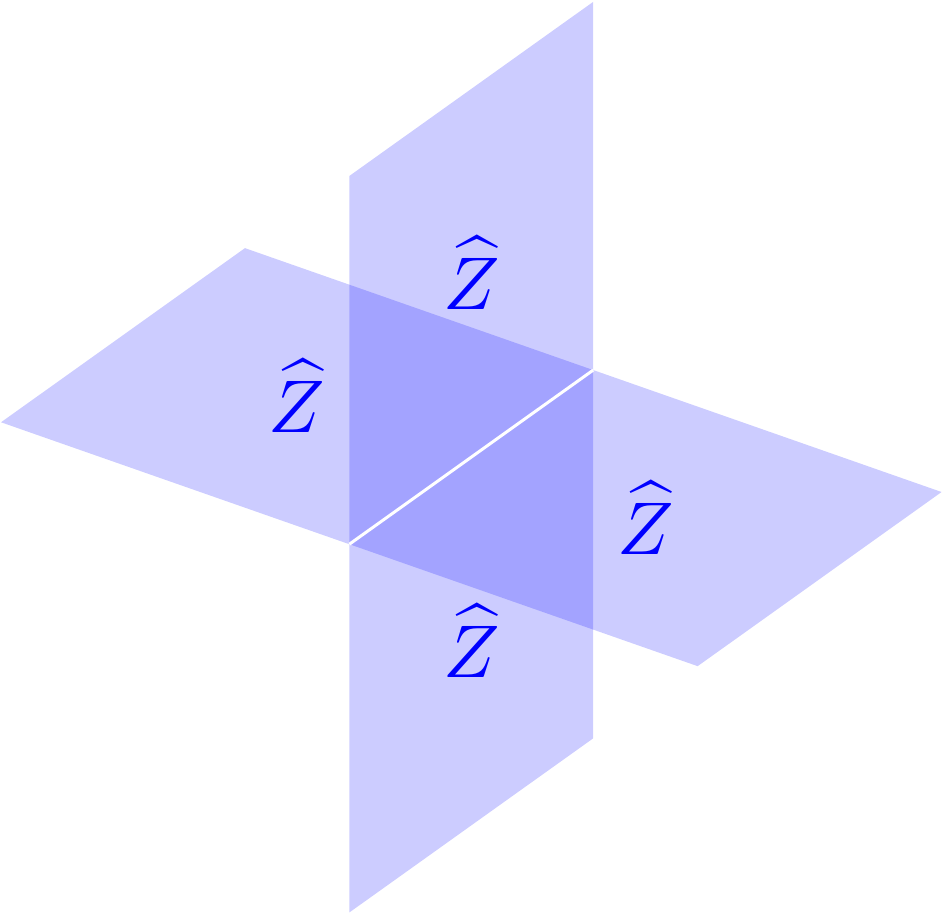}} \hfill \raisebox{-0.5\height}{\includegraphics[scale=0.22]{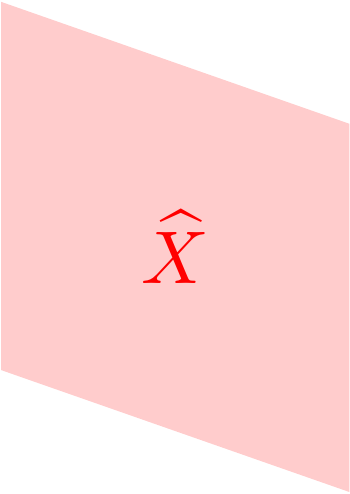}} \hfill \raisebox{-0.5\height}{\includegraphics[scale=0.22]{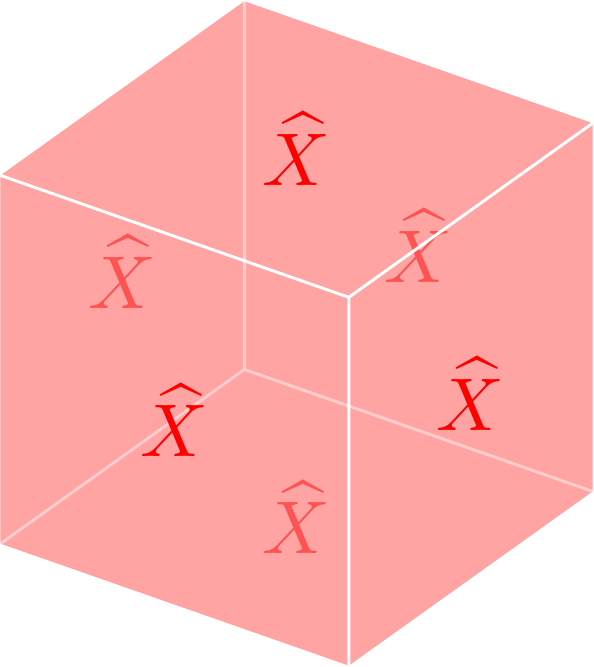}} \hfill ~
\\~\\
\hfill ~~~~~~(a)~~~~~~~~ \hfill ~~(b)~ \hfill ~~~(c)~ \hfill \hfill
\caption{The local terms and the Gauss law of the gauged Hamiltonian \eqref{gaugedH} in the new variables. (a) The dual magnetic flux term on a link (second term of \eqref{dualH}), (b) the dual electric field term on a plaquette (first term of \eqref{dualH}), and (c) the dual Gauss law operator on a cube (third term of \eqref{dualH}).\label{fig:dualHterms}}
\end{figure}

Next, we define a new set of gauge-invariant local operators that commute with the Gauss law $G_\ell$:
\ie
\widehat X_p  = \sigma^x_p \prod_{\ell \in p} Z_\ell~,\qquad
\widehat Z_p  =\sigma^z_p~.
\fe
The gauged Hamiltonian written in terms of the new variables is
\ie\label{dualH}
\widehat H  = -  J \sum_p \widehat X_p -  h \sum_\ell \prod_{p\ni \ell} \widehat Z_p - \widehat g \sum_c \prod_{p\in c }\widehat X_p~.
\fe
(See Figure \ref{fig:dualHterms} for an illustration of these terms.) If we dualize the lattice, this becomes the Hamiltonian of a lattice $\bZ_2$  gauge field theory. The last term is interpreted as the Gauss law for the hatted gauge fields imposed dynamically. 
The gauged Hamiltonian has a dual, non-topological 1-form symmetry, generated by
\ie
\widehat \eta(\Sigma) = \prod_{p\in \Sigma} \widehat X_p ~,
\fe
where $\Sigma$ is a closed 2-cycle of the original lattice. 
 Note that this dual symmetry exists whether we impose the Gauss law \eqref{gauss} for the 1-form gauge field exactly or energetically.

Finally, we consider the limit $\widehat g\to \infty $, which sets  the 2-form gauge fields to be topological. 
Furthermore, the dual 1-form symmetry $\widehat \eta$ becomes topological, i.e., it depends on $\Sigma$ only through its homological class.  
We refer to the gauging procedure with $\widehat g\to\infty$ as the \textit{topological gauging} of the 1-form symmetry.

In this limit, we retrieve the original Hamiltonian on the dual lattice before gauging with $J\leftrightarrow h$. 
We conclude that the $\mathbb{Z}_2$ lattice gauge theory is self-dual under the topological gauging when $J=h$.  
This is the Wegner duality for the 3+1d $\mathbb{Z}_2$ lattice gauge theory \cite{Wegner:1971app}, generalizing the Kramers-Wannier duality for the 1+1d Ising lattice model.

To summarize, we see that the topological gauging of the 1-form symmetry maps the local terms in the Hamiltonian as
\ie\label{Wegner-duality}
&X_\ell \rightsquigarrow \prod_{p\ni \ell} \widehat Z_p~,\qquad\prod_{\ell\in p}Z_\ell \rightsquigarrow \widehat X_p ~,
\fe
which is the Wegner duality transformation.

\section{Non-invertible Wegner operator in 3+1d}\label{sec:3d-noninv}

When $J=h$, the Hamiltonian \eqref{tH} for the lattice $\bZ_2$ gauge theory:
\ie\label{tHSD}
\widetilde H =   -  J  \left( \sum_p \prod_{\ell \in p}  Z_\ell    +  \sum_\ell X_\ell \right)~, 
\fe
 is invariant under the Wegner duality transformation \eqref{Wegner-duality}, 
\ie\label{transformation}
X_\ell \rightsquigarrow  \prod_{\ell' \in \mathfrak t(\ell)} Z_{\ell'}~,\qquad
\prod_{\ell \in p} Z_\ell  \rightsquigarrow X_{\mathfrak{t}(p)}~,
\fe
up to a half lattice translation $\mathfrak{t}$:
\ie
\mathfrak{t} (x,y,z) = (x+\tfrac12 ,y+\tfrac12, z+\tfrac12)~,
\fe
where $(x,y,z)\in (\mathbb{Z}/L_x\mathbb{Z} \, , \, \mathbb{Z}/L_y\mathbb{Z} \, ,\,  \mathbb{Z}/L_z\mathbb{Z})$ labels a site on the lattice, and so on.

However, this transformation \eqref{transformation} cannot be implemented by an invertible operator on a closed periodic torus. Suppose there were such an invertible operator $U$ such that $UX_\ell U^{-1}  =  \prod_{\ell' \in \mathfrak t(\ell)} Z_{\ell'}$. Then consider the conjugation of the 1-form symmetry operator $\eta(\widehat \Sigma)$ by $U$:
 \ie
 U \eta(\widehat \Sigma) U^{-1} 
 =
U \left( \prod_{\ell\in \widehat \Sigma} X_\ell \right) U^{-1}
 = \prod_{\ell \in \widehat \Sigma} \prod_{\ell' \in \mathfrak t(\ell)} Z_{\ell'} =1~.
 \fe
 This implies the 1-form symmetry operator $\eta$ is trivial, which is a contradiction.

\begin{figure}
\centering
$\mathsf D$ \raisebox{-0.5\height}{\includegraphics[scale=0.22]{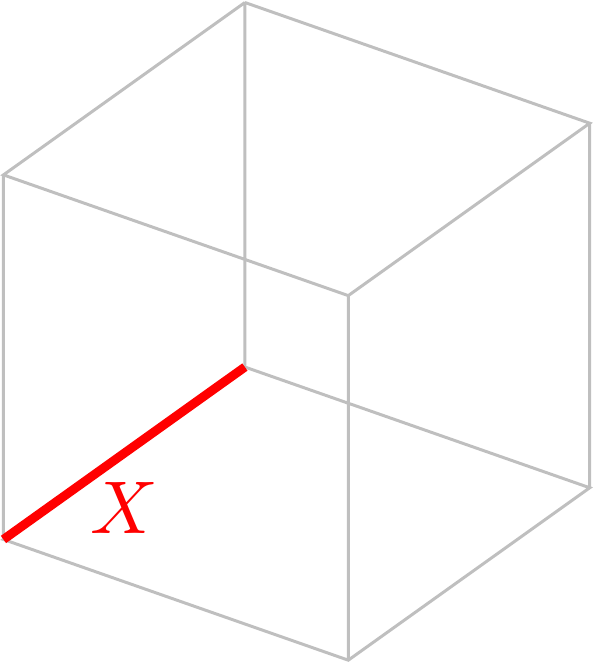}} $=$ \raisebox{-0.5\height}{\includegraphics[scale=0.22]{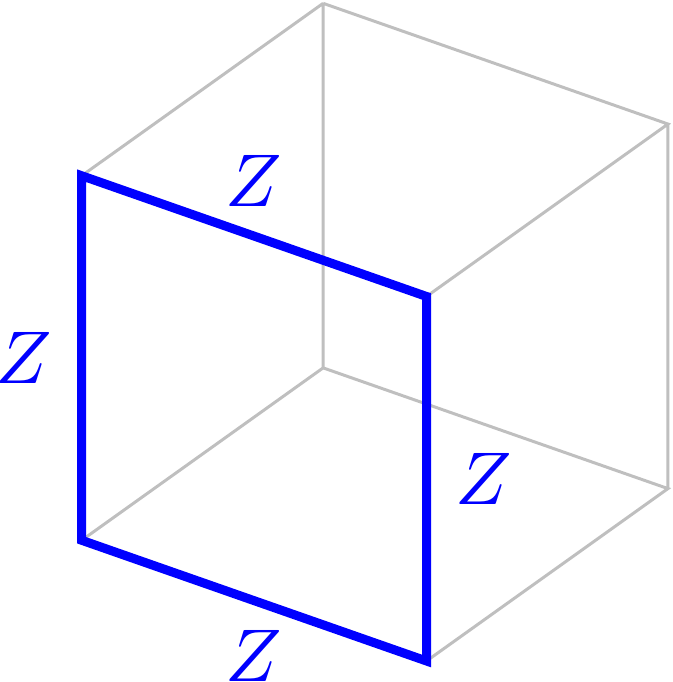}} $\mathsf D~,\quad\qquad \mathsf D$ \raisebox{-0.5\height}{\includegraphics[scale=0.22]{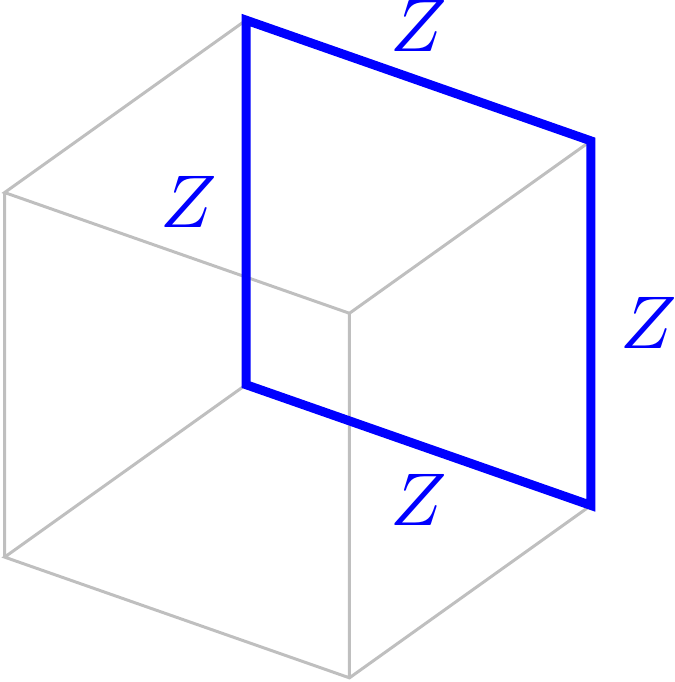}} $=$ \raisebox{-0.5\height}{\includegraphics[scale=0.22]{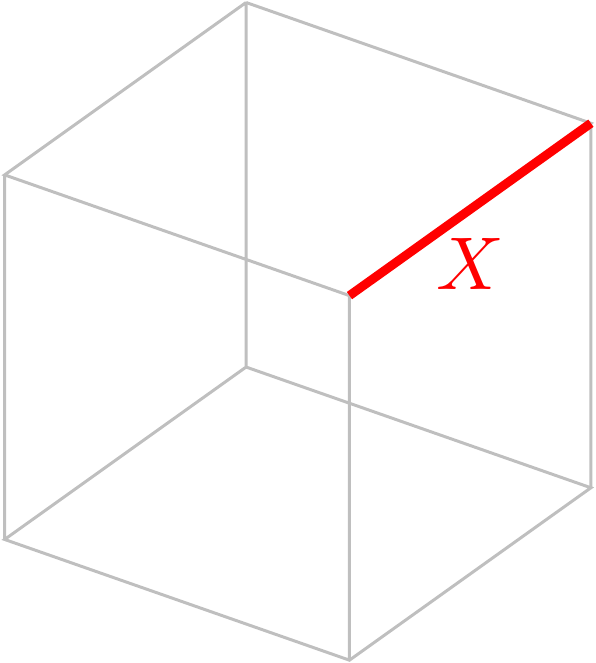}} $\mathsf D~.$
\caption{Action of the duality operator $\mathsf D$ on the terms of the Hamiltonian $\widetilde H$.\label{fig:D-action}}
\end{figure}

Rather, the transformation \eqref{transformation} is implemented by a non-invertible operator $\mathsf D$ in the following sense
\ie\label{D-action}
\mathsf D X_\ell = \left( \prod_{\ell' \in \mathfrak t(\ell)} Z_{\ell'} \right) \mathsf D~,\qquad \mathsf D \prod_{\ell \in p} Z_\ell = X_{\mathfrak t(p)} \mathsf D~.
\fe
See Figure \ref{fig:D-action} for an illustration of the action of $\mathsf D$ on the terms of the Hamiltonian \eqref{tH}.
It follows from \eqref{D-action} that
\ie\label{D-eta,D-G}
 \mathsf D \eta(\widehat \Sigma) = \eta(\widehat \Sigma) \mathsf D = \mathsf D~,
\fe
which in particular implies $\mathsf D G_s = G_s \mathsf D =\mathsf D$. 
This means that the operator $\mathsf D$ acts within the constrained Hilbert space $\widetilde {\mathcal H}$, i.e., the subspace of $ {\cal H}$ with $G_s=1$.

From these relations, for $J=h$, we see that this operator $\mathsf D$ commutes with the Hamiltonian $H$ in \eqref{H}:
\ie
H =   -  J  \left( \sum_p \prod_{\ell \in p}  Z_\ell    +  \sum_\ell X_\ell \right) 
 -  g\sum_s \prod_{\ell\ni s}X_\ell
\fe
 even when the Gauss law is only imposed energetically. 
That is,  $\mathsf D H  =H \mathsf D$.
We will henceforth view $\mathsf D$ as an operator acting on the tensor product Hilbert space ${\cal H} = \bigotimes_\ell {\cal H}_\ell$.

Moreover, we have
\ie\label{D^2-action}
\mathsf D^2 X_\ell = X_{\mathfrak t^2(\ell)} \mathsf D^2~,\qquad \mathsf D^2 \prod_{\ell \in p} Z_\ell = \left(\prod_{\ell \in \mathfrak t^2(p)} Z_\ell \right)\mathsf D^2~.
\fe
The algebra \eqref{D-eta,D-G} and \eqref{D^2-action} suggests that $\mathsf D^2$ is proportional to a projection onto the Gauss-law-preserving, $\mathbb Z_2^{(1)}$-symmetric states times $T_{1,1,1}$, which is the generator of translation in the $(1,1,1)$ direction.

In the following subsections, we will make the last statement more precise. Specifically, we give an explicit realization of the duality operator $\mathsf D$ in terms of ZX-diagrams, and compute its fusion algebra with itself and other operators using ZX-calculus.

\subsection{Non-invertible duality symmetry as a tensor network operator}

In this subsection, we give a ZX-diagrammatic representation of the duality operator $\mathsf D$. 
Our operator $\mathsf D$ in the 3+1d Hamiltonian lattice gauge theory is the counterpart of the duality defect for the 4d Euclidean lattice gauge theory of \cite{Koide:2021zxj}.

First, we define an auxiliary Hilbert space on the plaquettes $\mathcal H_\text{p}$ in the same way as the Hilbert space on the links defined above, which we now denote as $\mathcal H_\text{l}$ for clarity. Consider the following maps between $\mathcal H_\text{l}$ and $\mathcal H_\text{p}$:
\ie\label{3d-duality-map}
\mathsf D_{\text{p}\leftarrow\text{l}} := 2^{4V}~\raisebox{-0.5\height}{\includegraphics[scale=0.17]{Dp-l}}~,\qquad
\mathsf D_{\text{l}\leftarrow\text{p}} := 2^{4V}~\raisebox{-0.5\height}{\includegraphics[scale=0.17]{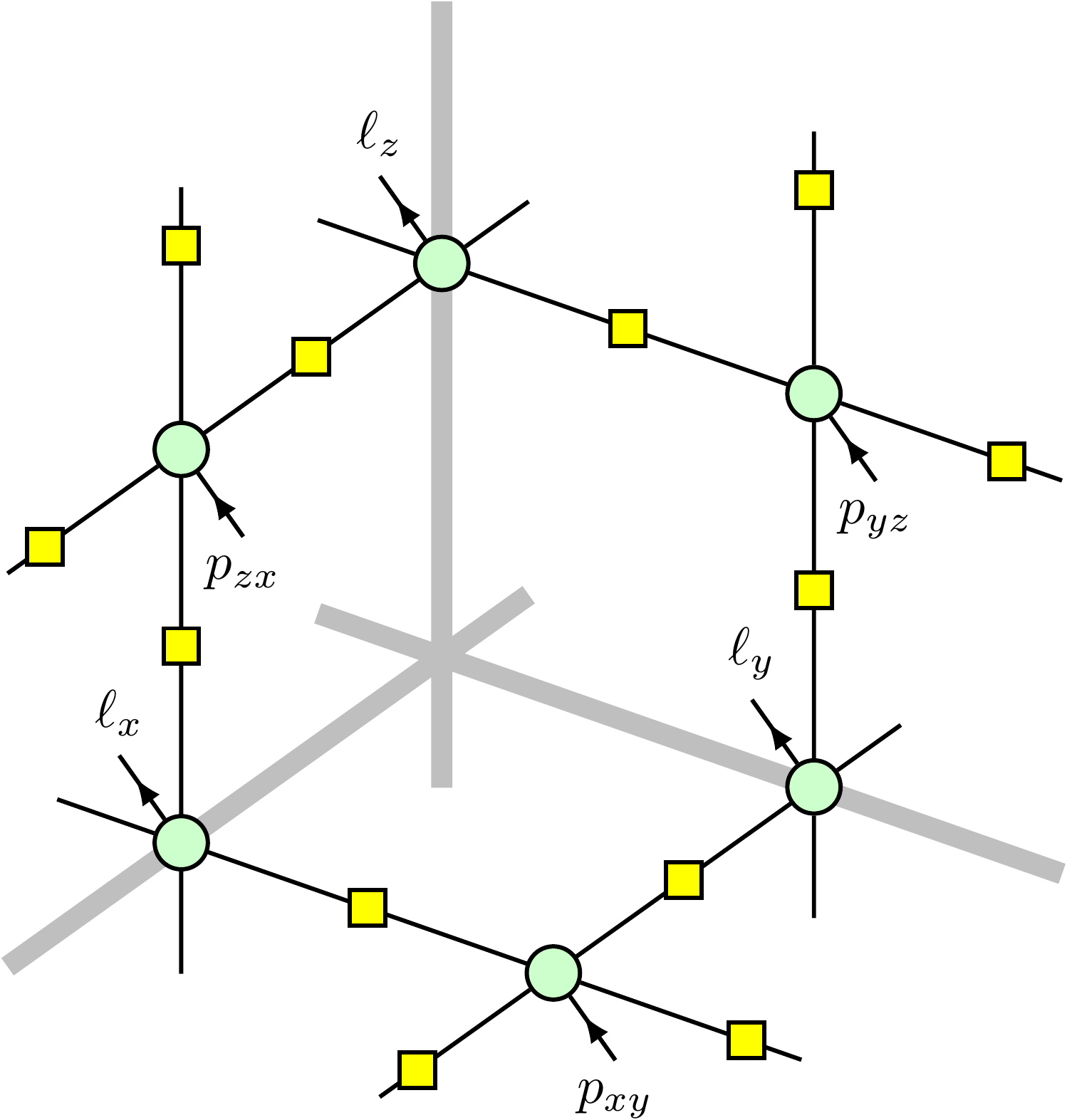}}~,
\fe
where $V=L_x L_y L_z$ is the number of unit cells in the lattice, the ingoing (outgoing) arrows represent inputs (outputs), the thick gray lines represent the links of the cubic lattice (in particular, the green dots are not the sites of the lattice), and the diagram is repeated and glued in a periodic way in the three spatial directions. These maps  have been considered in \cite{Yoshida:2015cia}.

We also define the \emph{half-translation} map $T_{\text{p}\leftarrow\text{l}}$ which maps a qubit in $\mathcal H_\text{l}$ on the link $\ell$ to a qubit in $\mathcal H_\text{p}$ on the plaquette $\mathfrak t(\ell)$. The other half-translation map $T_{\text{l}\leftarrow\text{p}}$ is defined similarly.

With these preparations, the duality operator is defined as
\ie\label{D-def}
\mathsf D := T_{\text{l}\leftarrow\text{p}}\mathsf D_{\text{p}\leftarrow\text{l}} = \mathsf D_{\text{l}\leftarrow\text{p}}T_{\text{p}\leftarrow\text{l}}~.
\fe
We can also write the translation generator in the $(1,1,1)$ direction as
\ie\label{T-def}
T_{1,1,1} = T_{\text{l}\leftarrow\text{p}}T_{\text{p}\leftarrow\text{l}}~.
\fe
Similarly, there are auxiliary versions of these operators that act on $\mathcal H_\text{p}$:
\ie\label{D-plaq}
\widehat T_{1,1,1} = T_{\text{p}\leftarrow\text{l}}T_{\text{l}\leftarrow\text{p}}~,\qquad \widehat{\mathsf D} := T_{\text{p}\leftarrow\text{l}} \mathsf D_{\text{l}\leftarrow\text{p}} = \mathsf D_{\text{p}\leftarrow\text{l}} T_{\text{l}\leftarrow\text{p}}~.
\fe

Equation \eqref{D-def} expresses the Wegner duality operator as a \textit{tensor network operator}, also known as the \textit{projected entangled pair operator} (PEPO) \cite{Cirac:2020obd}. This is a higher dimensional generalization of the MPO presentation of the 1+1d Kramers-Wannier operator.\footnote{The relation between $\mathsf{D}$ and quantum cellular automata will be discussed in \cite{QCA}.}

\subsection{Action on operators}
Let us show that $\mathsf D$ defined in \eqref{D-def} is indeed the duality operator, i.e., it satisfies \eqref{D-action}, using ZX-calculus. It turns out to be more convenient to consider the action of $\mathsf D$ on a more general operator
\ie
\prod_\ell X_\ell^{m_\ell} = \raisebox{-0.5\height}{\includegraphics[scale=0.17]{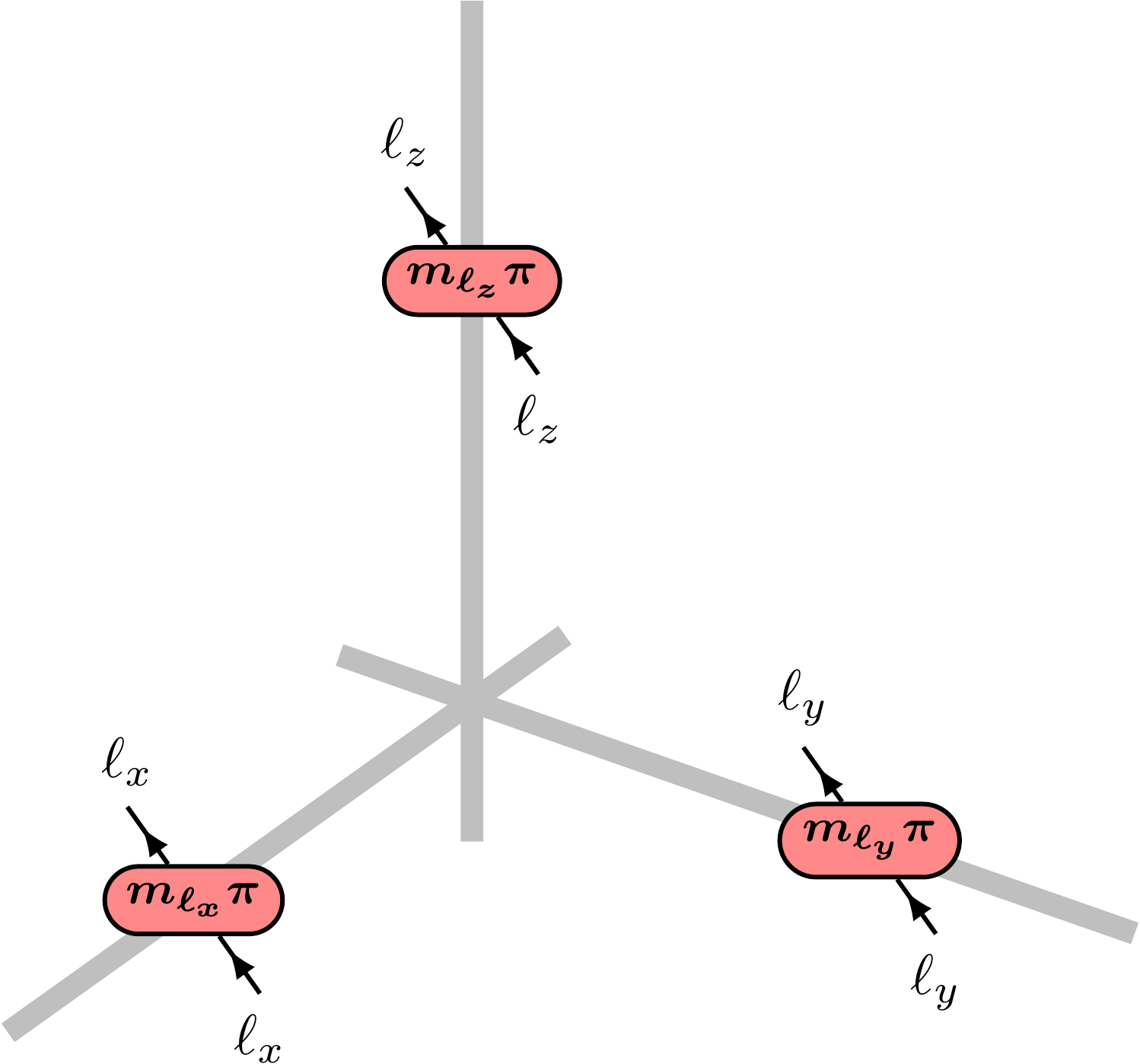}}~,\qquad m_\ell \in \{0,1\}~.
\fe
We have
\ie
\mathsf D_{\text{p}\leftarrow\text{l}} \prod_\ell X_\ell^{m_\ell} = 2^{4V}~\raisebox{-0.5\height}{\includegraphics[scale=0.17]{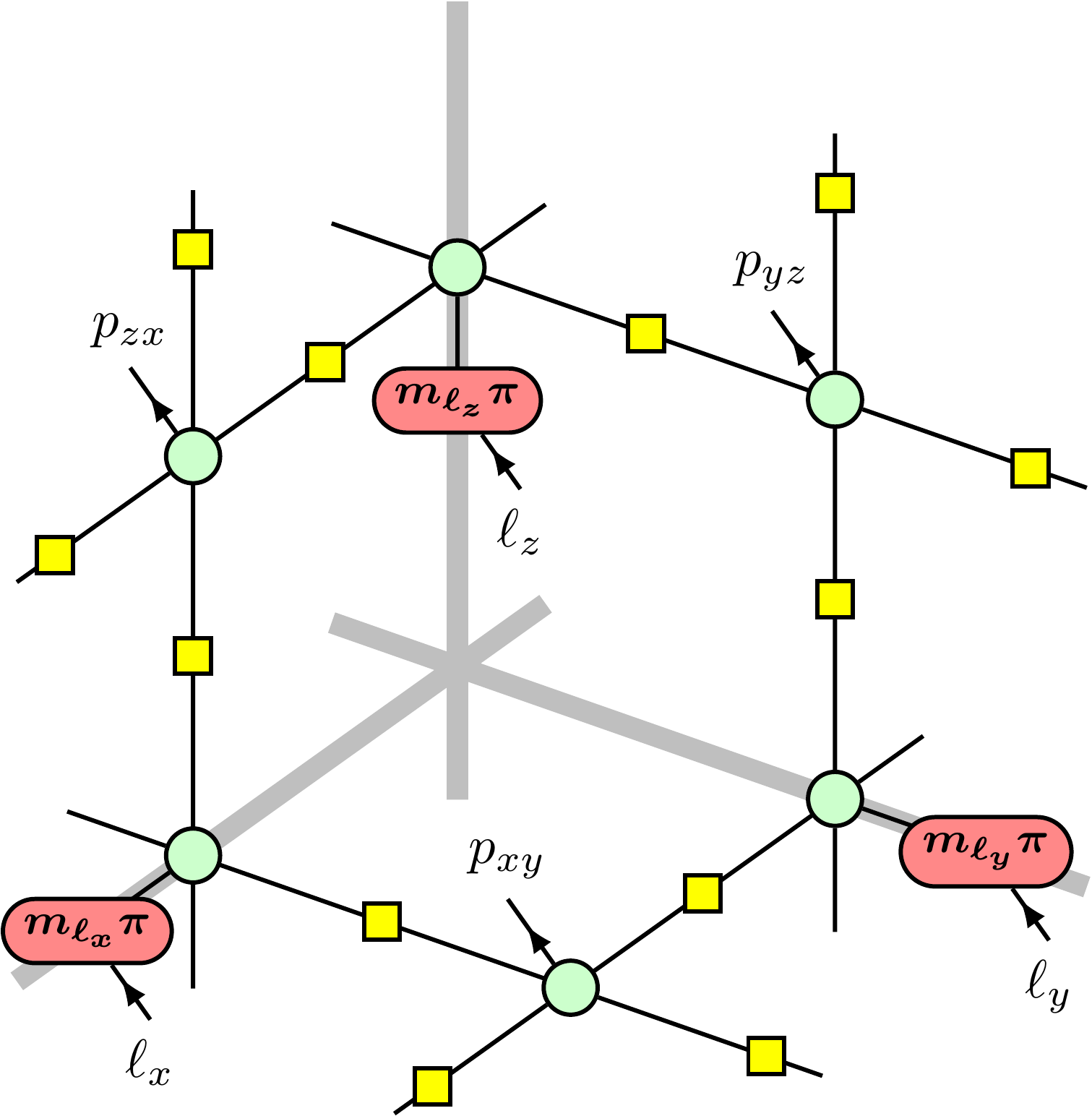}}~.
\fe
Using $\pi$ commutation \ref{pi} through the green dots on the links or identity removal \ref{i}, followed by color change \ref{cc'}, and finally spider fusion \ref{sf} through the green dots on the plaquettes, we get
\ie
\mathsf D_{\text{p}\leftarrow\text{l}} \prod_\ell X_\ell^{m_\ell} = 2^{4V}~\raisebox{-0.5\height}{\includegraphics[scale=0.17]{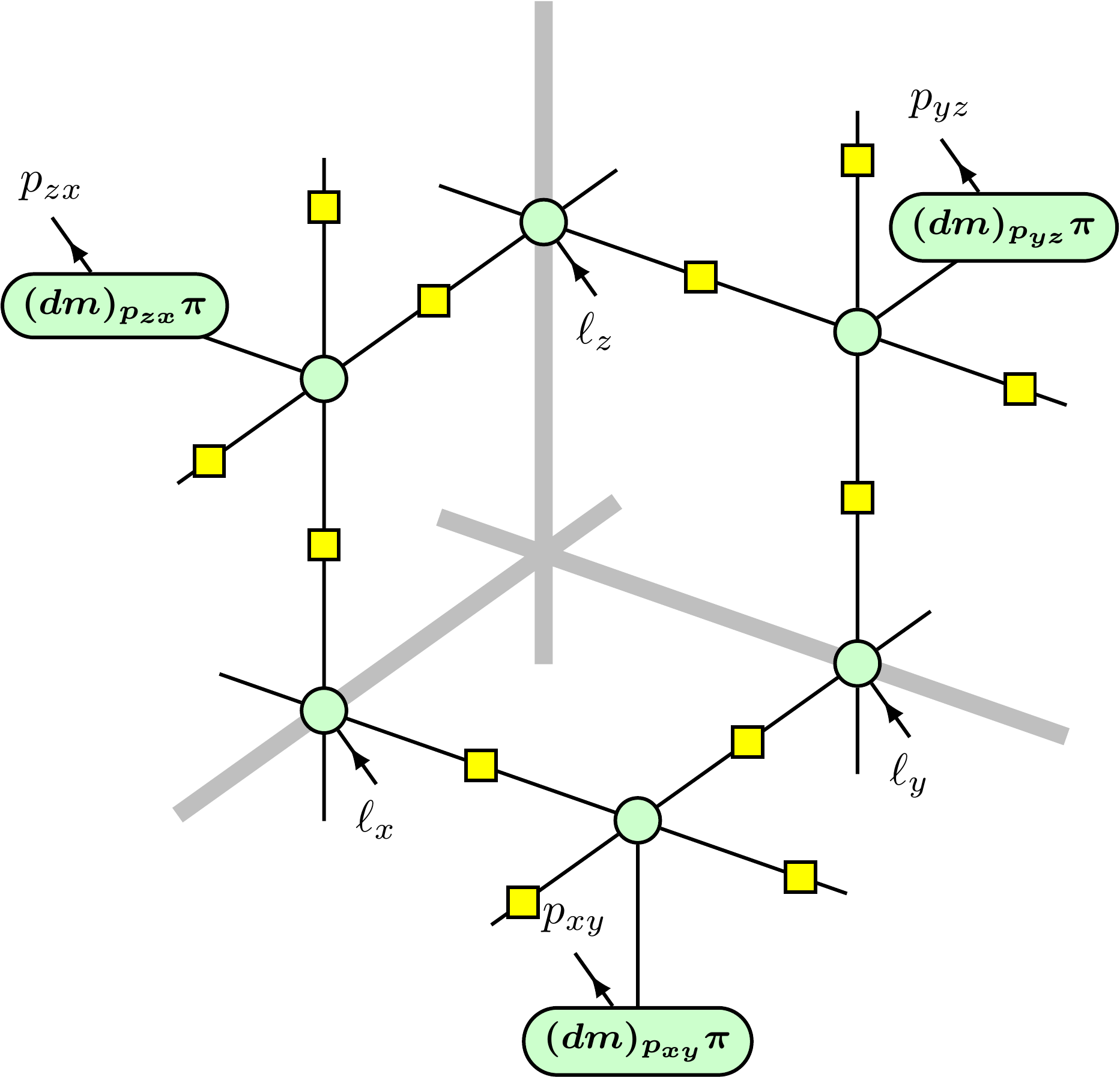}} = \prod_p \widehat Z_p^{(dm)_p} \mathsf D_{\text{p}\leftarrow\text{l}}~,
\fe
where $(dm)_p := \sum_{\ell \in p} m_\ell$ is the sum of $m_\ell$ over the links $\ell$ around the plaquette $p$ and $\widehat Z_p$ is a Pauli operator on the auxiliary Hilbert space $\mathcal H_\text{p}$. Multiplying $T_{\text{l}\leftarrow\text{p}}$ on the left on both sides of this equation, we get
\ie\label{DprodX=prodZD}
\mathsf D \prod_\ell X_\ell^{m_\ell} = T_{\text{l}\leftarrow\text{p}} \prod_p \widehat Z_p^{(dm)_p} \mathsf D_{\text{p}\leftarrow\text{l}} = \prod_{\ell'} Z_{\ell'}^{(dm)_{\mathfrak t^{-1}(\ell')}} \mathsf D = \prod_\ell \left(\prod_{\ell' \in \mathfrak t(\ell)}Z_{\ell'}\right)^{m_\ell} \mathsf D~,
\fe
where, in the second equality, we used $T_{\text{l}\leftarrow\text{p}} \widehat Z_p = Z_{\mathfrak t(p)} T_{\text{l}\leftarrow\text{p}}$, and in the last equality, we used $\ell \in \mathfrak t^{-1}(\ell') \iff \ell' \in \mathfrak t(\ell)$.\footnote{Recall that $\mathfrak t(p)$ is the link obtained by translating from the plaquette $p$ by $(\tfrac12,\tfrac12,\tfrac12)$, and $\mathfrak t(\ell)$ is the plaquette obtained by translating from the link $\ell$ by $(\tfrac12,\tfrac12,\tfrac12)$.} 
This in particular implies the first equation of \eqref{D-action}. 
Similarly, one can show that $\mathsf D \prod_{\ell \in p} Z_\ell = X_{\mathfrak t(p)} \mathsf D$, or more generally,
\ie\label{DprodZ=prodXD}
\mathsf D \prod_p \left(\prod_{\ell \in p} Z_\ell\right)^{\hat m_p} = \prod_p X_{\mathfrak t(p)}^{\hat m_p}~\mathsf D~.
\fe 
Therefore, $\mathsf D$ is indeed the duality operator which satisfies \eqref{D-action}.

\begin{figure}
\centering
$\mathsf D\quad $ \raisebox{-0.5\height}{\includegraphics[scale=0.22]{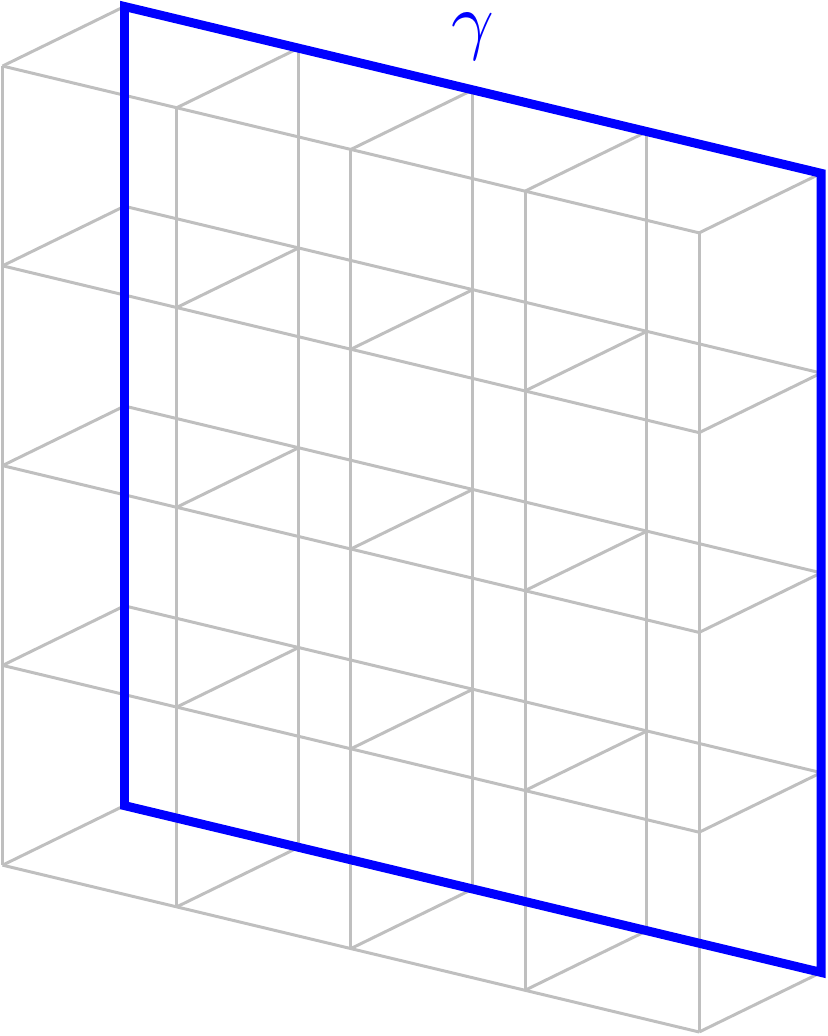}} $\quad =\quad $ \raisebox{-0.5\height}{\includegraphics[scale=0.22]{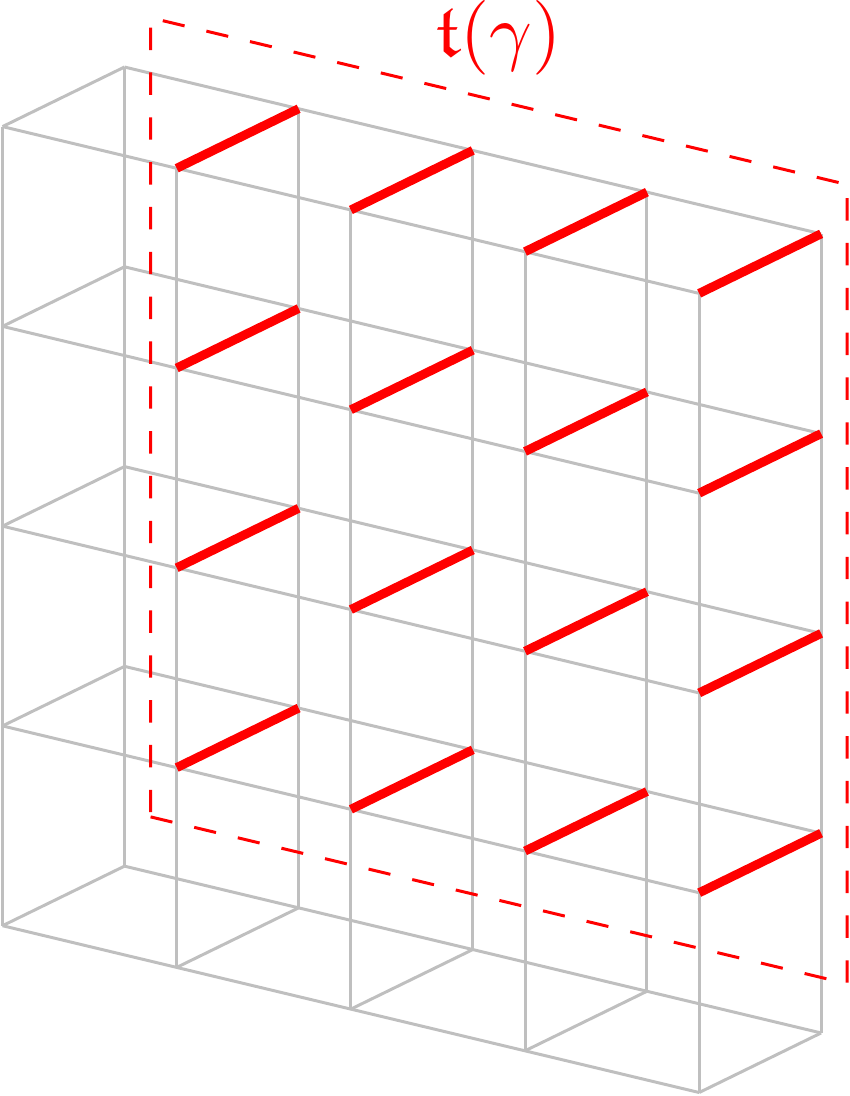}} $\quad \mathsf D$
\caption{The duality operator $\mathsf D$ maps a (contractible) Wilson loop along $\gamma$ to a truncated $\mathbb Z_2$ 1-form symmetry operator on the open dual surface $\mathfrak t(\Sigma)$ with boundary $\mathfrak t(\gamma)$. Here, the blue (red) links correspond to the Pauli $Z$ ($X$) operators.\label{fig:D-action-W}}
\end{figure}

Let us look at some special cases of the action of $\mathsf D$.
\begin{itemize}
\item Let $\gamma$ be a closed contractible curve bounding the open surface $\Sigma$. In the special case where $\hat m_p = 1$ for $p\in \Sigma$ and $0$ otherwise, \eqref{DprodZ=prodXD} reduces to
\ie
\mathsf D \prod_{\ell \in \gamma} Z_\ell = \prod_{\ell \in \mathfrak t(\Sigma)}  X_\ell~\mathsf D~,
\fe
where $\mathfrak t(\Sigma)$ is an open dual surface with boundary $\mathfrak t(\gamma)$. This is illustrated in Figure \ref{fig:D-action-W}. We recognize the operator on the left hand side as the Wilson line operator along the curve $\gamma$\footnote{The Wilson line operator, which is charged under the $\mathbb Z_2$ 1-form symmetry, is well-defined even when $\gamma$ is non-contractible.}
\ie\label{Wilson-op}
W(\gamma) := \prod_{\ell \in \gamma} Z_\ell~,
\fe
whereas the operator on the right hand side is a truncated $\mathbb Z_2$ 1-form symmetry operator along the open dual surface $\mathfrak t(\Sigma)$ with boundary $\mathfrak t(\gamma)$. This equation is, therefore, the lattice counterpart of the continuum result in \cite{Choi:2021kmx,Choi:2022jqy}. In particular, see Figure 1 of \cite{Choi:2022jqy}.

\item In the special case where $m_\ell$ is such that $(dm)_p = 0 \mod 2$ for all $p$, i.e., if $m_\ell$ is a flat $\mathbb Z_2$ gauge field configuration, then \eqref{DprodX=prodZD} reduces to $\mathsf D \prod_\ell X_\ell^{m_\ell} = \mathsf D$. 
This is indeed the case for $G_s$ and $\eta(\widehat \Sigma)$, so we have $\mathsf D G_s = \mathsf D \eta(\widehat \Sigma) = \mathsf D$. 
Similarly, one can show that $G_s \mathsf D = \eta(\widehat \Sigma) \mathsf D = \mathsf D$. 
These reproduce the algebra in \eqref{D-eta,D-G} using ZX-calculus.

\end{itemize}

\subsection{Operator algebra and the condensation operator}

Here we summarize the operator algebra of the Wegner duality operator $\mathsf D$ and the 1-form symmetry operator $\eta$. 
The most interesting one is
\ie\label{D-fusion}
&\mathsf D^2    = {1\over 2^V} \,T_{1,1,1}\sum_{{\widehat \Sigma} } \eta(\widehat \Sigma) ~,
\fe
where the sum is over all (contractible and non-contractible, and not necessarily connected) 2-cycles $\widehat \Sigma$  on the dual lattice. 
It is the 3+1d generalization of $\mathsf D^2 =(1+\eta)T$ \eqref{1+1dD2} found in \cite{Seiberg:2023cdc,Seiberg:2024gek}. 
We derive this algebra in Appendix \ref{app:D2} using ZX-calculus. 

The other equations are\footnote{The first line is obtained as follows: first, the ZX-diagram of $(\mathsf D_{\text{p}\leftarrow\text{l}})^\dagger$ is obtained by reversing the arrows (exchanging inputs/outputs) in the diagram of $\mathsf D_{\text{p}\leftarrow\text{l}}$ (since there are no nontrivial phases in any of the spiders, flipping the signs of the phases does not change the diagram). It is clear that the resulting diagram is the same as that of $\mathsf D_{\text{l}\leftarrow\text{p}}$, so $(\mathsf D_{\text{p}\leftarrow\text{l}})^\dagger = \mathsf D_{\text{l}\leftarrow\text{p}}$. It follows that
\ie
\mathsf D^\dagger = (\mathsf D_{\text{p}\leftarrow\text{l}})^\dagger (T_{\text{l}\leftarrow\text{p}})^\dagger = \mathsf D_{\text{l}\leftarrow\text{p}} (T_{\text{l}\leftarrow\text{p}})^\dagger = \mathsf D_{\text{l}\leftarrow\text{p}} T_{\text{p}\leftarrow\text{l}} (T_{\text{p}\leftarrow\text{l}})^\dagger (T_{\text{l}\leftarrow\text{p}})^\dagger = \mathsf D \, T_{1,1,1}^{-1}~.
\fe
Similarly, it can be shown that $\mathsf D^\dagger = T_{1,1,1}^{-1}\,\mathsf D$.}
\ie
&\mathsf D^\dagger=  \mathsf D \, T_{1,1,1}^{-1} = T_{1,1,1}^{-1}\, \mathsf D~,\\
&\eta(\widehat\Sigma)^2=1~,\qquad\eta (\widehat \Sigma) \, \mathsf D = \mathsf D \,\eta(\widehat \Sigma) = \mathsf D~.
\fe

It is useful to define the condensation operator:
\ie\label{C-sum}
{\mathsf C}= {1\over 2^V} \sum_{{\widehat \Sigma} } \eta(\widehat \Sigma)~.
\fe
We can decompose the sum over all 2-cycles into a sum over the non-contractible 2-cycles times a sum over all contractible surfaces:
\ie\label{sumofeta}
{1\over 2^V} \sum_{\widehat \Sigma} \eta(\widehat \Sigma)
= \frac12 \prod_{i<j}(1+\eta_{ij})
\prod_{s\neq s_0} \left(\frac{1+G_s}{2}\right)~,
\fe 
where $s_0$ is an arbitrary site and $G_s=\prod_{\ell\ni s}X_\ell$. Here, $\eta_{ij}$ is the 1-form symmetry operator on the $ij$-plane, defined as
\ie
\eta_{ij} =  \eta(\widehat \Sigma_{ij}) = \prod_{\ell \in \widehat \Sigma_{ij}} X_\ell~,
\fe
Using $\prod_s G_s=1$, we can include the factor $1+G_{s_0}\over2$ for free to obtain
\ie\label{C-Pauli-def}
\mathsf C = \frac12 
(1+\eta_{xy})(1+\eta_{yz})(1+\eta_{zx})
\prod_{s} \left(\frac{1+G_s}{2}\right) 
~.
\fe
In Appendix \ref{app:condensation}, we further show that $\mathsf C$ can be diagrammatically represented as
\ie\label{C-def}
\mathsf C = 
{2^{7V/2}}~\raisebox{-0.5\height}{\includegraphics[scale=0.17]{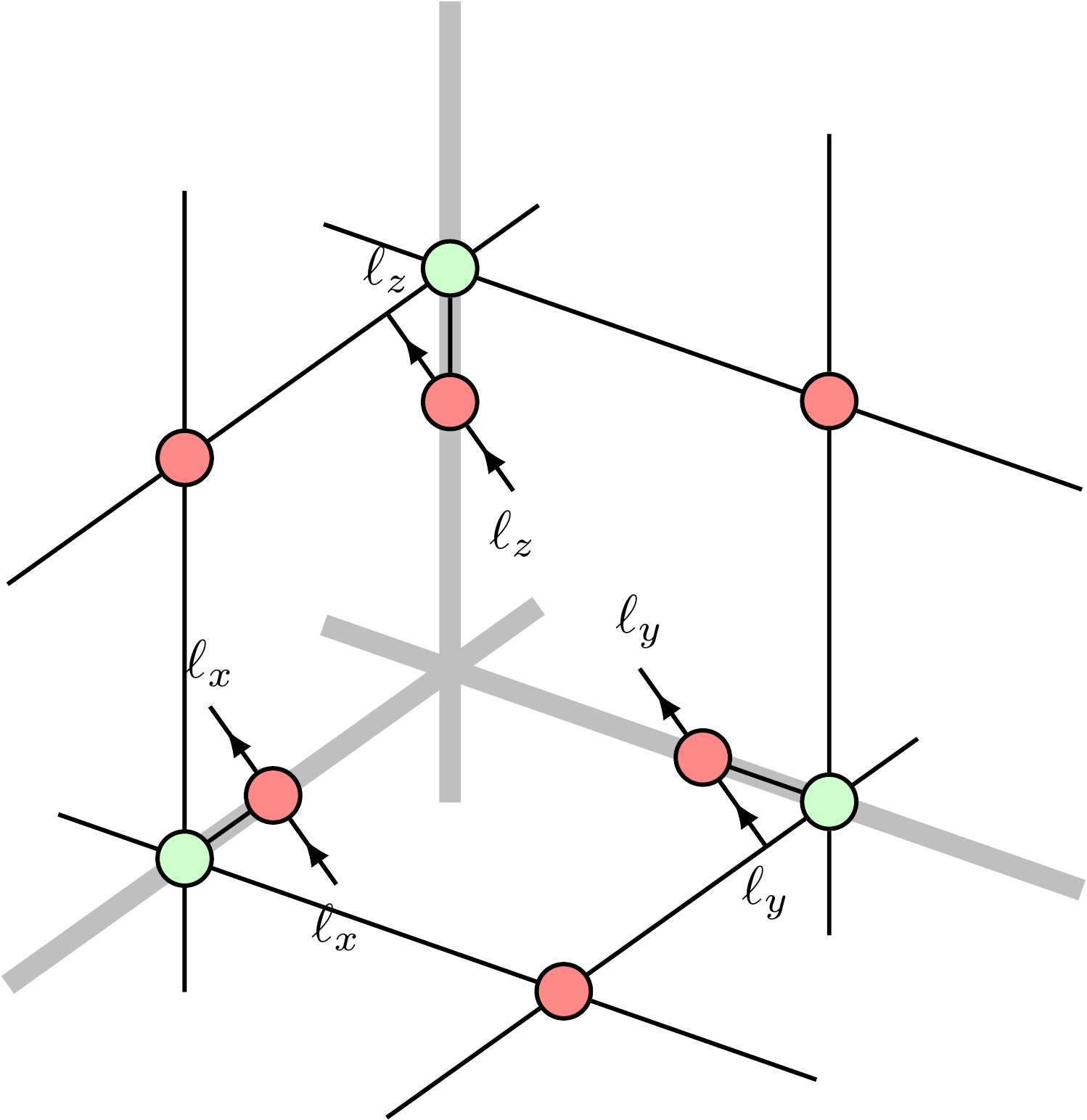}} ~.
\fe

In terms of the condensation operator, \eqref{D-fusion} can be rewritten as
\ie\label{DD2}
&\mathsf D^2 =\mathsf C \,T_{1,1,1}~.
\fe
Using \eqref{D-eta,D-G} and \eqref{C-Pauli-def}, we also find
\ie\label{C-fusion}
 \mathsf C^2 = 4\mathsf C~,\qquad\mathsf D \mathsf C =\mathsf C \mathsf D = 4\mathsf D~.
\fe
Furthermore, 
\ie
\mathsf C \, \eta(\widehat\Sigma) = \eta(\widehat\Sigma) \, \mathsf C = \mathsf C~,
\fe
 follows immediately from writing $\mathsf C = \mathsf D^2 T_{1,1,1}^{-1}$ and \eqref{D-eta,D-G}. (Here, we have used the fact that $T_{1,1,1}$ commutes with all the operators.) 
This in particular implies $\mathsf C G_s  = G_s \mathsf C = \mathsf C$.

Let us now compare the lattice algebra with the continuum one. 
In the thermodynamic limit, $T_{1,1,1}\sim 1$ becomes the identity operator on the low-lying states, and \eqref{D-fusion} reduces to the algebra of the duality operator $\mathcal D(M)$ in continuum field theory\cite{Choi:2021kmx,Kaidi:2021xfk,Choi:2022zal}:\footnote{The continuum duality operator is self-adjoint, ${\cal D}(M)= {\cal D}(M)^\dagger $, but on the lattice $\mathsf D^\dagger$ and $\mathsf D$ differ by a lattice translation $T_{1,1,1}$. In the continuum, the duality operator associated with a $\bZ_N^{(1)}$ symmetry with $N>2$ is not self-adjoint; $\cal D$ and ${\cal D}^\dagger$ differ by a charge conjugation.}
\ie\label{contDD}
\mathcal D(M) ^2= {1\over |H^0 (M; \mathbb{Z}_2)| }
\sum_{\Sigma\in H_2(M;\mathbb{Z}_2) } \eta(\Sigma) ~.
\fe
Here, $M$ is the spatial 3-manifold, which in our case is $T^3$.

Next, we compare the condensation operators in the continuum and on the lattice. 
When restricted to the constrained Hilbert space $\widetilde{\mathcal H}$ where the Gauss law is strictly enforced $G_s=1$, the lattice condensation operator \eqref{C-Pauli-def} reduces to 
 $ \frac12 (1+\eta_{xy})(1+\eta_{yz})(1+\eta_{zx})$. 
  This  agrees with the definition of the condensation operator  $\mathcal C(M)$ in  continuum field theory  \cite{Choi:2022zal} (see also \cite{Roumpedakis:2022aik})
\ie\label{contC}
\mathcal C (M) = {1\over |H^0 (M;\mathbb{Z}_2)| }
\sum_{\Sigma\in H_2(M;\mathbb{Z}_2)} \eta(\Sigma)~.
\fe

Finally, the full algebra of the condensation and duality operators in the continuum is \cite{Choi:2022zal} 
\ie\label{contDD2}
{\cal D}(M)^2 = {\cal C}(M)~,\qquad  {\cal C}(M)^2=  ({\cal Z}_2)_0 \, {\cal C}(M)~,\\
{\cal D}(M) \times {\cal C}(M) = {\cal C}(M)\times {\cal D}(M) = ({\cal Z}_2)_0 \, {\cal D}(M)~,
\fe
where $({\cal Z}_2)_0$ stands for the partition function of the untwisted 2+1d $\bZ_2$ gauge theory (i.e., the low energy limit of the 2+1d toric code) on $M$. 
In our case, $M=T^3$, and $({\cal Z}_2)_0$=4, reflecting the four anyon lines. 
This algebra agrees with the lattice operator algebra in \eqref{DD2} and \eqref{C-fusion} in the thermodynamic limit, where the lattice translation $T_{1,1,1}\sim 1$ becomes the identity operator on the low-lying states. 
See \cite{Shao:2023gho} for a review of the duality and condensation operators/defects in continuum field theory.

\subsection{Lattice symmetries}

The presence of the lattice translation $T_{1,1,1}$ in \eqref{D-fusion} leads to an interesting action of lattice symmetries on the duality operator $\mathsf D$. The ZX-diagram of $\mathsf D$ makes the translation invariance manifest, i.e., 
\ie
\mathsf D T_{\b a} = T_{\b a} \mathsf D~,\qquad \b a\in \mathbb Z^3~.
\fe
Similarly, $\mathsf C$ is translationally invariant.

The remaining lattice symmetries correspond to the point group of the cubic lattice,\footnote{Technically, the rotational part of this group exists only when $L_x = L_y = L_z$ or when the lattice is infinite.} which is the octahedral group, denoted as $O_h$. It is isomorphic to $S_4 \times \mathbb Z_2$, where $S_4$ is associated with orientation-preserving symmetries (rotations) and $\mathbb Z_2$ is associated with the reflection (parity). It is easy to see that the ZX-diagrams of the duality maps $\mathsf D_{\text{p}\leftarrow\text{l}}$ and $\mathsf D_{\text{l}\leftarrow\text{p}}$, and the condensation operator $\mathsf C$ are invariant under rotations and reflections. So the nontrivial action on $\mathsf D$ comes from the transformation of the half-translation maps $T_{\text{p}\leftarrow\text{l}}$ and $T_{\text{l}\leftarrow\text{p}}$.

The parity operator $P$ acts as
\ie
P T_{\b a} P = T_{\b a}^\dagger = T_{-\b a}~,\qquad P \mathsf D P = \mathsf D T_{1,1,1}^{-1} = \mathsf D^\dagger~,
\fe
where the latter follows from the fact that the half-translation map $T_{\text{p}\leftarrow\text{l}}$, associated with the half-translation $\mathfrak t$, transforms into $(T_{\text{l}\leftarrow\text{p}})^\dagger = T_{\text{p}\leftarrow\text{l}} T_{1,1,1}^{-1}$, associated with $\mathfrak t^{-1}$, under parity $(x,y,z) \mapsto (-x,-y,-z)$. A similar statement holds for the other half-translation map $T_{\text{l}\leftarrow\text{p}}$.

Let us consider rotations now. It is well-known that $S_4$ acts freely and transitively on the space-diagonals of a cube. It follows that, given a rotation $R$, there is a unique $\b \varepsilon = (\varepsilon_x,\varepsilon_y,\varepsilon_z)\in \{\pm1\}^3$ such that $R T_{1,1,1} R^{-1} = T_{\b \varepsilon}$.\footnote{One can think of $\b \varepsilon$ as labelling the space-diagonal along that direction. Note that $\b \varepsilon$ and $-\b \varepsilon$ label the same space-diagonal.} We define $\mathsf D_{\b \varepsilon} := R \mathsf D R^{-1}$; in particular, $\mathsf D_{1,1,1} = \mathsf D$. Then,
\ie
\mathsf D_{\b \varepsilon}^2 = R \mathsf D^2 R^{-1} = R T_{1,1,1} \mathsf C R^{-1} = T_{\b \varepsilon} \mathsf C~,
\fe
where we used the fact that the ZX-diagram of $\mathsf C$ is rotationally invariant. In fact, $\mathsf D_{\b \varepsilon}$ is related to $\mathsf D$ by a translation,\footnote{The half translation map $T_{\text{p}\leftarrow\text{l}}$ transforms into $T^{\b \varepsilon}_{\text{p}\leftarrow\text{l}}$ under the rotation $R$. Here, $T^{\b \varepsilon}_{\text{p}\leftarrow\text{l}}$ maps the physical qubit on link $\ell$ to the auxiliary qubit on plaquette $\mathfrak t_{\b \epsilon}(\ell)$, where $\mathfrak t_{\b \epsilon}(\b x) = \b x + \frac12 \b \varepsilon$ is the half-translation in the $\b \varepsilon$ direction. The desired relation then follows from the simple observation $T^{\b \varepsilon}_{\text{p}\leftarrow\text{l}} = T_{\text{p}\leftarrow\text{l}} T_{\frac12(\b \varepsilon-\b 1)}$. Similar statements apply to the other half-translation map $T_{\text{l}\leftarrow\text{p}}$.}
\ie
\mathsf D_{\b \varepsilon} = \mathsf D T_{\frac12(\b \varepsilon-\b 1)}~,
\fe
where $\b 1 = (1,1,1)$. In particular, $\mathsf D_{-1,-1,-1} = \mathsf D^\dagger$.

Consider the special case where the rotation $R$ fixes the space diagonal along $(1,1,1)$---such rotations form the $S_3$ subgroup of $S_4$. This can happen in two ways:
\begin{enumerate}
\item $R$ is the $2\pi/3$-rotation about the $(1,1,1)$ direction, which generates the $\mathbb Z_3$ subgroup of the stabilizer group $S_3$. Then, $R T_{1,1,1} R^{-1} = T_{1,1,1}$ and $R \mathsf D R^{-1} = \mathsf D$.
\item $R$ is the $\pi$-rotation about $(1,-1,0)$, $(0,1,-1)$, or $(-1,0,1)$ direction. Then, $R T_{1,1,1} R^{-1} = T_{-1,-1,-1}$ and $R \mathsf D R^{-1} = \mathsf D^\dagger$.
\end{enumerate}
Note that $\mathsf D$ is mapped to $\mathsf D^\dagger$ by both the parity $P$, which is an orientation-reversing operation, and the rotation $R$ of the second-type above, which is an orientation-preserving operation. This is the lattice counterpart of the continuum fact that $\mathcal D(M)$ is the same as its orientation reversal.

\subsection{Action on the toric code ground states}\label{sec:3d-groundstates}

The phase diagram of the lattice $\mathbb Z_2$ gauge theory \eqref{tH} is well understood \cite{PhysRevD.20.1915}: the $\mathbb Z_2$ 1-form symmetry is spontaneously broken (Higgs phase) for $J>h$, whereas it is preserved (confining phase) for $J<h$. The broken phase is captured by the 3+1d topological $\mathbb Z_2$ gauge theory at low energies. At the self-dual point $J=h$, there is a first order phase transition.
 
Let us say a few words about the ground states in these phases. In the limit $J \gg h$ (or when $h=0$), the magnetic flux term dominates. On a cubic lattice with periodic boundary conditions, this leads to $2^3 = 8$ ground states,
\ie\label{TC}
|\xi\> := 2^{(V-1)/2}\prod_{i<j} \eta_{ij}^{\xi_{ij}} \prod_{s} \left(\frac{1+G_s}{2}\right) |0\cdots0\>~,
\fe
where $\xi = (\xi_{xy},\xi_{yz},\xi_{zx}) \in \{0,1\}^3$ 
 with $\xi_{ij}$ defined modulo 2. 
We recognize that $|\xi\>$ are   the ground states of the 3+1d toric code.\footnote{Indeed, starting with the Hamiltonian $H$ \eqref{H}, which has the same ground states as $\tilde H$ \eqref{tH} for $g>0$, and setting $h=0$ gives the Hamiltonian of the 3+1d toric code. The normalization of $|\xi\>$ is chosen such that $\< \xi | \xi' \> = \delta_{\xi,\xi'}$.} 
 On the other hand, in the limit $J\ll h$ (or when $J=0$), the electric field term dominates, so there is a unique ground state,
\ie\label{plus2}
|{+}\cdots{+}\>~.
\fe
At the self-dual point $J=h$, there are nine ground states in the thermodynamic limit.  
These ground states are not the above nine states, but are related to them by a finite depth local unitary (FDLU).
Incidentally, these eight toric code ground states $\ket{\xi}$ and the product state $\ket{+\cdots+}$ are the exact ground states of the Hamiltonian in \eqref{deformtH} at $\lambda=1$ discussed below.

Irrespective of the model, one can ask how the symmetry operators $\eta_{ij}$ and $\mathsf D$ act on these ground states. 
Clearly, the 1-form symmetries $\eta_{ij}$ permute the toric code ground states among themselves and leave the product state invariant:
\ie
\prod_{i<j}\eta_{ij}^{\xi_{ij}'} |\xi\> = |\xi+\xi' \>~,\qquad \eta_{ij} |{+}\cdots{+}\> = |{+}\cdots{+}\>~.
\fe
More interestingly, the non-invertible symmetry exchanges the toric code ground states with the product state:
\ie\label{Daction}
\mathsf D |\xi\> = \frac1{\sqrt2} |{+}\cdots{+}\>~,\qquad \mathsf D |{+}\cdots{+}\> = \frac1{\sqrt2} \sum_{\xi \in \{0,1\}^3}|\xi\>~.
\fe
We interpret the above equations as the spontaneous breaking of the non-invertible Wegner symmetry $\mathsf D$ and the 1-form symmetry $\eta$.  We derive \eqref{Daction} using ZX-calculus in Appendix \ref{app:D-action-states}.

The basis in  \eqref{TC} and \eqref{plus2} corresponds to the ground states in the nine superselection sectors of the model in \eqref{deformtH} in the thermodynamic limit. 
Alternatively, for $\zeta = (\zeta_{xy},\zeta_{yz},\zeta_{zx}) \in \{0,1\}^3$, we define
\ie
|\zeta\> &:= \frac1{2\sqrt2} \sum_{\xi \in \{0,1\}^3} (-1)^{\sum_{i<j}\zeta_{ij}\xi_{ij}} |\xi\>~.
\fe
Then, the basis that diagonalizes the symmetry operators can be written as\footnote{Comparing with the discussion in Section \ref{sec:1d-groundstates}, the basis $|\xi\>$ is the analog of $|0\cdots0\>, |1\cdots1\>$ in 1+1d, while the basis $|\zeta\>$ is the analog of $|\text{GHZ}^\pm\>$.}
\ie
&\frac1{\sqrt2} |{+}\cdots{+}\> \pm \frac1{\sqrt2} |\zeta=0\>~,\quad &&\mathsf D = \pm 2~,\quad \eta_{ij} = 1~,
\\
&|\zeta\ne0\>~,\quad &&\mathsf D = 0~,\quad \eta_{ij} = (-1)^{\zeta_{ij}}~.
\fe
See Appendix \ref{app:TC} for more discussions on the two bases $|\xi\>$ and $|\zeta\>$ for the toric code ground states.

Using \eqref{C-Pauli-def}, we see that the toric code ground state $|\zeta=0\>$ can be obtained by acting the condensation operator on a product state:\footnote{On the other hand, the state $|\xi=0\>$ can be obtained by acting a condensation operator for a 2-form symmetry on the product state $|{+}\cdots{+}\>$. See Appendix \ref{app:TC}, in particular, \eqref{xi=0-zx}.}
\ie
|\zeta=0\>&
=\frac{2^{V/2}}{2} \mathsf C |0\cdots 0\>~.
\fe
This generalizes the relation $|\text{GHZ}^+\> ={1\over \sqrt{2}} (1+\eta)|0\cdots0\>$ in 1+1d. 
The duality operator $\mathsf{D}$ exchanges this toric code ground state $|\zeta=0\>$ and the product state $|{+}\cdots{+}\>$:
\ie
\mathsf{D} |\zeta=0\> = 2 |{+}\cdots{+}\>~,\qquad
\mathsf{D} |{+}\cdots{+}\>  = 2 |\zeta=0\>~.
\fe
This is the 3+1d analog of the relation \eqref{Dexchange2d} in 1+1d.

\subsection{LSM-type constraint and duality-preserving deformations}

Here we argue that, in the thermodynamic limit $L_i\to\infty$,  any Hamiltonian that commutes with the non-invertible operator $\mathsf D$ cannot be in a trivially gapped phase with a unique ground state without  long range entanglement.
It is either gapless,  gapped with multiple superselection sectors, or gapped with topological order.

This constraint is reminiscent of the Lieb-Schultz-Mattis (LSM) theorem \cite{LSM}, and is a 3+1d generalization of the constraint for the 1+1d Kramers-Wannier symmetry in \cite{Levin:2019ifu,Seiberg:2024gek}. 
Our lattice argument largely follows from   the corresponding theorem in the 3+1d continuum field theory proven in \cite{Choi:2021kmx}, which was interpreted as an 't Hooft anomaly of the non-invertible global symmetry.\footnote{A non-invertible global symmetry is sometimes said to have an 't Hooft anomaly if it is incompatible with a trivially gapped phase \cite{Chang:2018iay,Thorngren:2019iar,Choi:2021kmx} (see also \cite{Choi:2022zal,Choi:2022rfe,Kaidi:2023maf,Zhang:2023wlu,Cordova:2023bja,Antinucci:2023ezl}). See \cite{Komargodski:2020mxz,Choi:2023xjw,Choi:2023vgk} for the relation between this definition of anomalies and the obstruction to gauging.}

Our setup is a 3d spatial cubic lattice with a qubit on each link. 
We assume that the Gauss law $G_s = \prod_{\ell \ni s }X_\ell=1$ is enforced strictly as an operator equation on every site, so that the total Hilbert space is   the constrained Hilbert space $\widetilde{\cal H}$, which is a subspace of the tensor product Hilbert space $\mathcal H$. 
We consider a general translationally invariant Hamiltonian $\widetilde H$ on $\widetilde{\cal H}$ that is invariant under the topological $\bZ_2$ 1-form symmetry $\eta (\widehat\Sigma ) = \prod_{\ell \in\widehat \Sigma} X_\ell$.

We furthermore assume that the Hamiltonian commutes with the non-invertible operator $\mathsf D$, i.e., $\mathsf D{ \widetilde H}  ={\widetilde H} \mathsf D$. 
Up to a lattice translation,  $\mathsf D$ implements the gauging of $\bZ_2^{(1)}$.  (See  \eqref{D-action} and  \eqref{Wegner-duality}.)  
Therefore, any Hamiltonian that commutes with $\mathsf D$ must be invariant under gauging the $\bZ_2^{(1)}$ symmetry.

\begin{figure}
\centering
\raisebox{-0.5\height}{\includegraphics[scale=0.3]{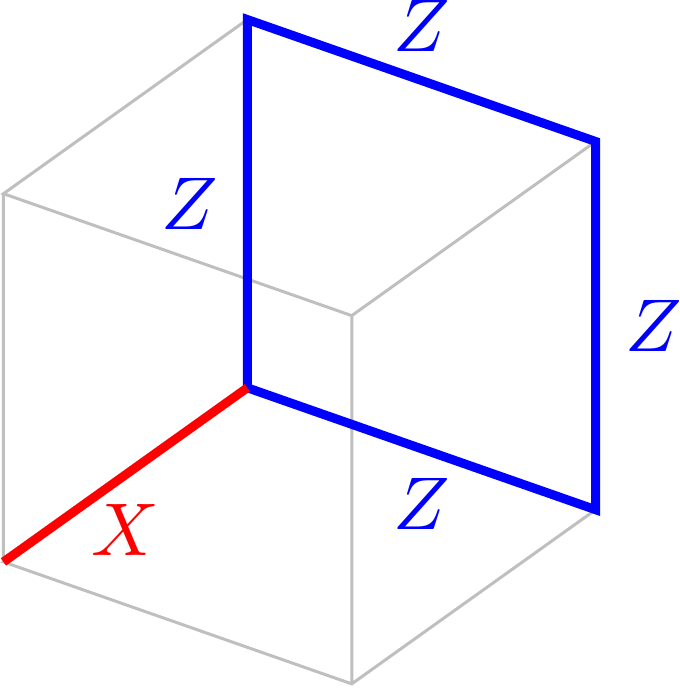}}
\caption{The deformation term in $\widetilde H_\lambda$ \eqref{deformtH}. As indicated, the red (blue) links represent the Pauli $X$ ($Z$) operators on links. We add such a term for every link and plaquette that are orthogonal to each other and meet at a site. For example, each cube contains 24 such terms.\label{fig:deformHterm}}
\end{figure}

As a concrete example, consider a one-parameter deformation of the lattice $\mathbb{Z}_2$ gauge theory Hamiltonian  \eqref{tHSD} \cite{deform}:
\ie\label{deformtH}
\widetilde H_\lambda = -J \left( \sum_p \prod_{\ell\in p}Z_\ell + \sum_\ell X_\ell 
- \frac{\lambda}{8} \sum_{\ell,p: \ell \perp p} X_\ell \prod_{\ell'\in p} Z_{\ell'} \right)~,
\fe
where $\ell\perp p$ (or $p\perp \ell$) means the link $\ell$ is orthogonal to the plaquette $p$ and they meet at a site. See Figure \ref{fig:deformHterm} for an illustration of this deformation. This deformation is analogous to the deformations $X_{i-1} Z_i Z_{i+1}$ and $Z_{i-1} Z_i X_{i+1}$ in the 1+1d model \eqref{1d-deformedHlambda} in \cite{OBrien:2017wmx}. 
For any $\lambda$, this deformed Hamiltonian preserves the $\mathbb Z_2$ 1-form symmetry and the Wegner duality symmetry $\mathsf D$.\footnote{To see this, note that $\ell\perp p \iff \mathfrak t(\ell)\perp \mathfrak t(p)$.}
It will be shown in \cite{deform}  that, when $\lambda = 1$, the deformed model has an exact nine-fold ground state degeneracy even in finite volume.

We are now ready to argue for the main statement. 
Let us assume that the thermodynamic limit of this Hamiltonian is trivially gapped with no long range entanglement. 
Then in the low energy limit it must be described by a bosonic invertible field theory for the  topological  phase protected by a $\bZ_2^{(1)}$ global symmetry. 
However, it was shown in \cite{Choi:2021kmx} that there is no such $\bZ_2^{(1)}$ SPT phase that is invariant under gauging the $\bZ_2^{(1)}$ symmetry.\footnote{Generally, a bosonic 3+1d $\bZ_N^{(1)}$ SPT phase is labeled by  an integer $p$ modulo $2N$ if $N$ is even, and $p$ modulo $N$ if $N$ is odd \cite{Kapustin:2013uxa,Gaiotto:2014kfa}.  Under gauging the $\bZ_N^{(1)}$ symmetry, the phase remains invertible if $\gcd(p,N)=1$. Under this condition, gauging maps $p\to -1/p$ for even $N$, and $p\to -1/4p$ for odd $N$  \cite{Choi:2021kmx}. In particular, when $N=2$, the $p=1$ and $p=3$ SPT phases are exchanged under gauging $\bZ_2^{(1)}$, and there is no bosonic SPT phase invariant under the gauging. In \cite{Apte:2022xtu}, the authors further show that for some values of $N$, there is no fermionic topological quantum field theory (TQFT) (with a unique local operator) that is invariant under gauging $\bZ_N^{(1)}$. However, their result does not exclude the possibility of a bosonic TQFT invariant under gauging $\bZ_2^{(1)}$.}
This proves that the Hamiltonian cannot be in a trivially gapped phase.

What are the gapped phases that are compatible with the non-invertible Wegner symmetry $\mathsf D$? 
A trivially gapped phase is exchanged with a deconfined $\bZ_2$ gauge theory phase under gauging $\bZ_2^{(1)}$, and therefore a first order transition point of these two phases is one such example. 
This phase is realized at the self-dual point $J=h$ for the lattice $\bZ_2$ gauge theory \eqref{tH}, and it has  1+8 superselection sectors on a spatial three-torus in the infinite volume limit. 
It is also realized in the deformed Hamiltonian \eqref{deformtH} at $\lambda=1$, where these nine states are the exact ground states even in finite volume. 
Finally, in the continuum, this phase is realized by the pure SO(3) Yang-Mills gauge theory with $\mathcal{N}=1$ supersymmetry \cite{Aharony:2013hda}, where the anomalous R-symmetry is resurrected as a non-invertible symmetry  \cite{Kaidi:2021xfk}.\footnote{We thank Nathan Seiberg for discussions on this point.}

There can be more general gapped phases compatible with $\mathsf D$, which we leave for future investigations.

\section{Non-invertible symmetries based on graphs}\label{sec:graph}

In this section, we extend the discussion of non-invertible duality operator to a large class of lattice models based on arbitrary bipartite graphs. We refer to them as \emph{generalized transverse-field Ising models} (generalized TFIM) due to their resemblance to the usual transverse-field Ising model. 
(See \cite{Wegner:1971app,Vijay:2016phm,Williamson:2016jiq,Shirley:2018vtc,Tantivasadakarn:2020lhq,Tantivasadakarn:2021wdv,Liu:2022cxc,Rakovszky:2023fng,Rakovszky:2024iks}  for related usages of the term ``generalized Ising model''. See also \cite{Okuda:2024azp} for duality \emph{maps} between different models.)
 We give the condition under which there is a duality symmetry, and derive the condition for which the duality operator does not mix with spatial symmetry. We also discuss several well-known lattice models as examples of this general model.

 \subsection{Hamiltonian of the generalized transverse-field Ising model}

We consider an arbitrary tensor product Hilbert space of qubits. Denote the set of all qubits $V$ such that the total Hilbert space is $\mathcal H := \bigotimes_{v\in V} \mathbb C^2$.\footnote{Throughout this section, we use $V$ to denote a set of vertices which host physical qubits. This is not to be confused with $V=L_xL_yL_z$, the number of sites on the cubic lattice in Section \ref{sec:3d-noninv}.} Note that for the current discussion, we do not need to make any assumption yet about geometric locality.

Consider an additional set $\widehat V$ and define a bipartite graph $\mathcal G$ on $\widehat V \sqcup V$, with edge set $E \subseteq \widehat V\times V$. Without loss of generality, we assume that $\mathcal G$ is connected. Its adjacency matrix takes the form
\ie
\begin{pmatrix}
0_{|\widehat V|\times |\widehat V|} & \sigma_{|\widehat V|\times |V|}\\
\sigma_{|V|\times |\widehat V|}^\intercal & 0_{|V|\times |V|}
\end{pmatrix}~,
\fe
where $\sigma$ is called the biadjacency matrix,\footnote{In the quantum information literature, $\sigma$ and $\mathcal G$ are called the parity check matrix and the corresponding Tanner graph, respectively.} given by $\sigma_{\hat vv} = 1$ if $(\hat v,v)\in E$ and $0$ otherwise, and its incidence matrix takes the form
\ie
\begin{pmatrix}
\hat \tau_{|\widehat V|\times |E|}
\\
\tau_{|V|\times |E|}
\end{pmatrix}~,
\fe
where $\hat \tau_{\hat ve} = 1$ if $e$ is incident to $\hat v$ and $0$ otherwise, and $\tau$ is defined similarly. Note that
\ie\label{sigma-tau}
\sigma = \hat\tau \tau^\intercal~.
\fe
Moreover, the entries in all these matrices are $0$ or $1$.

We may use $\sigma$ to construct a generalized transverse-field Ising Hamiltonian as follows. First, we define generalized Ising terms
\ie
B_{\hat v} = \prod_{v \in V}Z_{v}^{\sigma_{\hat vv}}~,
\fe
for $\hat v\in \widehat V$ (i.e., for each row $\hat v$ of $\sigma$, the positions of the Pauli $Z$'s are given by the entries in that row). Then, we can define the Hamiltonian
\ie
H =  - J \sum_{\hat v \in \widehat V} B_{\hat v} -h \sum_{v\in V} X_v~.
\fe
To summarize, an element of $V$ represents a physical qubit while an element of $\widehat V$ represents a generalized Ising interaction term.

The invertible symmetries of this Hamiltonian are $\mathbb Z_2$ symmetries. Let us define the operator 
\ie
\eta_a = \prod_{v\in V} X_v^{a_v}~,
\fe
 for a vector $a$ with components $a_v \in \{0,1\}$ and see what choices of $a$ give symmetries of this Hamiltonian. For this, we only need to check whether it commutes with the Ising terms. The commutation of a Z-type and an X-type Pauli operators is $ \prod_w Z_w^{b_w} \prod_v X_v^{a_v} = (-1)^{\sum_v a_v b_v}\prod_w Z_w^{b_w} \prod_v X_v^{a_v}$. Therefore, the commutation between $B_{\hat v}$ and $\eta_a$ is $(-1)^{\sum_v \sigma_{\hat vv}a_v}$, and we see that $\eta_a$ is a symmetry of the Hamiltonian exactly when $a \in \ker \sigma$. To conclude, the total invertible symmetry group is $\mathbb Z_2^{\dim \ker \sigma}$.

\subsection{Spatial symmetries from preserving automorphisms}

A spatial symmetry  of the generalized TFIM on the bipartite graph $\mathcal G$ is associated with  a \textit{preserving automorphism} $\pi$ of $\mathcal G$,\footnote{An \emph{automorphism} $\alpha$ of a graph is a bijection from its vertex-set to itself such that $\alpha(v)$ and $\alpha(w)$ have an edge between them if and only if $v$ and $w$ have an edge between them. For a bipartite graph, an automorphism is said to be \emph{preserving} if it fixes the two parts of the vertex-set, i.e., $\alpha(V) = V$ and $\alpha(\widehat V) = \widehat V$, whereas it is said to be \emph{reversing} if it exchanges the two parts of the vertex-set, i.e., $\alpha(V) = \widehat V$ and $\alpha(\widehat V) = V$.} which satisfies
\ie
\sigma_{\pi(\hat v) \pi(v)} = \sigma_{\hat v v}~.
\fe
Note that $\pi|_V$ is a permutation of the vertices in $V$. This spatial symmetry is implemented by a permutation operator $T_\pi$ that maps the qubit at vertex $v$ to the qubit at vertex $\pi(v)$. In particular,
\ie
T_\pi X_v T_\pi^{-1} = X_{\pi(v)}~,\qquad T_\pi Z_v T_\pi^{-1} = Z_{\pi(v)}~.
\fe
It follows that
\ie
T_\pi B_{\hat v} T_\pi^{-1} = T_\pi \prod_v Z_v^{\sigma_{\hat v v}} T_\pi^{-1} = \prod_v Z_{\pi(v)}^{\sigma_{\hat v v}} = \prod_v Z_{\pi(v)}^{\sigma_{\pi(\hat v) \pi(v)}} = \prod_v Z_v^{\sigma_{\pi(\hat v) v}} = B_{\pi(\hat v)}~.
\fe
Therefore, the Hamiltonian commutes with $T_\pi$, i.e.,
\ie
T_\pi H = H T_\pi~.
\fe

Since the composition of two preserving automorphisms is also a preserving automorphism, we have
\ie
T_\pi T_{\pi'} = T_{\pi \circ \pi'}~.
\fe
The operator $T_\pi$ acts on $\eta_a$ as
\ie
T_\pi \eta_a T_\pi^{-1} = T_\pi \prod_v X_v^{a_v} T_\pi^{-1} = \prod_v X_{\pi(v)}^{a_v} = \prod_v X_v^{a_{\pi^{-1}(v)}} = \eta_{\pi \cdot a}~,
\fe
where we defined $(\pi\cdot a)_v = a_{\pi^{-1}(v)}$. As a check, note that $\pi \cdot a \in \ker \sigma$ because $\sum_v \sigma_{\hat v v} (\pi\cdot a)_v = \sum_v \sigma_{\pi^{-1}(\hat v) v} a_v = 0$.

\subsection{Non-invertible duality operators from reversing automorphisms}
Let us now construct a map $\mathsf D_{\hat{\text{v}}\leftarrow \text{v}}: \mathcal H \rightarrow \widehat{\mathcal H}$, where $\widehat{\mathcal H} := \bigotimes_{\hat v\in \widehat V} \mathbb C^2$, as a ZX-diagram where we replace vertices of $\mathcal G$ with green spiders, place Hadamards on the edges, and add input legs into $V$ and output legs out of $\widehat V$.
\tikzexternalenable
After a color change \ref{cc'}, the ZX-diagram representing this map is given by
\tikzsetnextfilename{graph-D}
\ie
\mathsf D_{\hat{\text{v}}\leftarrow \text{v}} := 2^{\kappa} ~ \begin{ZX}[circuit]
\zxN{} & \zxN-{v\in V} & \zxN-{\hat v\in \widehat V} & \zxN{}\\[-8pt]
\zxN{} \rar & \zxZ{} \ar[d,3 vdots] \ar[dr] & \zxX{} \ar[dl,"\sigma" {description,font=\normalsize}] \ar[d,3 vdots] \rar[H] & \zxN{}\\[8pt]
\zxN{} \rar & \zxZ{} & \zxX{} \rar[H] & \zxN{}\\
\end{ZX}~,
\fe
where $\kappa$ is a real number to be fixed later and the middle part of the diagram is understood as follows: for each green dot labelled by $v\in V$ and each red dot labelled by $\hat v\in \widehat V$, the number of wires connecting them is $\sigma_{\hat vv}$. One can similarly define the map $\mathsf D_{\text{v}\leftarrow \hat{\text{v}}}$, whose ZX-diagram is obtained by flipping the ZX-diagram of $\mathsf D_{\hat{\text{v}}\leftarrow \text{v}}$ horizontally, i.e., $\mathsf D_{\text{v}\leftarrow \hat{\text{v}}} = (\mathsf D_{\hat{\text{v}}\leftarrow \text{v}})^\dagger$.

The map $\mathsf D_{\hat{\text{v}}\leftarrow \text{v}}$ implements the gauging of the $\mathbb Z_2$ symmetries generated by $\eta_a$'s. (Such a map was studied in \cite{Cobanera:2011wn,Haegeman:2014maa,Vijay:2016phm,Williamson:2016jiq,Yoshida:2015cia,Yoshida:2015xsv,Williamson:2017uzx,Kubica:2018lhn,Shirley:2018vtc,Radicevic:2019vyb,Tantivasadakarn:2019qdi,Tantivasadakarn:2020lhq,Tantivasadakarn:2021vel,Tantivasadakarn:2022hgp,Dolev:2021ofc,Rayhaun:2021ocs,Rakovszky:2023fng,Okuda:2024azp}.) Indeed, commuting the Hamiltonian $H$ through $\mathsf D_{\hat{\text{v}}\leftarrow \text{v}}$ gives the gauged Hamiltonian $\widehat H$, i.e., 
\ie
\mathsf D_{\hat{\text{v}}\leftarrow \text{v}} H = \widehat H \mathsf D_{\hat{\text{v}}\leftarrow \text{v}}~,
\fe
where
\ie\label{graph-gaugedH}
\widehat H = - h \sum_{v \in V} \widehat B_v - J \sum_{\hat v\in \widehat V} \widehat X_{\hat v}~,\qquad \widehat B_v = \prod_{\hat v \in \widehat V} \widehat Z_{\hat v}^{\sigma_{\hat vv}}~.
\fe
Evidently, the gauged Hamiltonian $\widehat H$ is also a generalized TFIM, albeit on a different Hilbert space $\widehat{\mathcal H}$.

So far we have defined a map between two Hilbert spaces ${\cal H}$ and  $\widehat{\cal H}$. 
To turn it into an operator that acts within one Hilbert space, we need to identify these two Hilbert spaces in some way. 

More precisely, this is achieved  by choosing a \textit{reversing automorphism} $\rho$ of $\mathcal G$,\footnote{Two necessary conditions for the existence of a reversing automorphism are (i) $|V| = |\widehat V|$ and (ii) the \emph{degree-sequences} of vertices in $V$ and $\widehat V$ match. However, they are not sufficient---a bipartite graph which satisfies these conditions but does not have a reversing automorphism can be found in \cite{Brualdi:1980}.} which satisfies
\ie
\label{eq:sigmacondition}
\sigma_{\rho(v) \rho(\hat v)} = \sigma_{\hat v v}~.
\fe
Define the map $T^\rho_{\hat{\text{v}}\leftarrow \text{v}} : \mathcal H \rightarrow \widehat{\mathcal H}$ that maps the physical qubit at vertex $v$ to the auxiliary qubit at vertex $\rho(v)$. One can similarly define the map $T^\rho_{\text{v} \leftarrow \hat{\text{v}}}$. Since $\rho$ is reversing, $\rho^2$ is preserving, and we have
\ie
T^\rho_{\text{v} \leftarrow \hat{\text{v}}} T^\rho_{\hat{\text{v}}\leftarrow \text{v}} = T_{\rho^2}~.
\fe
More generally, we have
\ie
T^\rho_{\text{v} \leftarrow \hat{\text{v}}} T^{\rho'}_{\hat{\text{v}}\leftarrow \text{v}} = T_{\rho\circ \rho'}~,
\fe
which is well-defined because the composition of two reversing automorphisms is a preserving automorphism. Similarly, we have
\ie\label{rev-pre-compose}
T_\pi T^\rho_{\text{v} \leftarrow \hat{\text{v}}} = T^{\pi \circ \rho}_{\text{v} \leftarrow \hat{\text{v}}}~,\qquad T^\rho_{\hat{\text{v}}\leftarrow \text{v}} T_\pi = T^{\rho \circ \pi}_{\hat{\text{v}}\leftarrow \text{v}}~,
\fe
which are well-defined because the composition of a reversing automorphism and a preserving automorphism is a reversing automorphism.

For every reversing automorphism $\rho$ of $\mathcal{G}$, we define a non-invertible duality operator $\mathsf D_\rho : {\cal H}\rightarrow {\cal H}$ as
\ie
\mathsf D_\rho := T^\rho_{\text{v} \leftarrow \hat{\text{v}}} \mathsf D_{\hat{\text{v}}\leftarrow \text{v}} = \mathsf D_{\text{v}\leftarrow \hat{\text{v}}} T^\rho_{\hat{\text{v}}\leftarrow \text{v}}~,
\fe
where the equality of the two factorizations follows from \eqref{eq:sigmacondition}. From \eqref{rev-pre-compose}, we have
\ie
T_\pi \mathsf D_\rho  = \mathsf D_{\pi \circ \rho}~,\qquad \mathsf D_\rho T_\pi = \mathsf D_{\rho \circ \pi}~.
\fe

\subsection{Action on operators}
 $\mathsf D_\rho$ acts on the $\mathbb{Z}_2$-symmetric operators as
\ie
\mathsf D_\rho X_v = T^\rho_{\text{v} \leftarrow \hat{\text{v}}} \prod_{\hat v\in \widehat V} \widehat Z_{\hat v}^{\sigma_{\hat vv}} \mathsf D_{\hat{\text{v}}\leftarrow \text{v}} = \prod_{\hat v\in \widehat V} Z_{\rho(\hat v)}^{\sigma_{\hat vv}} \mathsf D_\rho = \prod_{\hat v\in \widehat V} Z_{\rho(\hat v)}^{\sigma_{\rho(v)\rho(\hat v)}} \mathsf D_\rho = B_{\rho(v)} \mathsf D_\rho~,
\fe
and similarly,
\ie
\mathsf D_\rho B_{\hat v} = X_{\rho(\hat v)} \mathsf D_\rho~.
\fe
This can be seen easily using ZX-calculus by pushing a red spider with a $\pi$ phase through a vertex on either side of the ZX-diagram of $\mathsf D_{\hat{\text{v}}\leftarrow \text{v}}$. Thus, it is clear that $\mathsf D_\rho$ commutes with the Hamiltonian when $J=h$. Moreover, we have
\ie
\mathsf D_\rho \eta_a =  \mathsf D_\rho \prod_{v\in V} X_v^{a_v} =  \prod_{v\in V} B_{\rho(v)}^{a_v} \mathsf D_\rho = \prod_{v\in V} \prod_{u \in V} Z_u^{\sigma_{\rho(v) u}a_v} \mathsf D_\rho = \prod_{v\in V} \prod_{u \in V} Z_u^{\sigma_{\rho^{-1}(u) v}a_v} \mathsf D_\rho = \mathsf D_\rho~.
\fe
where used $a \in \ker \sigma$. Similarly, we have $\eta_a \mathsf D_\rho = \mathsf D_\rho$.

\subsection{Operator algebra}
Let us look at the action of $\mathsf D_\rho^2$ on symmetric operators:
\ie
\mathsf D_\rho^2 X_v &= \mathsf D_\rho  B_{\rho(v)} \mathsf D_\rho = X_{\rho^2(v)}\mathsf D_\rho^2~,
\\
\mathsf D_\rho^2 B_{\hat v} &= \mathsf D_\rho X_{\rho(\hat v)} \mathsf D_\rho = B_{\rho^2(\hat v)}\mathsf D_\rho^2~.
\fe
So one might expect that
\ie
\mathsf D_\rho^2 \propto T_{\rho^2} \sum_{a \in \ker \sigma} \eta_a~.
\fe
Indeed, we verify this using ZX-calculus. First, we have
\ie
\mathsf D_\rho^2 = \mathsf D_{\text{v}\leftarrow \hat{\text{v}}} T^\rho_{\hat{\text{v}}\leftarrow \text{v}} \mathsf D_{\text{v}\leftarrow \hat{\text{v}}} T^\rho_{\hat{\text{v}}\leftarrow \text{v}} = \mathsf D_{\text{v}\leftarrow \hat{\text{v}}} \mathsf D_{\hat{\text{v}}\leftarrow \text{v}} T_{\rho^2} = T_{\rho^2} \mathsf D_{\text{v}\leftarrow \hat{\text{v}}} \mathsf D_{\hat{\text{v}}\leftarrow \text{v}}~.
\fe
Now,\tikzsetfigurename{graph-Dsquare}
\ie
\mathsf D_{\text{v}\leftarrow \hat{\text{v}}} \mathsf D_{\hat{\text{v}}\leftarrow \text{v}} \overset{\ref{hc},\ref{sf}}{=} 2^{2\kappa} ~ \begin{ZX}[circuit]
\zxN{} & \zxN-{v\in V} & \zxN-{\hat v\in \widehat V} & \zxN-{v'\in V} &\\[-8pt]
\zxN{} \rar & \zxZ{} \ar[d,3 vdots] \ar[dr] & \zxX{} \ar[dl,"\sigma" {description,font=\normalsize}] \ar[d,3 vdots] \ar[dr] & \zxZ{} \ar[d,3 vdots] \ar[dl,"\sigma^\intercal" {description,font=\normalsize}] \rar & \zxN{}\\[8pt]
\zxN{} \rar & \zxZ{} & \zxX{} & \zxZ{} \rar &\\
\end{ZX}
\overset{\ref{sf}}{=} 2^{2\kappa} ~ \begin{ZX}[circuit,mbr=4.5]
\zxN{} & \zxN{} & \zxN{} & \zxN-{\hat v\in \widehat V} &
\\[-8pt]
\zxN{} & \zxN{} & \zxN{} \ar[dr] \ar[d,3 vdots] & \zxX{} \ar[d,3 vdots] \ar[dl,"\hat \tau"{description,font=\normalsize}] & \zxN{}
\\[8pt]
\zxN{} & \zxN{} & \zxN{} & \zxX{} & \zxN{} &
\\[8pt]
\zxN{} \rar & \zxZ{} \ar[d,3 vdots] \ar[dr] & \zxX{} \ar[uu,('] \ar[dl,"\tau^\intercal" {description,font=\normalsize}] \ar[d,3 vdots] \ar[dr] & \zxZ{} \ar[d,3 vdots] \ar[dl,"\tau" {description,font=\normalsize}] \rar & \zxN{}
\\[8pt]
\zxN{} \rar & \zxZ{} & \zxX{} \ar[uu,('] & \zxZ{} \rar & \zxN{}
\\[-8pt]
\zxN{} & \zxN-{v\in V} & \zxN-{e\in E} & \zxN-{v'\in V} & \zxN{}
\end{ZX}~,
\fe
where in the second equality, we used \eqref{sigma-tau}.
Now, for each $v\in V$, we apply the generalized bialgebra \ref{gb} on the following part of the diagram:
\ie
\begin{ZX}[circuit,mbr=4]
\zxN{} & \zxN{} & \zxN{} \ar[d,3 vdots] & \zxN{} & \zxN{}\\[8pt]
\zxN{} & \zxN{} & \zxN{} & \zxN{} & \zxN{}\\
\zxN{} & \zxN{} & \zxX{} \ar[dd,3 vdots] \ar[uu,('] & \zxN{} & \zxN{}\\[-8pt]
\zxN-{v} \rar & \zxZ{} \ar[ur] \ar[dr] & \zxN{} & \zxZ{} \ar[ul] \ar[dl] \rar & \zxN-{v}\\[-8pt]
\zxN{} & \zxN{} & \zxX{} \ar[uuu,('] & \zxN{} & \zxN{}\\[-8pt]
\zxN{} & \zxN{} & \zxN-{e\in E_v} & \zxN{} & \zxN{}
\end{ZX}
\overset{\ref{gb}}{=} \left(\frac{1}{\sqrt2}\right)^{d_v-1} ~\begin{ZX}[circuit,mbr=2]
\zxN{} & \zxZ{} \rightManyDots{e\in E_v}\\
\zxN-{v} \rar & \zxX{} \ar[u,('] \rar & \zxN-{v}
\end{ZX}~,
\fe
where $E_v$ is the set of edges incident to $v$ and $d_v = |E_v|$ is the degree of the vertex $v$. Using this and noting that $E = \bigsqcup_{v\in V} E_v$, we get\footnote{The diagram in the first line is well-defined because for each $e$, there is only one $v\in V$ and one $\hat v \in \widehat V$ that it is incident to. In other words, there are no dangling wires and no junctions of wires at any $e\in E$.}
\ie
\mathsf D_{\text{v}\leftarrow \hat{\text{v}}} \mathsf D_{\hat{\text{v}}\leftarrow \text{v}} &= 2^{2\kappa}\left(\frac{1}{\sqrt2}\right)^{\sum_{v\in V} (d_v-1)} ~ \begin{ZX}[circuit,mbr=4.5]
\zxN{} & \zxN{} & \zxN-{e\in E} & \zxN-{\hat v\in \widehat V}
\\[-8pt]
\zxN{} & \zxZ{} \ar[d,3 vdots] \ar[dr] & \zxN{} \ar[d,3 vdots] \ar[dl,"\tau^\intercal"{description,font=\normalsize}] \ar[dr] & \zxX{} \ar[d,3 vdots]\ar[dl,"\hat\tau"{description,font=\normalsize}]
\\[8pt]
\zxN{} & \zxZ{} & \zxN{} & \zxX{} & \zxN{}
\\[8pt]
\zxN{} \rar & \zxX{} \ar[d,3 vdots] \ar[uu,('] \ar[r] & \zxN{} & \zxN{}
\\[8pt]
\zxN{} \rar & \zxX{} \ar[uu,('] \ar[r] & \zxN{} & \zxN{}
\\[-8pt]
\zxN{} & \zxN-{v\in V} & \zxN{} & \zxN{}
\end{ZX}
\\
&= 2^{2\kappa}\left(\frac{1}{\sqrt2}\right)^{|E| - |V|} ~ \begin{ZX}[circuit,mbr=4.5]
\zxN{} & \zxN{} & \zxN-{\hat v \in \widehat V} &
\\[-8pt]
\zxN{} & \zxZ{} \ar[d,3 vdots] \ar[dr] & \zxX{} \ar[d,3 vdots] \ar[dl,"\sigma"{description,font=\normalsize}]
\\[8pt]
\zxN{} & \zxZ{} & \zxX{} &
\\[8pt]
\zxN{} \rar & \zxX{} \ar[d,3 vdots] \ar[uu,('] \ar[r] & \zxN{} & \zxN{}
\\[8pt]
\zxN{} \rar & \zxX{} \ar[uu,('] \ar[r] & \zxN{}
\\[-8pt]
\zxN{} & \zxN-{v\in V} & \zxN{}
\end{ZX}
=: \mathsf C~,
\fe
where we used $|E| = \sum_{v\in V} d_v$, and defined the condensation operator $\mathsf C$. Therefore, we have
\ie
\mathsf D_\rho^2 = T_{\rho^2} \mathsf C~.
\fe
More generally, for two reversing automorphisms $\rho$ and $\rho'$, one can show that
\ie
\mathsf D_\rho \mathsf D_{\rho'} = T_{\rho \circ \rho'} \mathsf C~.
\fe
\tikzexternaldisable

Note that $\mathsf D_\rho$ mixes with the spatial symmetry generated by $T_{\rho^2}$, generalizing the observation in \cite{Seiberg:2023cdc,Seiberg:2024gek}. 
In the special case when the reversing automorphism $\rho$ is an \textit{involution}---i.e., $\rho^2 = 1$, where $1$ is the identity automorphism---then we see that the duality operator $\mathsf D_\rho$ does not mix with any spatial symmetry. We will see examples of this below.

We can find the Pauli representation of $\mathsf C$ as follows. Using \eqref{projection-identity} for all the green dots, we have
\ie
\mathsf C &= 2^{2\kappa}\left(\frac{1}{\sqrt2}\right)^{|E| - |V|} \frac{(\sqrt2)^{2\sum_{v\in V} 1}}{(\sqrt2)^{\sum_{v\in V} (d_v+1)}} \left(\bigotimes_{\hat v\in \widehat V} \<0|_{\hat v}\right) \left[ \prod_{v\in V} \left( 1+ X_v \prod_{\hat v\in \widehat V} \widehat X_{\hat v}^{\sigma_{\hat vv}} \right) \right] \left(\bigotimes_{\hat v\in \widehat V} |0\>_{\hat v}\right)
\\
&= \frac{2^{2\kappa}}{2^{|E| - |V|}} \sum_{a\in \{0,1\}^{|V|}}~\prod_{v\in V} X_v^{a_v}  \prod_{\hat v\in \widehat V} \delta_{\sum_{v\in V} \sigma_{\hat vv} a_v,0}
\\
&= \frac{2^{2\kappa}}{2^{|E| - |V|}} \sum_{a\in \ker \sigma} \eta_a~,
\fe
and hence,
\ie
\mathsf D_\rho^2 = \frac{2^{2\kappa}}{2^{|E| - |V|}}~T_{\rho^2} \sum_{a\in \ker \sigma} \eta_a~.
\fe
It follows that 
\ie
\mathsf D_\rho \mathsf C =\mathsf{C}\mathsf{D}_\rho= \frac{2^{2\kappa} \cdot 2^{\dim \ker \sigma}}{2^{|E| - |V|}}\,\mathsf D_\rho~.
\fe
We will say more about the choice of $\kappa$ in the examples below.  It turns out that, in all the examples that are translationally invariant, there is a natural choice of $\kappa$ such that all extensive factors cancel in the fusion of $\mathsf D$ and $\mathsf C$. The remaining coefficient is precisely the fusion coefficient expected from the continuum.

The following is a summary of the algebra of all symmetry operators:
\ie
&\eta_a \eta_{a'} = \eta_{a+a'}~,\qquad T_\pi T_{\pi'} = T_{\pi \circ \pi'}~,\qquad T_\pi \eta_a = \eta_{\pi \cdot a} T_\pi~,
\\
&\mathsf D_\rho \eta_a = \eta_a \mathsf D_\rho = \mathsf D_\rho~,\qquad T_\pi \mathsf D_\rho  = \mathsf D_{\pi \circ \rho}~,\qquad \mathsf D_\rho T_\pi = \mathsf D_{\rho \circ \pi}~,
\\
&\mathsf D_\rho \mathsf D_{\rho'} = T_{\rho \circ \rho'} \mathsf C~,\qquad \mathsf C = \frac{2^{2\kappa}}{2^{|E| - |V|}} \sum_{a\in \ker \sigma} \eta_a~,
\fe
where $a,a'\in \ker \sigma$, $\pi,\pi'$ are preserving automorphisms, $(\pi\cdot a)_v = a_{\pi^{-1}(v)}$, and $\rho,\rho'$ are reversing automorphisms. The relation between the symmetries of the Hamiltonian $H$ and the properties of the bipartite graph $\mathcal G$ are summarized in Table \ref{tbl:graph-sym}.

\renewcommand{\arraystretch}{1.5}
\begin{table}
\centering
\begin{tabular}{|c|c|c|}
\hline
Symmetry & Graph-theoretic property & Fusion rule
\tabularnewline
\hline\hline
Internal $\mathbb Z_2$'s ($\eta_a$) & $a\in\ker \sigma$ & $\eta_a \eta_{a'} = \eta_{a+a'}$
\tabularnewline
\hline
Spatial ($T_\pi$) & Preserving automorphism ($\pi$) & $T_\pi T_{\pi'} = T_{\pi \circ \pi'}$
\tabularnewline
\hline
Duality ($\mathsf D_\rho$) & Reversing automorphism ($\rho$) & $\mathsf D_\rho \mathsf D_{\rho'} \sim T_{\rho\circ \rho'} \sum_{a\in\ker \sigma} \eta_a$
\tabularnewline
\hline
\end{tabular}
\caption{The one-one correspondence between the symmetries of the generalized TFIM and the properties of the associated bipartite graph. When $\rho$ is a reversing involution, i.e., $\rho^2 = 1$, then $\mathsf D_\rho$ does not mix with spatial symmetries.\label{tbl:graph-sym}}
\end{table}
\renewcommand{\arraystretch}{1}

\subsection{Examples}

\subsubsection{1+1d Ising model}
\begin{figure}
\centering
\hfill \raisebox{-0.5\height}{\includegraphics[scale=0.22]{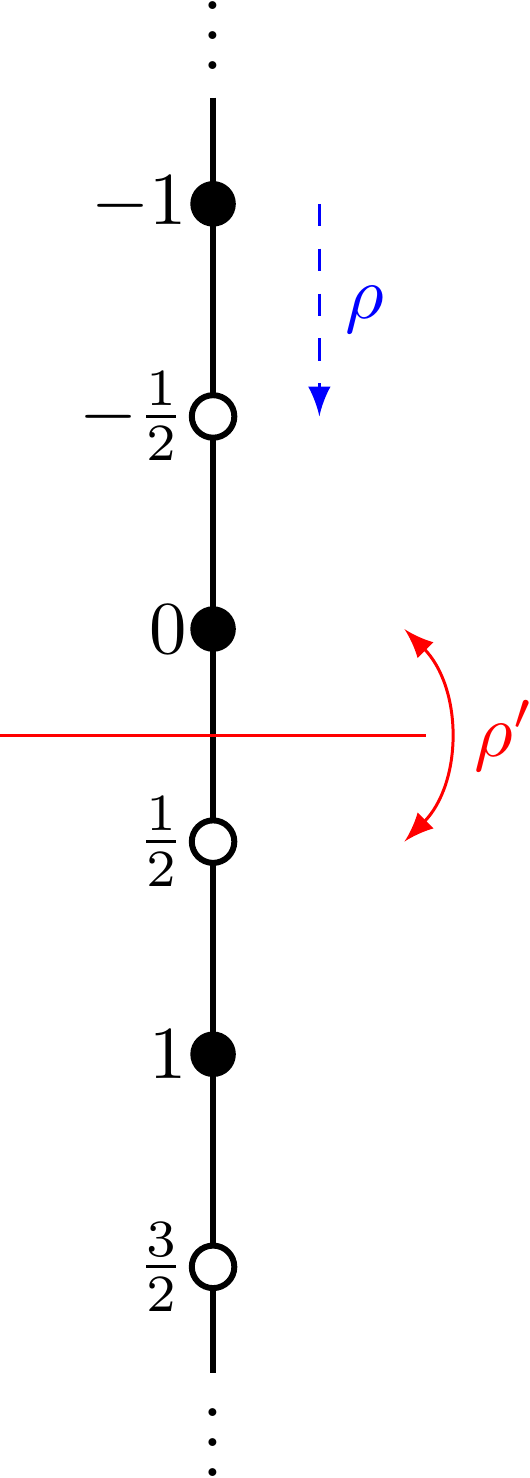}} \hfill \raisebox{-0.5\height}{\includegraphics[scale=0.22]{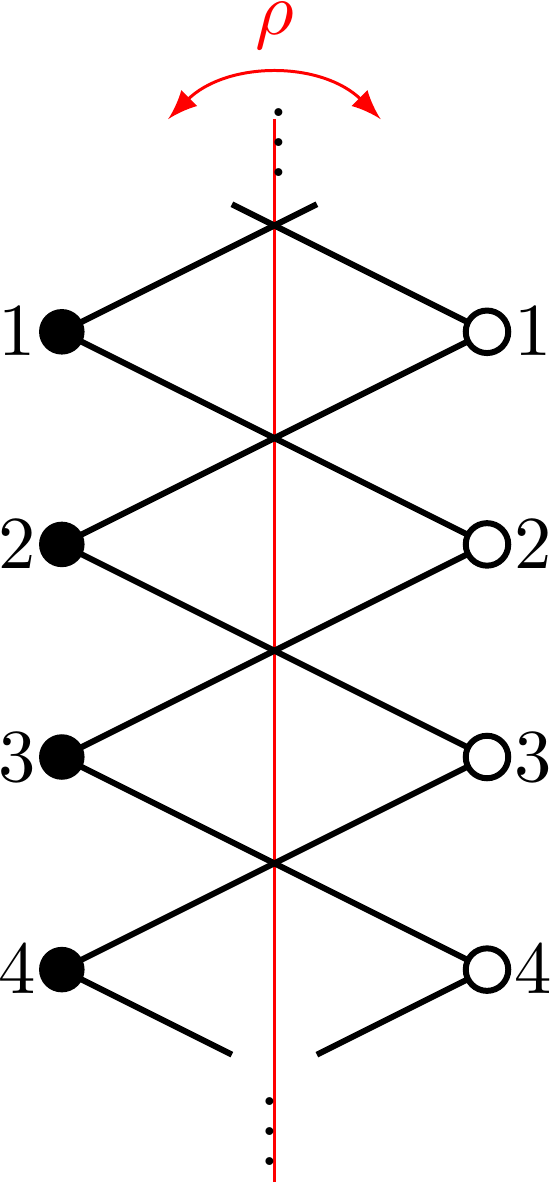}} \hfill \raisebox{-0.5\height}{\includegraphics[scale=0.22]{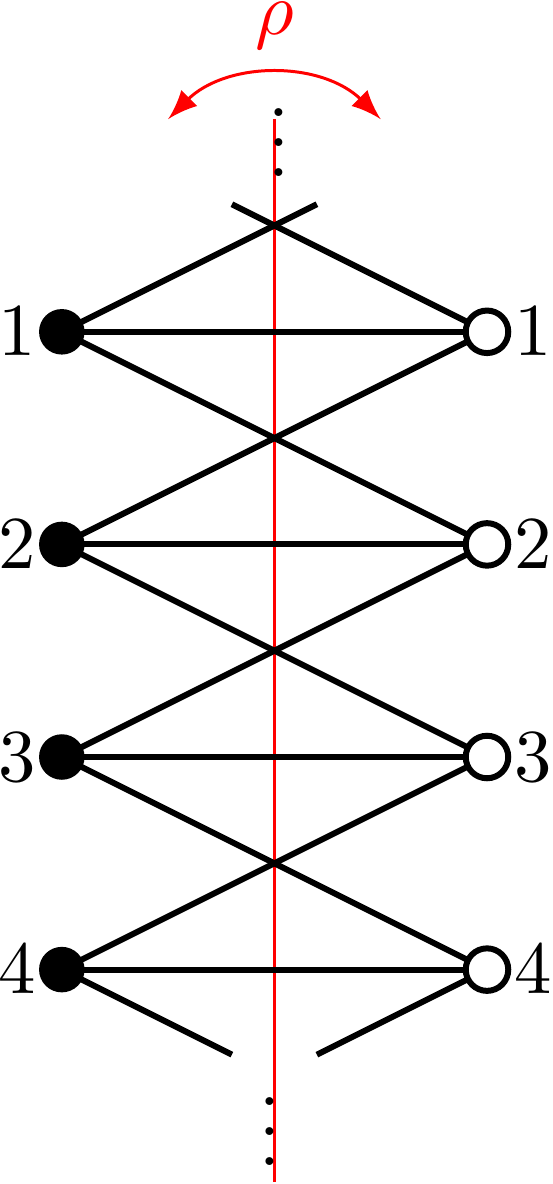}} \hfill \,
\\~\\
\hfill (a)~~~~~ \hfill ~~~(b)~~~ \hfill ~~~~(c)~~ \hfill \,
\caption{The bipartite graphs for the 1+1d examples: (a) the transverse-field Ising model \eqref{graph-Ising-H}, (b) the Ashkin-Teller model at $\lambda=0$ \eqref{graph-AT-H}, and (c) the three-spin Ising model \eqref{graph-3spin-H}. The vertex-sets $V$ and $\widehat V$ correspond to the solid and hollow vertices, respectively. The blue (dashed) arrow indicates a reversing automorphism that is not an involution and the red arrows indicate reversing involutions.\label{fig:graph}}
\end{figure}

In this case, $V$ is the set of $L$ sites on a periodic chain and $\widehat V$ is the set of links. We use an integer $i$ to label a site and a half-integer $i+\tfrac12$ to label the link between sites $i$ and $i=1$. The Hamiltonian is
\ie\label{graph-Ising-H}
H = -J \sum_i Z_i Z_{i+1} - h \sum_i X_i~.
\fe
The associated bipartite graph is shown in Figure \ref{fig:graph}(a). Its biadjacency matrix is given by
\ie
\sigma_{i (j+\frac12)} = \delta_{i,j} + \delta_{i,j+1}~.
\fe
The reversing automorphism $\rho$, shown as blue dashed arrow in Figure \ref{fig:graph}(a), corresponds to a half-translation. This leads to the KW duality operator $\mathsf D$, which mixes with lattice translation as in \eqref{1+1dD2}. Since $|V| = L$, $|E| = 2L$, and $\dim \ker \sigma = 1$, we choose $\kappa = L/2$ to cancel all the extensive factors. This gives $\mathsf D \mathsf C = \mathsf C\mathsf D= 2 \mathsf D$, which is precisely what we obtained in Section \ref{sec:1d-op-alg}.

 The half-translation $\rho$ and all its odd powers are not the only reversing automorphisms of the bipartite graph in Figure \ref{fig:graph}(a).
In fact, there is a reversing involution $\rho'$, shown as red arrow in Figure \ref{fig:graph}(a), that corresponds to a reflection. This leads to a duality operator $\mathsf D' = \mathsf D P$ that does not mix with lattice translation. (Here, $P$ is the parity operator that maps the qubit at $i$ to the qubit at $-i$.) Indeed, using $\mathsf D P = P \mathsf D^\dagger$ and $T P = P T^\dagger$ (it is easy to see this using ZX-calculus), we have
\ie
(\mathsf D')^2 = \mathsf D P \mathsf D P = \mathsf D \mathsf D^\dagger = 1+\eta~.
\fe
While $\mathsf D'$ commutes with $\eta$, it does not commute with $T$:
\ie
\mathsf D' T = T^\dagger \mathsf D'~.
\fe
Moreover, $\mathsf D'$ maps local symmetric operators around the site $i$ to those around the site $-i$. Therefore, we refer to $\mathsf D'$ as a \emph{non-invertible parity (or reflection) operator}. 
(See \cite{Kaidi:2021xfk,Choi:2022rfe} for non-invertible parity and time-reversal symmetries in continuum field theory.)

\subsubsection{3+1d lattice $\mathbb Z_2$ gauge theory}
In this case, $V$ is the set of links on an $L_x \times L_y \times L_z$ periodic cubic lattice and $\widehat V$ is the set of plaquettes. We use $\ell$ to label a link and $p$ to label a plaquette. The Hamiltonian is
\ie
H = -J \sum_p \prod_{\ell\in p} Z_\ell - h \sum_\ell X_\ell~.
\fe
Note that we do not impose Gauss law here, or equivalently, we set $g=0$ in \eqref{H}. The biadjacency matrix in this case is given by
\ie
\sigma_{\ell p} = \sum_{\ell' \in p} \delta_{\ell,\ell'}~.
\fe
Since $|V| = 3L_x L_y L_z$, $|E| = 12L_x L_y L_z$, and $\dim \ker \sigma = L_x L_y L_z+2$, we choose $\kappa = 4L_x L_y L_z$ to cancel all the extensive factors. This gives $\mathsf D \mathsf C = \mathsf C\mathsf D= 4 \mathsf D$, which is precisely what we obtained in \eqref{C-fusion}.

There is a reversing automorphism given by the half-translation $\mathfrak t$ in the $(1,1,1)$ direction. The leads to the Wegner duality operator $\mathsf D$ \eqref{D-def}, which mixes with the lattice translation $T_{1,1,1}$.

There is also a reversing involution given by parity, or equivalently, a reflection followed by a $\pi$-rotation. This gives a duality operator $\mathsf D' = \mathsf D P$ that does not mix with lattice translation. (Here, $P$ is the usual parity operator that maps the qubit at link $\ell$ to the qubit at link $\mathfrak p(\ell)$, where
\ie
\mathfrak p(x,y,z) = (-x,-y,-z) \mod (L_x,L_y,L_z)~,
\fe
is the parity map on the lattice.) Since $\mathsf D'$ maps local symmetric operators around the link $\ell$ to those around the link $\mathfrak p(\ell)$, we refer to it as a non-invertible parity operator.

\subsubsection{1+1d Ashkin-Teller model}\label{sec:AT}
The Hamiltonian is \cite{PhysRev.64.178,FAN1972136,PhysRevB.24.230,PhysRevB.24.5229,Saleur_1987}
\ie\label{graph-AT-H}
H &= -J \sum_i \left( Z_{2i-1} Z_{2i+1} + Z_{2i} Z_{2i+2} + \lambda Z_{2i-1} Z_{2i} Z_{2i+1} Z_{2i+2} \right)
\\
&\qquad - h \sum_i \left( X_{2i} + X_{2i+1} + \lambda X_{2i} X_{2i+1} \right)~.
\fe
Assuming $L\in 2\mathbb Z$, this model has two $\mathbb Z_2$ symmetries generated by
\ie
\eta_\text{odd} = \prod_i X_{2i-1}~,\qquad \eta_\text{even} = \prod_i X_{2i}~.
\fe
When $J=h$, it is self-dual under the exchange
\ie
Z_{2i-1} Z_{2i+1} \leftrightsquigarrow X_{2i}~,\qquad Z_{2i} Z_{2i+2} \leftrightsquigarrow X_{2i+1}~.
\fe
When $\lambda = 0$, this model is equivalent to two copies of the TFIM on the odd and even sites, respectively.\footnote{It is also equivalent to the $\mathbb Z_2$ dipole Ising model \cite{Gorantla:2022eem}, so one can interpret this duality as the dipole KW duality \cite{Cao:2024qjj,dipoleKW}.} The phase diagram of the quantum Ashkin-Teller (AT) model is well-studied \cite{PhysRevB.24.5229,Bridgeman:2015nya}. Along $J=h$, for a range of $\lambda$, there is a critical line with continuously varying critical exponents described by the orbifold branch of the $c=1$ CFT. In particular, at $J=h$ and $\lambda=0$, the critical point is described by the Ising$^2$ point on the orbifold branch of the $c=1$ CFT, as expected from the above equivalence. Related discussions can be found in \cite{Bridgeman:2017etx,Choi:2024rjm}.

In this case, $V$ and $\widehat V$ are both the set of all sites, labelled by an integer $i$. The corresponding bipartite graph is shown in Figure \ref{fig:graph}(b). Its biadjacency matrix is given by
\ie
\sigma_{i j} = \delta_{i,j-1} + \delta_{i,j+1}~.
\fe
The generalized TFIM associated with this graph corresponds to the $\lambda=0$ case. 

There is a reversing involution $\rho$, shown as red arrow in Figure \ref{fig:graph}(b), that flips the graph about the red vertical line. Therefore, the duality operator $\mathsf D$ which implements the above exchange does not mix with lattice translation:
\ie
\mathsf D^2 = \mathsf C = (1+\eta_\text{odd}) (1+\eta_\text{even})~.
\fe
Here, since $|V| = L$, $|E| = 2L$, and $\dim \ker \sigma = 2$, we chose $\kappa = L/2$ to cancel all the extensive factors, and hence, $\mathsf D\mathsf C = \mathsf C \mathsf D= 4\mathsf D$.

We remark that the operators $\mathsf D$, $\eta_\text{odd}$, and $\eta_\text{even}$ together form the fusion category $\text{Rep}(D_8)$.  
Apart from the AT model, there are also other Hamiltonians which enjoy this symmetry, such as the 1+1d cluster Hamiltonian \cite{Seifnashri:2024dsd}.
This non-invertible symmetry in the continuum $c=1$ orbifold CFT was discussed in \cite{Thorngren:2021yso,Choi:2023vgk,Diatlyk:2023fwf,Perez-Lona:2023djo}.

\subsubsection{1+1d three-spin Ising model}
The Hamiltonian is
\ie\label{graph-3spin-H}
H = -J \sum_i Z_{i-1} Z_i Z_{i+1} - h \sum_i X_i~.
\fe
Assuming $L\in 3\mathbb Z$, this model has two $\mathbb Z_2$ symmetries generated by
\ie
\eta_1 = \prod_{i \notin 3\mathbb Z+1} X_i~,\qquad \eta_2 = \prod_{i\notin 3\mathbb Z+2} X_i~.
\fe
When $J=h$, it is self-dual under the exchange
\ie
Z_{i-1} Z_i Z_{i+1} \leftrightsquigarrow X_i~.
\fe
The three-spin Ising model, as well as the duality under the  Kramers-Wannier action, has been pointed out in \cite{Turban82,Maritan84,Penson82}, as well as recently constructed as a non-invertible symmetry operator \cite{Yan:2024eqz}. The Hamiltonian at $J=h$ has been studied numerically where the central charge was found to be approximately $1$, and the value of the critical exponents suggests that it lies on the orbifold branch of the $c=1$ CFT moduli space \cite{Kolb86,Igloi86,Alcaraz1987,Udupa:2023rjg}.

In this case, $V$ and $\widehat V$ are both the set of all sites, labelled by an integer $i$. The corresponding bipartite graph is shown in Figure \ref{fig:graph}(c). Its biadjacency matrix is given by
\ie
\sigma_{i j} = \delta_{i,j-1} + \delta_{i,j} + \delta_{i,j+1}~.
\fe
There is a reversing involution $\rho$, shown as red arrow in Figure \ref{fig:graph}(c), that flips the graph about the red vertical line. Therefore, the duality operator $\mathsf D$ which implements the above exchange does not mix with lattice translation:
\ie
\mathsf D^2 = \mathsf C = (1+\eta_1) (1+\eta_2)~.
\fe
Here, since $|V| = L$, $|E| = 3L$, and $\dim \ker \sigma = 2$, we chose $\kappa = L$ to cancel all the extensive factors, and hence, $\mathsf D\mathsf C =\mathsf C\mathsf D= 4\mathsf D$. These symmetries form a Rep$(D_8)$ fusion category.

\subsubsection{2+1d plaquette Ising model}
Consider a 2d $L_x\times L_y$ periodic square lattice with a qubit on each site labelled by $(i,j)$. The Hamiltonian is 
\ie\label{graph-PIM-H}
H = -J \sum_{i,j} Z_{i,j} Z_{i+1,j} Z_{i,j+1} Z_{i+1,j+1} - h \sum_{i,j} X_{i,j}~.
\fe
This model has a $\mathbb Z_2$ subsystem symmetry generated by
\ie
\eta^x_i = \prod_j X_{i,j}~,\qquad \eta^y_j = \prod_i X_{i,j}~.
\fe
They satisfy the constraint $\prod_i \eta^x_i = \prod_j \eta^y_j$ so there are only $L_x + L_y - 1$ independent generators. When $J=h$, this model is self-dual under the map
\ie
X_{i,j} \rightsquigarrow Z_{i,j} Z_{i+1,j} Z_{i,j+1} Z_{i+1,j+1} \rightsquigarrow X_{i+1,j+1}~,
\fe
which already hints at the duality operator mixing with lattice translation in the $(1,1)$ direction. This duality was studied in \cite{Cao:2023doz,ParayilMana:2024txy,Spieler:2024fby}.

\begin{figure}
\centering
\raisebox{-0.5\height}{\includegraphics[scale=0.2]{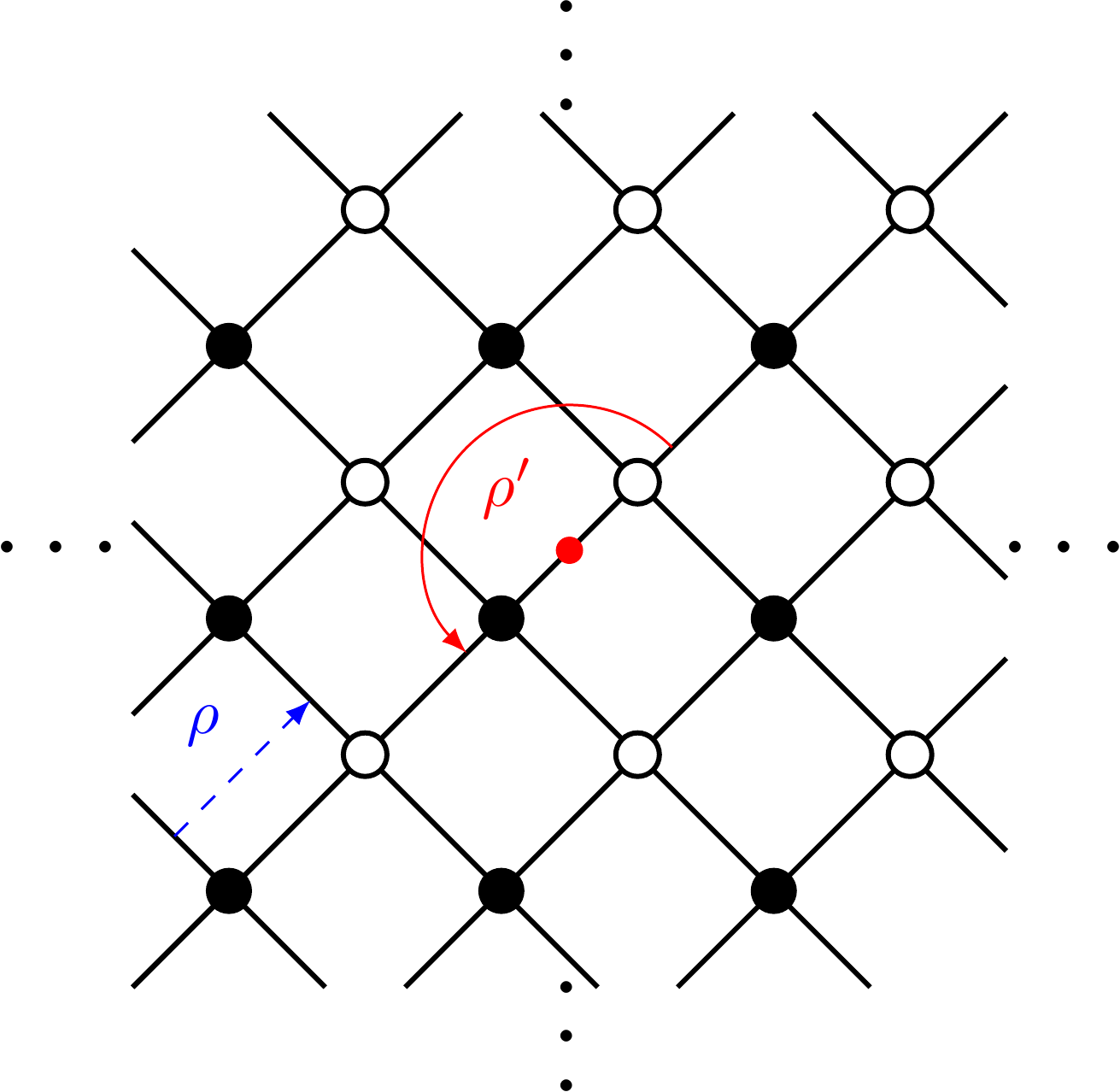}}
\caption{The bipartite graphs for the 2+1d plaquette Ising model \eqref{graph-PIM-H}. The vertex-sets $V$ and $\widehat V$ correspond to the solid and hollow vertices, respectively. The blue (dashed) arrow indicates a reversing automorphism that is not an involution, whereas the red arrow indicates a reversing involution.\label{fig:graph-PIM}}
\end{figure}

In this case, $V$ is the set of all sites and $\widehat V$ is the set of all plaquettes. We use $(i,j)$ to label a site and $(i+\tfrac12,j+\tfrac12)$ to label a plaquette bounded by the sites $(i,j)$, $(i+1,j)$, $(i,j+1)$, and $(i+1,j+1)$. The corresponding bipartite graph is shown in Figure \ref{fig:graph-PIM}. Its biadjacency matrix is given by
\ie
\sigma_{(i,j) (i'+\frac12,j'+\frac12)} = \delta_{i,i'} \delta_{j,j'} + \delta_{i,i'+1} \delta_{j,j'} + \delta_{i,i'} \delta_{j,j'+1} + \delta_{i,i'+1} \delta_{j,j'+1}~.
\fe

The reversing automorphism $\rho$, shown as blue dashed arrow in Figure \ref{fig:graph-PIM}, corresponds to the half-translation in $(1,1)$ direction. It leads to the subsystem KW duality operator $\mathsf D$, which mixes with the lattice translation in the $(1,1)$ direction as follows:
\ie
\mathsf D^2 = T_{1,1} \mathsf C~,\qquad \text{where}\qquad \mathsf C = \frac12\prod_i (1+\eta^x_i) \prod_j (1+\eta^y_j)~.
\fe
Since $|V| = L_x L_y$, $|E| = 4L_x L_y$, and $\dim \ker \sigma = L_x + L_y - 1$, we choose $\kappa = 3L_x L_y/2$ to cancel all the extensive factors. This gives $\mathsf D \mathsf C = \mathsf C\mathsf D= 2^{L_x + L_y - 1} \mathsf D$.

There is also a reversing involution $\rho'$, shown as red arrow in Figure \ref{fig:graph-PIM}, given by a $\pi$-rotation about the red point. The corresponding duality operator $\mathsf D' = \mathsf D P$ does not mix with lattice translation. (Here, $P$ is the usual parity operator that maps the qubit at site $(i,j)$ to the qubit at site $(-i,-j)$. Note that $P$ generates a $\pi$-rotation in the spatial plane, which is orientation preserving.) We refer to it as a non-invertible parity operator because it maps local symmetric operators around $(i,j)$ to those around $(-i,-j)$.

\subsubsection{Tensor product models}\label{sec:prod}

A bipartite graph $\mathcal G$ has a reversing involution if and only if it is of the form $\mathcal G_0 \times K_2$ \cite{Hammack2010},\footnote{Given two graphs $\mathcal G_1$ and $\mathcal G_2$, the \emph{tensor (Kronecker) product} $\mathcal G_1 \times \mathcal G_2$ is defined as the graph with vertex set $V(\mathcal G_1) \times V(\mathcal G_2)$ such that there is an edge between $(v_1,v_2)$ and $(w_1,w_2)$ if and only if there is an edge between $v_i$ and $w_i$ in $\mathcal G_i$ for $i=1,2$. The name is inspired by the fact that the adjacency matrix of $\mathcal G_1 \times \mathcal G_2$ is the tensor (Kronecker) product of the adjacency matrices of $\mathcal G_1$ and $\mathcal G_2$.} where $\mathcal G_0$ is a graph on $|V|$ vertices and $K_2$ is the complete graph on two vertices (i.e., a ``dumbbell"). (Given a reversing involution $\rho$ of $\mathcal G$, we can construct the graph $\mathcal G_0$ as follows: its vertex set is $V$ and there is an edge between $v$ and $w$ if and only if $\sigma_{\rho(v) w} = 1$. This definition is symmetric in $v$ and $w$ because $\sigma_{\rho(v) w} = \sigma_{\rho(w) v}$. Conversely, the map $\rho: (v_0,\epsilon) \mapsto (v_0,1-\epsilon)$, where $\epsilon = 0,1$ labels the vertices of $K_2$, gives a reversing involution of $\mathcal G_0 \times K_2$.)

All the examples above have reversing involutions so they fall into this category. Below, we consider a special case of this type which leads to a tensor product of decoupled generalized TFIMs. 

Consider the case where $\mathcal G_0$ is itself a bipartite graph with vertex-set $\widehat V_0\sqcup V_0$ and edge-set $E_0$. Let $\bar{\mathcal G}_0$ be the bipartite graph obtained by exchanging the two parts of $\mathcal G_0$, i.e., $\bar V_0 = \widehat V_0$ and $\widehat{\bar V}_0 = V_0$. Let $H_0$ be the Hamiltonian of the generalized TFIM based on $\mathcal G_0$. It is easy to see that the Hamiltonian $\bar H_0$ of the generalized TFIM based on $\bar{\mathcal G}_0$ corresponds to the gauged Hamiltonian $\widehat H_0$ \eqref{graph-gaugedH}.

Now, the bipartite graph $\mathcal G := \mathcal G_0 \times K_2$ is canonically isomorphic to the disjoint union of $\mathcal G_0$ and $\bar{\mathcal G}_0$, denoted as $\mathcal G_0 + \bar{\mathcal G}_0$.\footnote{An \emph{isomorphism} $\phi$ from $\mathcal G_1$ to $\mathcal G_2$ is a bijective map $\phi : V(\mathcal G_1) \rightarrow V(\mathcal G_2)$ such that $v_1$ and $w_1$ have an edge between them if and only if $\phi(v_1)$ and $\phi(w_1)$ have an edge between them. 
Disjoint union of graphs ${\cal G}_0+\bar{\cal G}_0$ is also known as \emph{graph sum}, which explains this notation. The canonical isomorphism from $\mathcal G_0 + \bar{\mathcal G}_0$ to $\mathcal G$ is given by the obvious inclusions: 
\ie
&V_0 \hookrightarrow V_0 \times \{0\}~,\qquad \bar V_0 = \widehat V_0 \hookrightarrow \widehat V_0 \times \{0\}~,
\\
&\widehat V_0 \hookrightarrow \widehat V_0 \times \{1\}~,\qquad \widehat{\bar V}_0 = V_0 \hookrightarrow V_0 \times \{1\}~.
\fe} 
Therefore, the generalized TFIM based on $\mathcal G$ is the tensor product of the generalized TFIMs based on $\mathcal G_0$ and $\bar{\mathcal G}_0$. In other words, it is the product of a generalized TFIM and its gauged version. More concretely, the total Hilbert space is $\mathcal H = \mathcal H_0 \otimes \bar{\mathcal H}_0$, and the Hamiltonian is
\ie
H = H_0 + \bar H_0~.
\fe

As explained above, $\mathcal G = \mathcal G_0 \times K_2$ has a reversing involution, and here, it exchanges the two components $\mathcal G_0$ and $\bar{\mathcal G}_0$. So there is a duality symmetry that swaps the two Hamiltonians $H_0$ and $\bar H_0$ in $H$. The corresponding \emph{non-invertible swap operator} $\mathsf D$ does not mix with any spatial symmetry; it satisfies
\ie
\mathsf D^2 = \mathsf C = \mathsf C_0 \bar{\mathsf C}_0~,
\fe
where $\mathsf C_0$ and $\bar{\mathsf C}_0$ are the condensation operators of the generalized TFIMs based on $\mathcal G_0$ and $\bar{\mathcal G}_0$, respectively.\footnote{While $\mathcal G_0$ does not necessarily have a duality symmetry, it always has a condensation operator $\mathsf C_0$, which is simply the sum of all the $\mathbb Z_2$ symmetry operators up to a normalization.}

The continuum field theory analog of this model is clear \cite{Choi:2024rjm,Hsin:2024aqb,Cui:2024cav}. Let ${\cal Q}_0$ be a general quantum  field theory (QFT) with a non-anomalous, generalized symmetry $G_0$. Then the tensor product QFT ${\cal Q}_0 \times {\cal Q}_0/G_0$ has a non-invertible swap symmetry since it is invariant under gauging $G_0\times \widehat G_0$, where $\widehat G_0$ is the dual symmetry of $G_0$. 

An example of this type is the Ashkin-Teller model of Section \ref{sec:AT} with $J=h$ and $\lambda = 0$. In this case, the bipartite graphs $\mathcal G_0$ and $\bar{\mathcal G}_0$ correspond to the Ising models on even and odd sublattices, respectively. One can also see from Figure \ref{fig:graph}(b) that there are two connected components, each of which is isomorphic to the bipartite graph in Figure \ref{fig:graph}(a).

Here is another example in 2+1d that was the focus of \cite{Choi:2024rjm}. Consider a periodic $L_x \times L_y$ square lattice with qubits on sites and links, labelled by $s$ and $\ell$, respectively. The Hamiltonian is
\ie
H =  - J \sum_\ell \prod_{s\in \ell} Z_{s} - h \sum_s X_{s} - \bar J \sum_s \prod_{\ell \ni s} \bar Z_{\ell} - \bar h \sum_\ell \bar X_{\ell}~.
\fe
One can recognize the first two terms as the Hamiltonian of the 2+1d Ising model on the original lattice and the last two terms as the Hamiltonian of the 2+1d lattice $\mathbb Z_2$ gauge theory on the dual lattice (albeit without the Gauss law constraint). Therefore, it has a $\mathbb Z_2^{(0)} \times \mathbb Z_2^{(1)}$ symmetry generated by
\ie
\eta_0 = \prod_s X_{s}~,\qquad \bar \eta_0(\gamma) = \prod_{\ell \in \gamma} \bar X_{\ell}~,
\fe
respectively, where $\gamma$ is a curve along the links of the original lattice. This model also has a non-invertible swap symmetry that exchanges
\ie
X_{s} \leftrightsquigarrow \prod_{\ell \ni s} \bar Z_{\ell}~,\qquad \prod_{s\in \ell} Z_{s} \leftrightsquigarrow \bar X_{\ell}~,
\fe
when $J = \bar h$ and $\bar J = h$. The corresponding non-invertible swap operator $\mathsf D$ does not mix with lattice translation; it satisfies
\ie
\mathsf D^2 = \mathsf C = \mathsf C_0 \bar{\mathsf C}_0~,\qquad \mathsf D \mathsf C =  \mathsf C\mathsf D=4 \mathsf D~,
\fe
where
\ie
\mathsf C_0 = 1+\eta_0~,\qquad \bar{\mathsf C}_0 = \frac1{2^A} \sum_\gamma \bar\eta_0(\gamma)~,
\fe
and $A = L_x L_y$.

\begin{figure}
\centering
\raisebox{-0.5\height}{\includegraphics[scale=0.17]{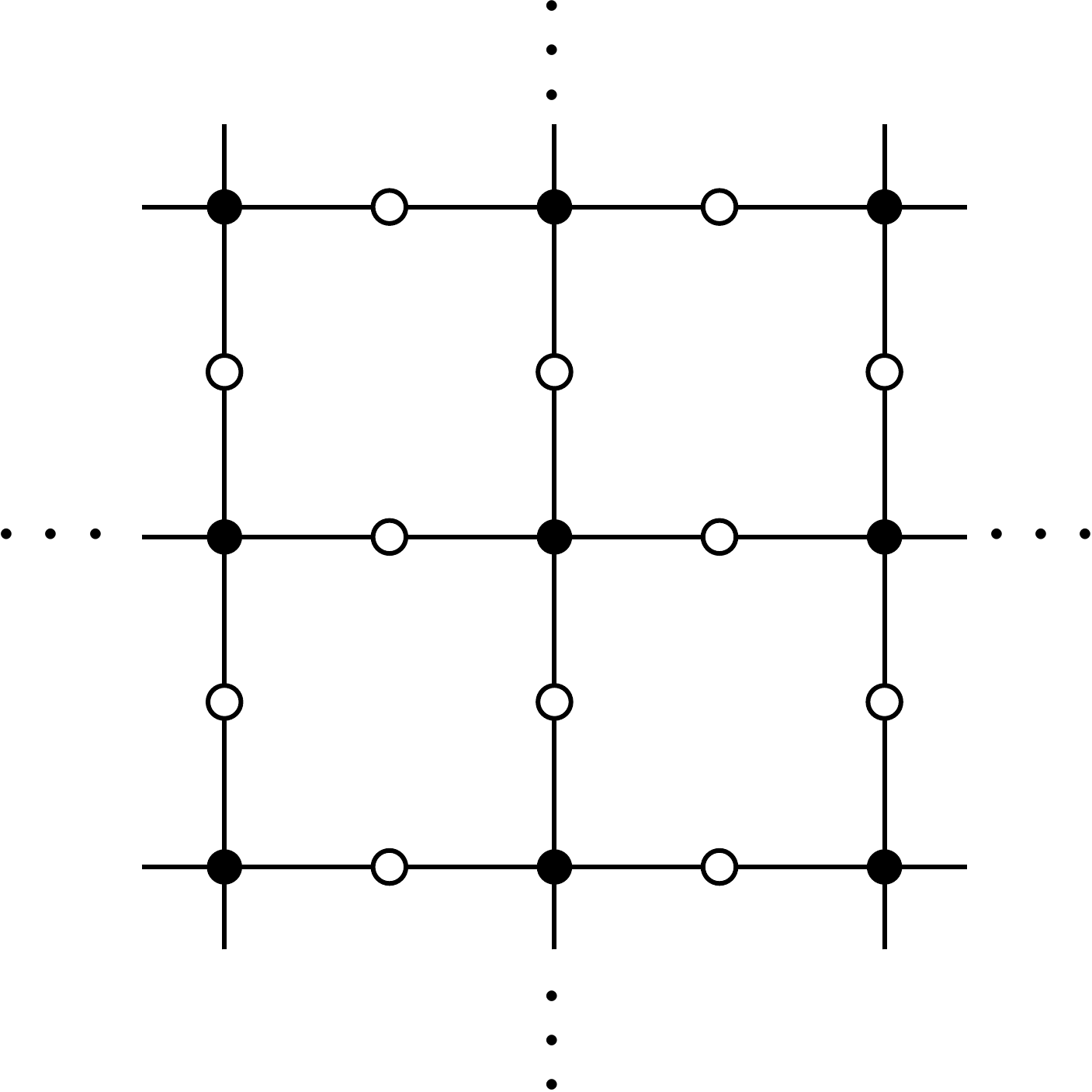}}
\caption{The bipartite graph $\mathcal G_0$ for the 2+1d Ising model. The vertex-sets $V_0$ and $\widehat V_0$ correspond to the solid and hollow vertices, respectively. The bipartite graph $\bar{\mathcal G}_0$ is obtained by replacing solid (hollow) vertices with hollow (solid) vertices.\label{fig:graph0}}
\end{figure}

This example fits into the above model: the vertex-sets $V_0$ and $\widehat V_0$ are the sites and links, respectively, and there is an edge between $s$ and $\ell$ if and only if $s\in \ell$. The resulting bipartite graph $\mathcal G_0$ is shown in Figure \ref{fig:graph0}. The generalized TFIM based on $\mathcal G_0$ is the 2+1d Ising model, whereas the one based on $\bar{\mathcal G}_0$ is the 2+1d lattice $\mathbb Z_2$ gauge theory. The duality operator associated with the  reversing involution of $\mathcal G = \mathcal G_0 \times K_2 \cong \mathcal G_0 + \bar{\mathcal G}_0$ is precisely the non-invertible swap operator $\mathsf D$. 
As an aside, note that $|\widehat V_0| = 2 |V_0|$, so there is no duality symmetry in the 2+1d Ising model or the 2+1d lattice $\mathbb Z_2$ gauge theory separately.

To complete the example, we note that $|V| = |V_0| + |\widehat V_0| = 3A$, $|E| = 2|E_0| = 8A$, and $\dim \ker \sigma = \dim \ker \sigma_0 + \dim \ker \bar \sigma_0 = 1 + (A+1) = A+2$. Therefore, choosing $\kappa = 2A$ to cancel all the extensive factors, we get $\mathsf D \mathsf C = \mathsf C\mathsf D = 4 \mathsf D$, which agrees with the coefficient above.

\subsubsection{On mixing with spatial symmetries}

In all the examples above, the bipartite graph has a reversing involution, so there is a duality operator that does not involve a nontrivial permutation of the qubits. 
In fact, in any $(d+1)$-dimensional self-dual generalized TFIM with translation symmetry, there exists a duality operator that does not mix with translations. 
(For instance, the non-invertible parity operator $\mathsf{D}'$ in the 1+1d Ising model.) 
Therefore, to find models where the duality operator necessarily mixes with a nontrivial permutation of the qubits, one must look beyond models with translation symmetry.

\begin{figure}
\centering
\raisebox{-0.5\height}{\includegraphics[scale=0.11]{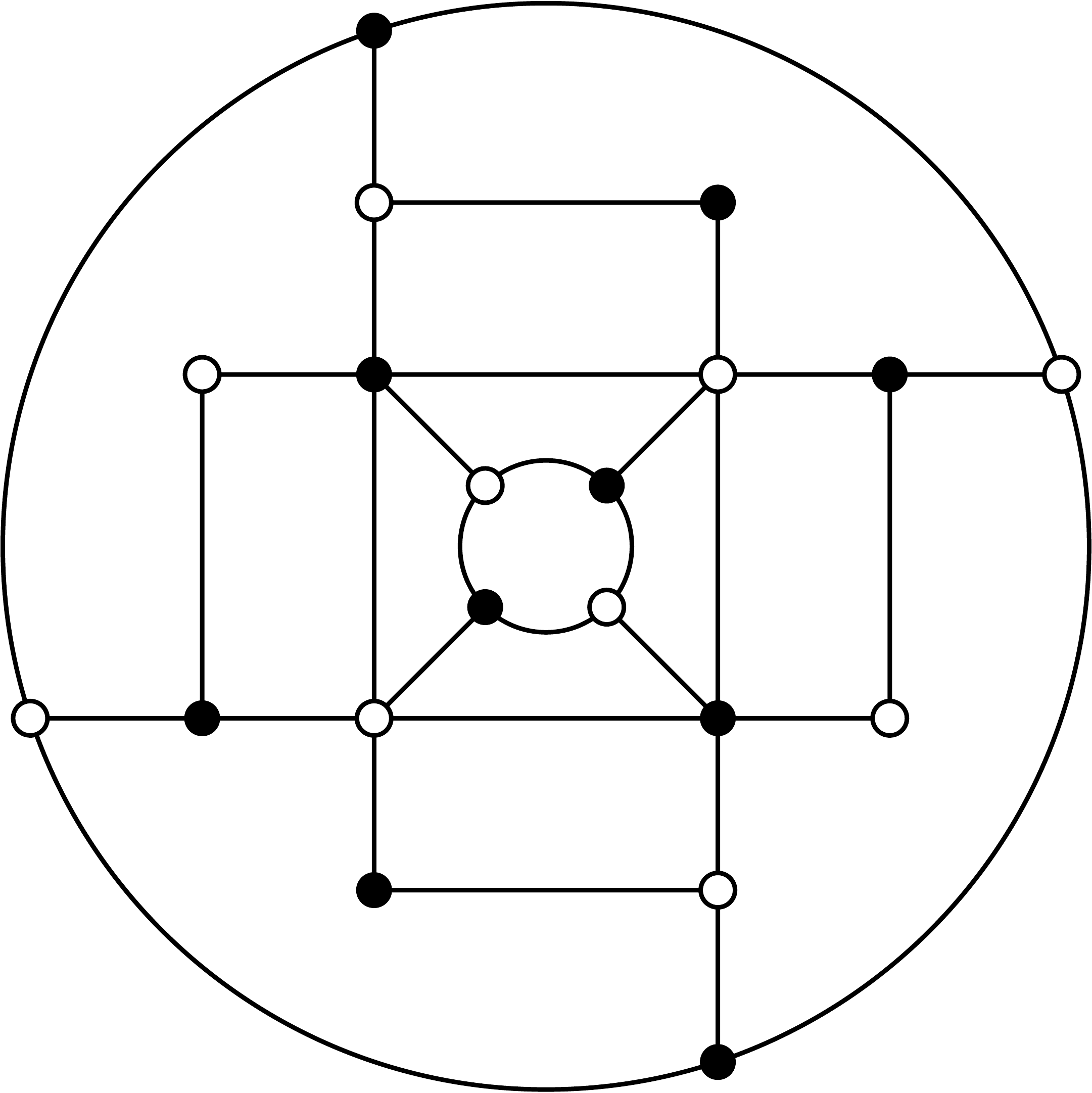}}
\caption{A bipartite graph with a reversing automorphism (90\textdegree~rotation) but no reversing involution.\label{fig:graph-millstone}}
\end{figure}

As an example, consider the bipartite graph shown in Figure \ref{fig:graph-millstone}.\footnote{PG would like to thank Richard Hammack for pointing out a related graph.} It has a reversing automorphism given by a 90\textdegree~rotation of the graph, but it does not have a reversing involution. So the generalized TFIM based on this graph has a duality symmetry that necessarily mixes with a nontrivial permutation of the qubits. Moreover, the duality operator is non-invertible because $\ker \sigma \ne 0$.

\section{Discussions and Outlook} \label{sec:discuss}

In this work, we used  ZX-diagrams to present non-invertible duality operators in many Hamiltonian lattice models, including the Kramers-Wannier operator in the 1+1d Ising model, the Wegner  operator in the 3+1d $\mathbb{Z}_2$ lattice gauge theory, and those in the  generalized TFIMs based on  graphs.  
In addition to the duality operator, we also provided  graphical presentations for the condensation operators. 
Their non-invertible algebra was computed using ZX-calculus.

In the generalized TFIM based on a bipartite graph $\cal G$, we construct  a non-invertible operator $\mathsf{D}_\rho$ associated with every reversing automorphism $\rho$ of $\cal G$. 
In the special case when $\rho$ is a reversing involution (i.e., if $\rho^2=1$), $\mathsf{D}_\rho$ does not mix with spatial symmetries. This is summarized in Table \ref{tbl:graph-sym}.

There are several  future directions:
\begin{itemize}
\item The generalization to lattice models of $n$-dimensional qunits with $n>2$ is particularly interesting since the 't Hooft anomaly of the non-invertible symmetry may depend on $n$ in a nontrivial way, such as in \cite{Choi:2021kmx}.  To this end, one can use an extension of ZX-calculus to qunits, which is known to be universal \cite{Wang_2014,Wang:2021hwx}, and also recently shown to be complete \cite{Poor:2024ozq}.
\item In this work, we focused on the \textit{operators} implementing the non-invertible symmetry. It is important to extend the discussion of \cite{Cheng:2022sgb,Seifnashri:2023dpa,Seiberg:2024gek} to study the corresponding \textit{defects}, which involve local modifications of the Hamiltonian. 
\item It would be interesting to find the diagrammatic presentation for SPT states protected by (non-anomalous) non-invertible symmetries, generalizing the results of \cite{Fechisin:2023dkj}.
\item Recently, there has been a growing interest in the Kennedy-Tasaki transformation \cite{kennedy1992hidden,PhysRevB.45.304,Li:2023ani,ParayilMana:2024txy,Seifnashri:2024dsd,Choi:2024rjm,Lu:2024ytl}, which is a twisted version of  the Kramers-Wannier transformation. In the continuum, it corresponds to gauging with a discrete torsion. It is natural consider extending our tensor network presentation to such maps.
\item What is the phase diagram for the duality-preserving deformed model in \eqref{deformtH} in 3+1d? Both the points $\lambda=0$ and $\lambda=1$ correspond to a first-order transition between the toric code phase and the product state, and it is plausible that there is no phase transition in between these two values of $\lambda$. This is unlike the situation for the analogous 1+1d model in \eqref{1d-deformedHlambda} where the $\lambda=0$ point is the Ising CFT while the $\lambda=1$ point is gapped, and there has to be a nontrivial phase transition in between \cite{OBrien:2017wmx}.
\end{itemize}

\section*{Acknowledgements}

We are grateful to Clay C\'ordova, Paul Fendley, Po-Shen Hsin, Ho Tat Lam, Abhinav Prem, Nathan Seiberg, Ruben Verresen, Carolyn Zhang for stimulating discussions. 
We thank Nathan Seiberg for comments on the draft. 
SHS thanks Yichul Choi, Yaman Sanghavi, and Yunqin Zheng for collaborations on a related project about non-invertible symmetries in 2+1d lattice gauge theory coupled to Ising matter \cite{Choi:2024rjm}. 
PG thanks Tzu-Chen Huang for collaborations on a related project about Wegner duality-preserving deformations of the 3+1d lattice $\mathbb Z_2$ gauge theory \cite{deform}.  
We also thank the authors of \cite{dipoleKW}, which discusses non-invertible symmetries associated with gauging modulated symmetries in 1+1d,  for  coordinated submissions. 
The work of PG was supported by the Simons Collaboration on Global Categorical Symmetries. 
The work of SHS was supported in part by NSF grant PHY-2210182. 
The work of NT was supported by the Walter Burke Institute for Theoretical Physics at Caltech.
This work was initiated at the Aspen Center for Physics during the workshops ``Traversing the Particle Physics Peaks: Phenomenology to Formal'' and ``New Frontiers for Quantum Dynamics" in 2023, which are supported by National Science Foundation grant PHY-2210452. The authors of this paper were ordered alphabetically.

\appendix

\section{ZX-calculus for the duality and condensation operators} \label{app:zx-3d}

In this appendix, we provide more details of the ZX-calculus presentation of the duality and condensation operators in the 3+1d lattice $\mathbb{Z}_2$ gauge theory.

\subsection{Condensation operator}\label{app:condensation}

Below we show that the two expressions \eqref{C-sum} and \eqref{C-def} for the condensation operator are equivalent.

\tikzexternalenable
First, recall that
\tikzsetnextfilename{projection}
\ie\label{projection-identity}
\begin{ZX}[circuit]
\zxZ[xshift=8pt]{m\pi} \ar[dr] \ar[ddr] &[-4pt] &\\[-8pt]
\zxN-{1} \ar[r] & \zxX{} \ar[d,3 vdots] \rar & \zxN-{1}\\
\zxN-{n}\ar[r] & \zxX{} \rar & \zxN-{n}\\
\end{ZX}
= \frac1{2^{n/2}} [1+(-1)^m X_1\cdots X_n]~,
\fe
\tikzexternaldisable
which follows from the discussion below \eqref{1d-Cn-def}. Using this identity for each green dot on the link in the ZX-diagram \eqref{C-def} of $\mathsf C$, we can write
\ie
\mathsf C &\overset{\ref{sf}}{=} 2^{7V/2}\raisebox{-0.5\height}{\includegraphics[scale=0.17]{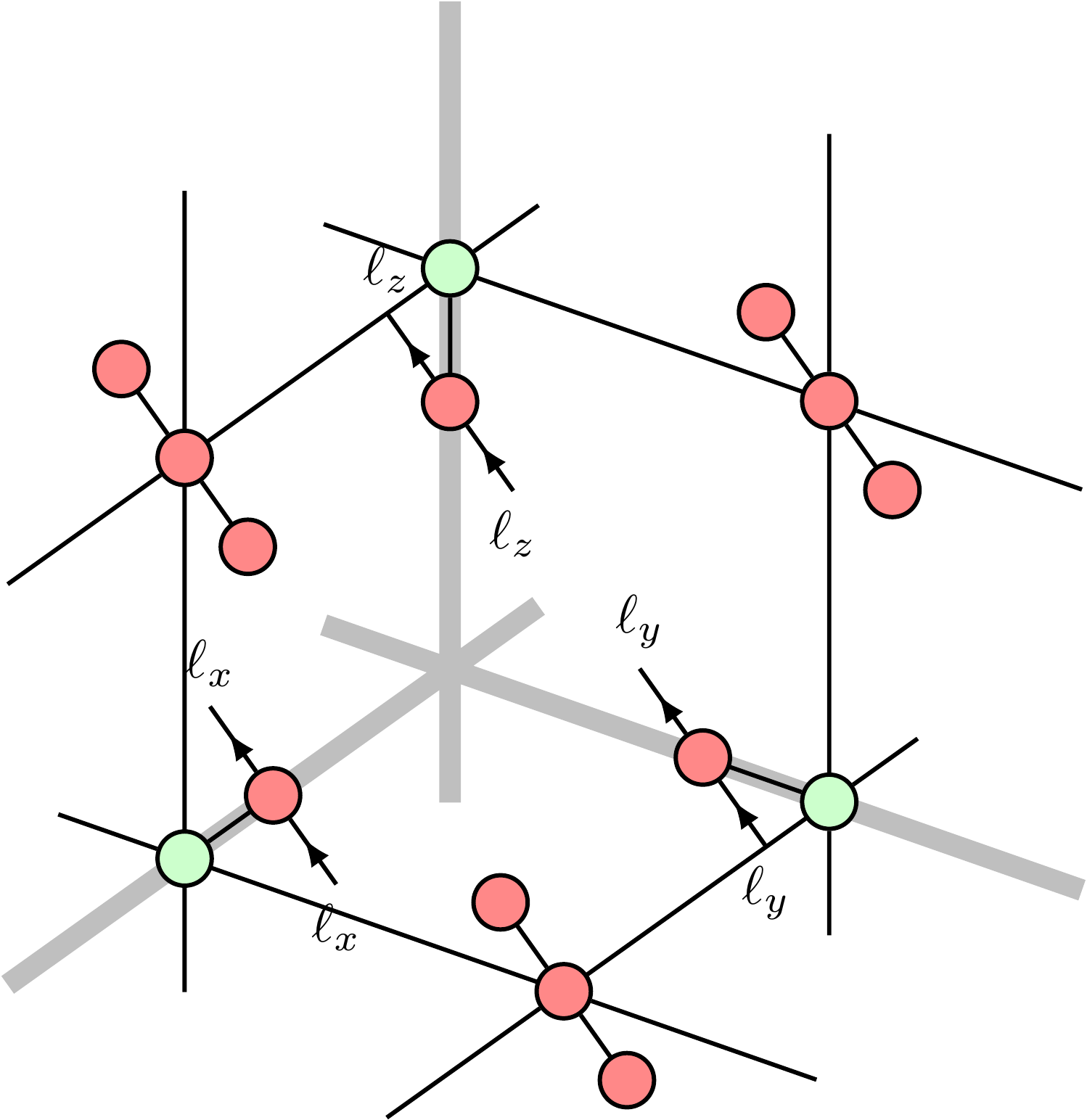}}
= \underbrace{\frac{2^{7V/2}\cdot 2^{3V}}{2^{15V/2}}}_{1/2^V}~\left(\bigotimes_p \<0|_p\right) \left[\prod_\ell \left(1+ X_\ell \prod_{p\ni \ell} \widehat X_p\right)\right] \left(\bigotimes_p |0\>_p\right)~,
\fe
where we put a hat on the auxiliary Pauli operator $\widehat X_p$ to distinguish it from the physical Pauli operator $X_\ell$. Expanding the operator in the middle, we get
\ie
\mathsf C
&= \frac1{2^V}\sum_{\{m_\ell = 0,1\}} \prod_\ell X_\ell^{m_\ell} \prod_p \delta_{(dm)_p,0}
&={1\over 2^V} \sum_{\widehat \Sigma} \eta(\widehat \Sigma)~,
\fe
where, in the third line, $\delta_{a,b}$ is the Kronecker delta function modulo $2$, i.e., $\delta_{a,b} = 1$ if $a-b=0\mod 2$ and $0$ if $a-b=1 \mod 2$. 
This is precisely the expression for $\mathsf C$ in \eqref{C-sum}. 

\subsection{Duality operator}\label{app:D2}

Here we derive the operator algebra involving $\mathsf{D}$ using ZX-calculus. 
We start with \eqref{D-fusion}, or equivalently, \eqref{DD2}:
\ie
\mathsf D^2 = \mathsf C\,T_{1,1,1}~.
\fe
First, using \eqref{D-def}, \eqref{D-plaq}, and \eqref{T-def}, we have
\ie
\mathsf D^2 = \mathsf D_{\text{l}\leftarrow\text{p}} T_{\text{p}\leftarrow\text{l}} \mathsf D_{\text{l}\leftarrow\text{p}} T_{\text{p}\leftarrow\text{l}} = \mathsf D_{\text{l}\leftarrow\text{p}} \mathsf D_{\text{p}\leftarrow\text{l}} T_{\text{l}\leftarrow\text{p}} T_{\text{p}\leftarrow\text{l}} = \mathsf D_{\text{l}\leftarrow\text{p}} \mathsf D_{\text{p}\leftarrow\text{l}} T_{1,1,1}~.
\fe
The product $\mathsf D_{\text{l}\leftarrow\text{p}}\mathsf D_{\text{p}\leftarrow\text{l}}$ can be computed   as follows. We have
\ie
\mathsf D_{\text{l}\leftarrow\text{p}}\mathsf D_{\text{p}\leftarrow\text{l}} = 2^{8V}~\raisebox{-0.5\height}{\includegraphics[scale=0.17]{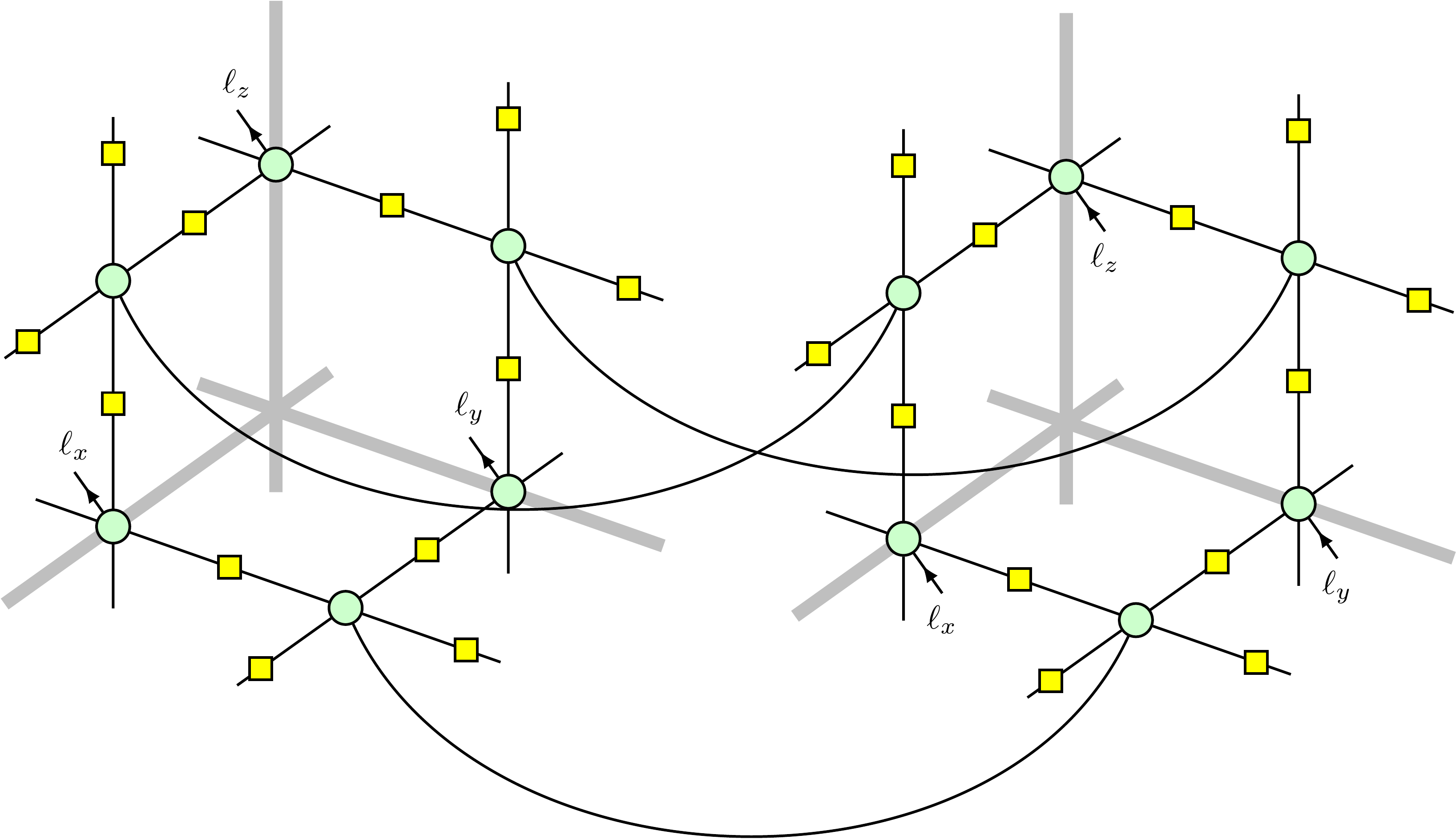}}~.
\fe
Using color change \ref{cc} and spider fusion \ref{sf}, we can simplify this diagram to  
\ie
\mathsf D_{\text{l}\leftarrow\text{p}}\mathsf D_{\text{p}\leftarrow\text{l}} = 2^{8V}~\raisebox{-0.5\height}{\includegraphics[scale=0.17]{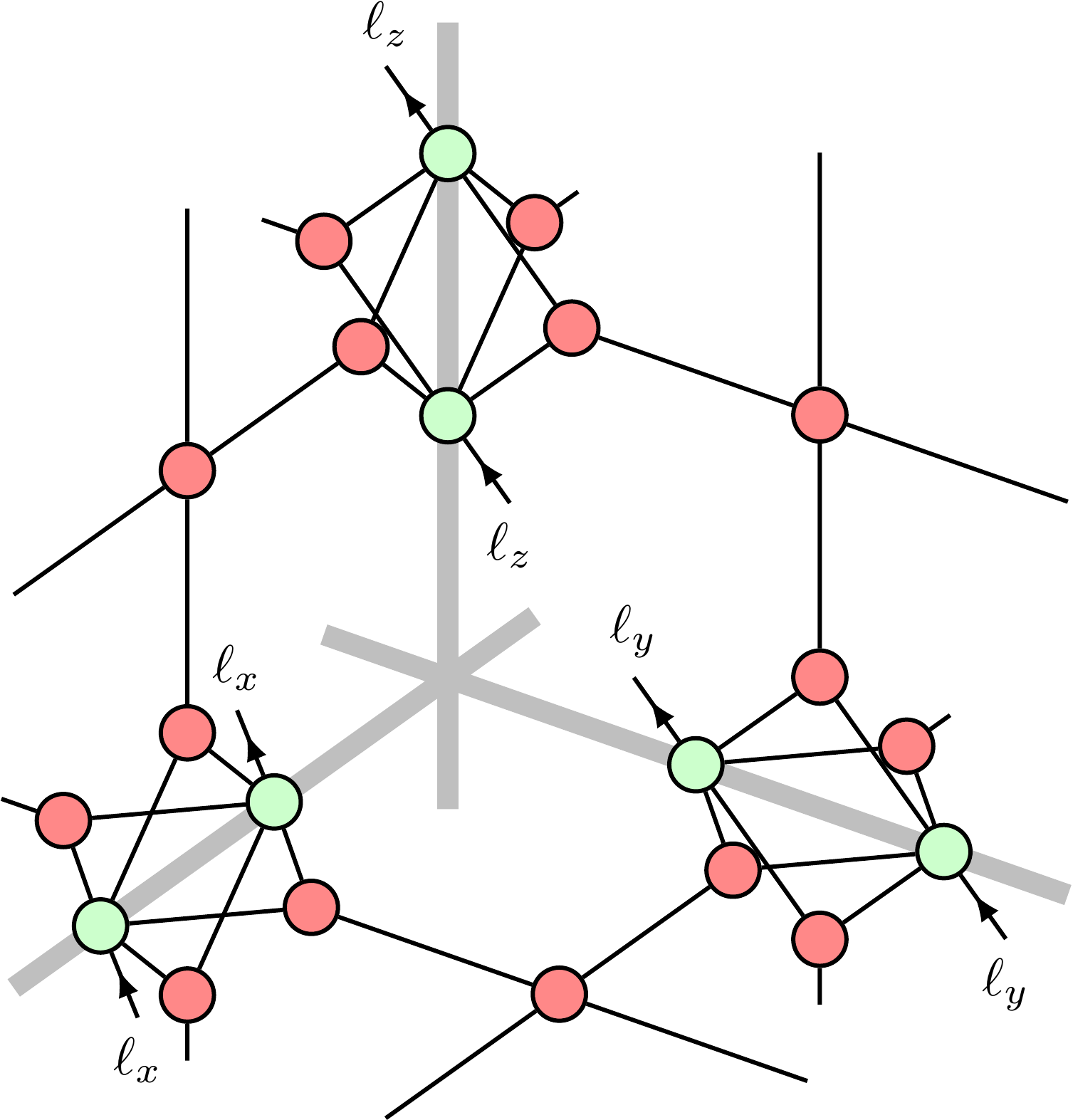}}~.
\fe
Now, we use the generalized bialgebra \ref{gb} to simplify it further to
\ie
\mathsf D_{\text{l}\leftarrow\text{p}}\mathsf D_{\text{p}\leftarrow\text{l}} = \underbrace{\frac{2^{8V}}{2^{9V/2}}}_{2^{7V/2}}~\raisebox{-0.5\height}{\includegraphics[scale=0.17]{C}} = \mathsf C~.
\fe
It follows that $\mathsf D^2 = \mathsf C\,T_{1,1,1}$.

It is also instructive to compute $\mathsf D \mathsf C$ using ZX-calculus. We have
\ie
\mathsf D_{\text{p}\leftarrow\text{l}} \mathsf C = \underbrace{2^{4V}\cdot 2^{7V/2}}_{2^{15V/2}}~\raisebox{-0.5\height}{\includegraphics[scale=0.17]{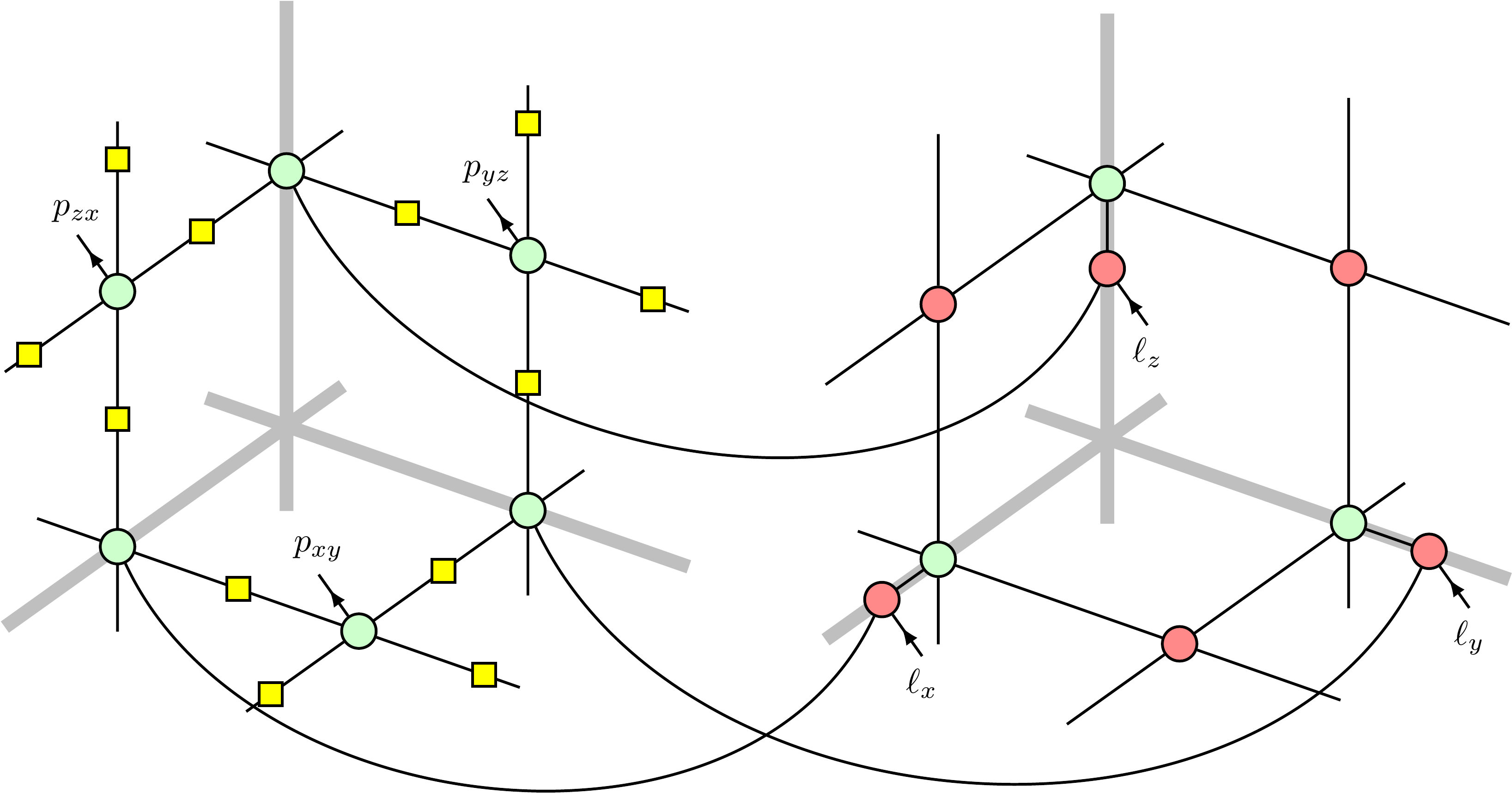}}~.
\fe
Applying the generalized bialgebra \ref{gb} along the curved wires, along with spider fusions \ref{sf}, we get
\ie
\mathsf D_{\text{p}\leftarrow\text{l}} \mathsf C = \underbrace{2^{15V/2}\cdot 2^{9V/2}}_{2^{12V}}~\raisebox{-0.5\height}{\includegraphics[scale=0.17]{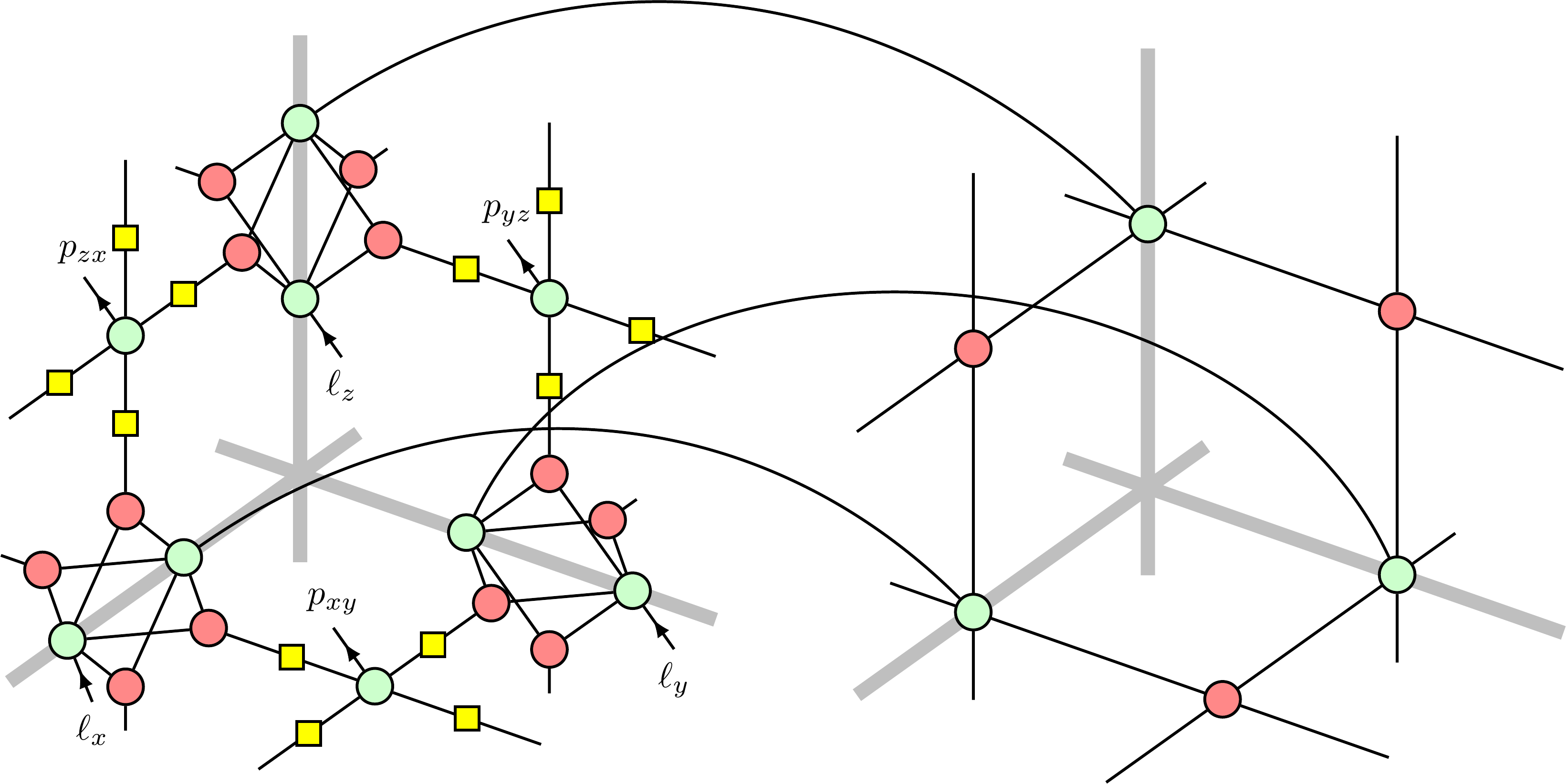}}~.
\fe
After a series of spider fusions \ref{sf} and color changes \ref{cc'}, we can modify the above diagram to
\ie
\mathsf D_{\text{p}\leftarrow\text{l}} \mathsf C = 2^{12V}~\raisebox{-0.5\height}{\includegraphics[scale=0.17]{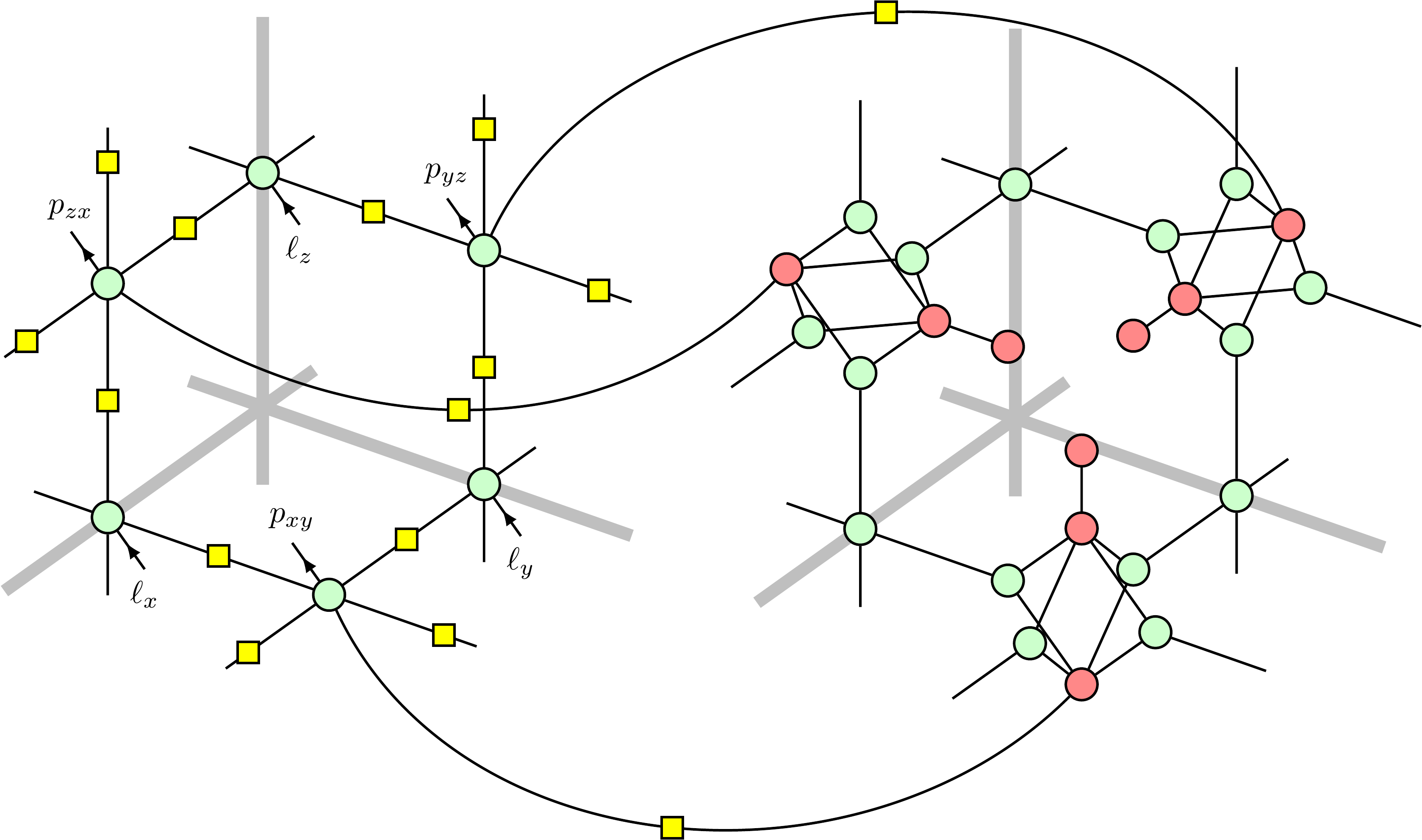}}~.
\fe
We now apply the generalized bialgebra \ref{gb} again to get
\ie
\mathsf D_{\text{p}\leftarrow\text{l}} \mathsf C = \underbrace{\frac{2^{12V}}{2^{9V/2}}}_{2^{15V/2}}~\raisebox{-0.5\height}{\includegraphics[scale=0.17]{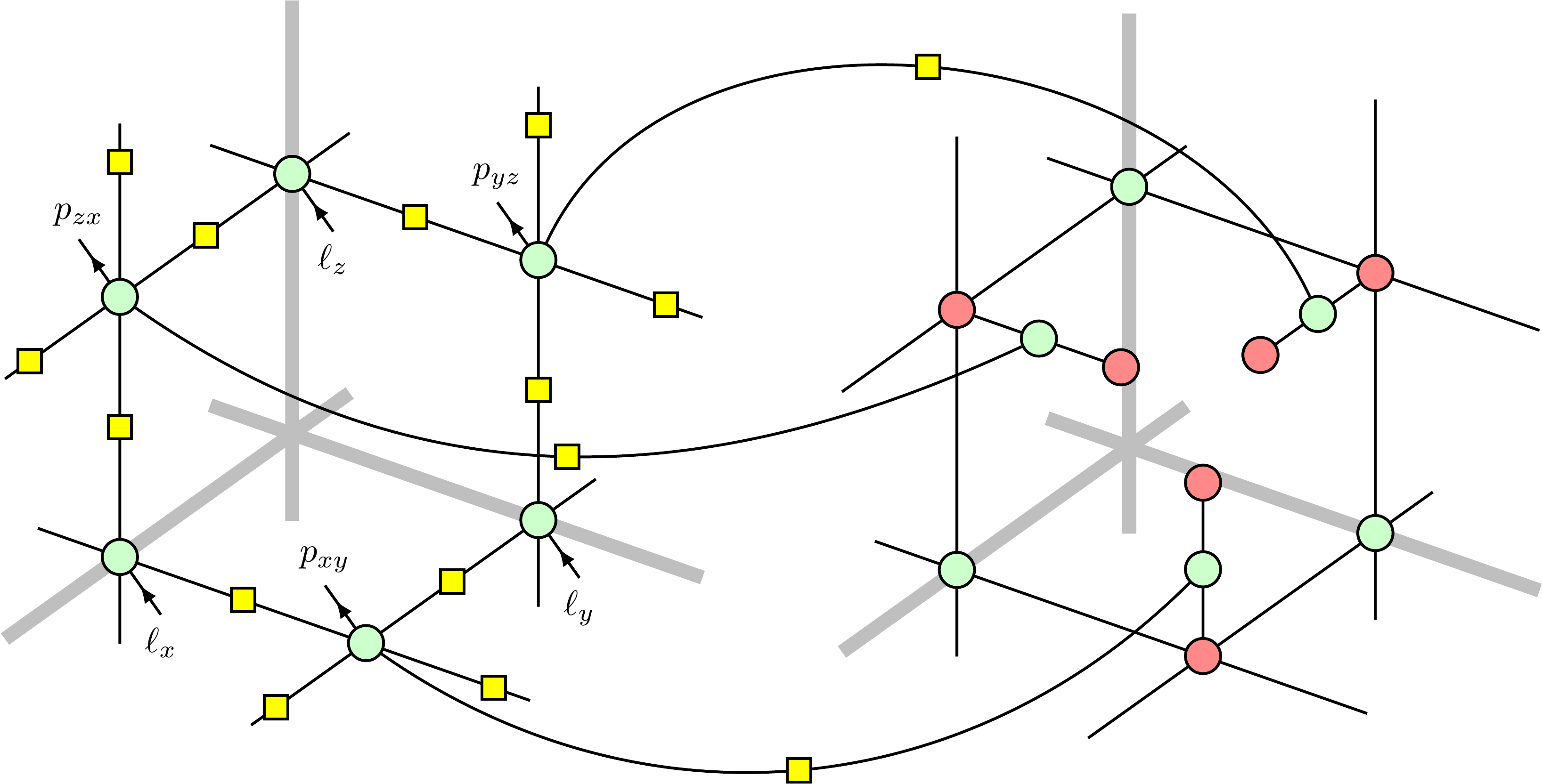}}~.
\fe
Using state copy \ref{sc}, spider fusion \ref{sf}, and color change \ref{cc'}, we finally get
\ie\label{DC=<C>D}
\mathsf D_{\text{p}\leftarrow\text{l}} \mathsf C &= \underbrace{\frac{2^{15V/2}}{2^{3V/2}}}_{2^{6V}}~\raisebox{-0.5\height}{\includegraphics[scale=0.17]{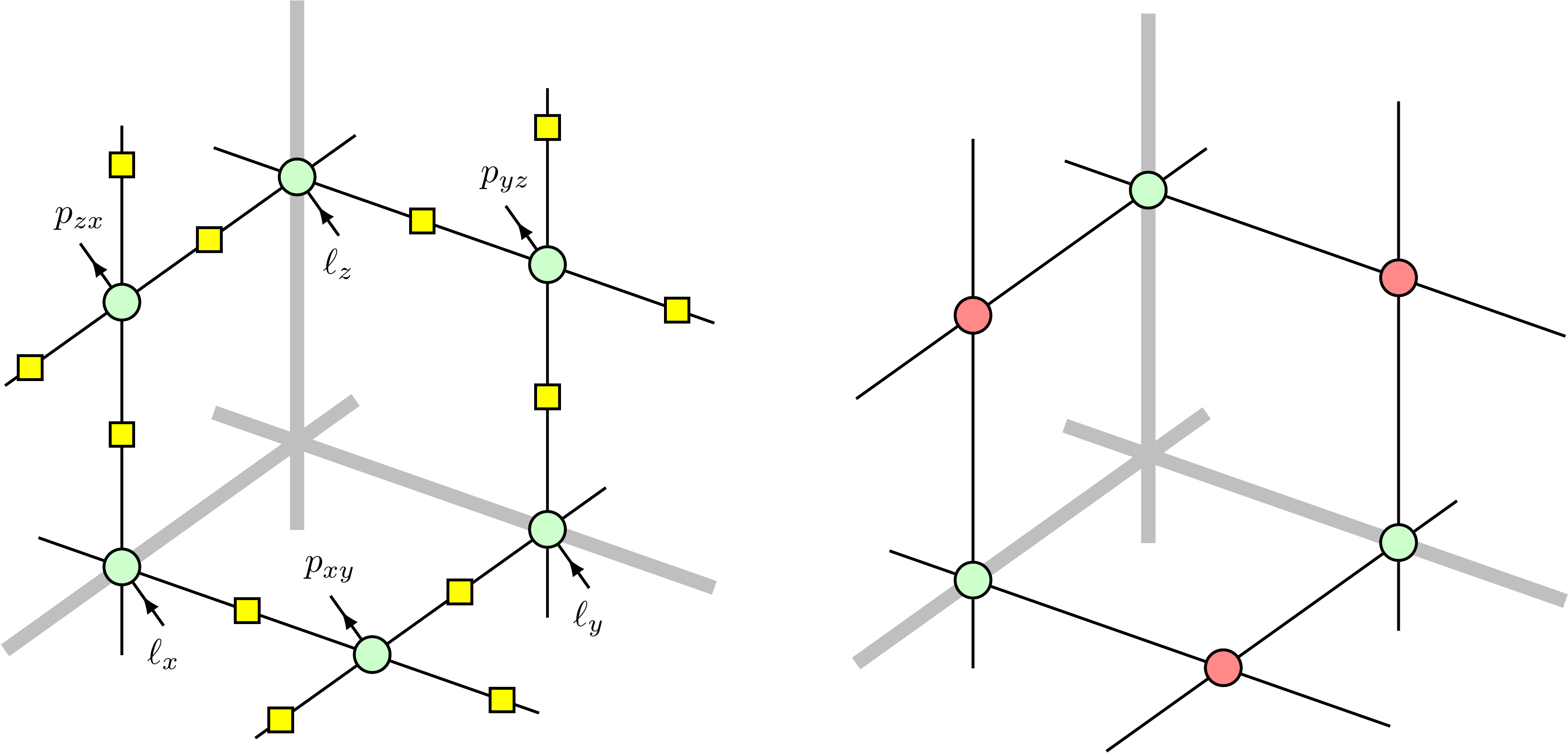}}
\\
&= \<{+}\cdots{+}| \mathsf C |{+}\cdots{+}\> \mathsf D_{\text{p}\leftarrow\text{l}}~,
\fe
where, up to scalar factors, we identified the left diagram as the operator $\mathsf D_{\text{p}\leftarrow\text{l}}$ and the right diagram as the matrix element $\<{+}\cdots{+}| \mathsf C |{+}\cdots{+}\>$. The latter can be seen as follows:
\ie
\<{+}\cdots{+}| \mathsf C |{+}\cdots{+}\> &= \underbrace{\frac{2^{7V/2}}{2^{3V}}}_{2^{V/2}}~\raisebox{-0.5\height}{\includegraphics[scale=0.17]{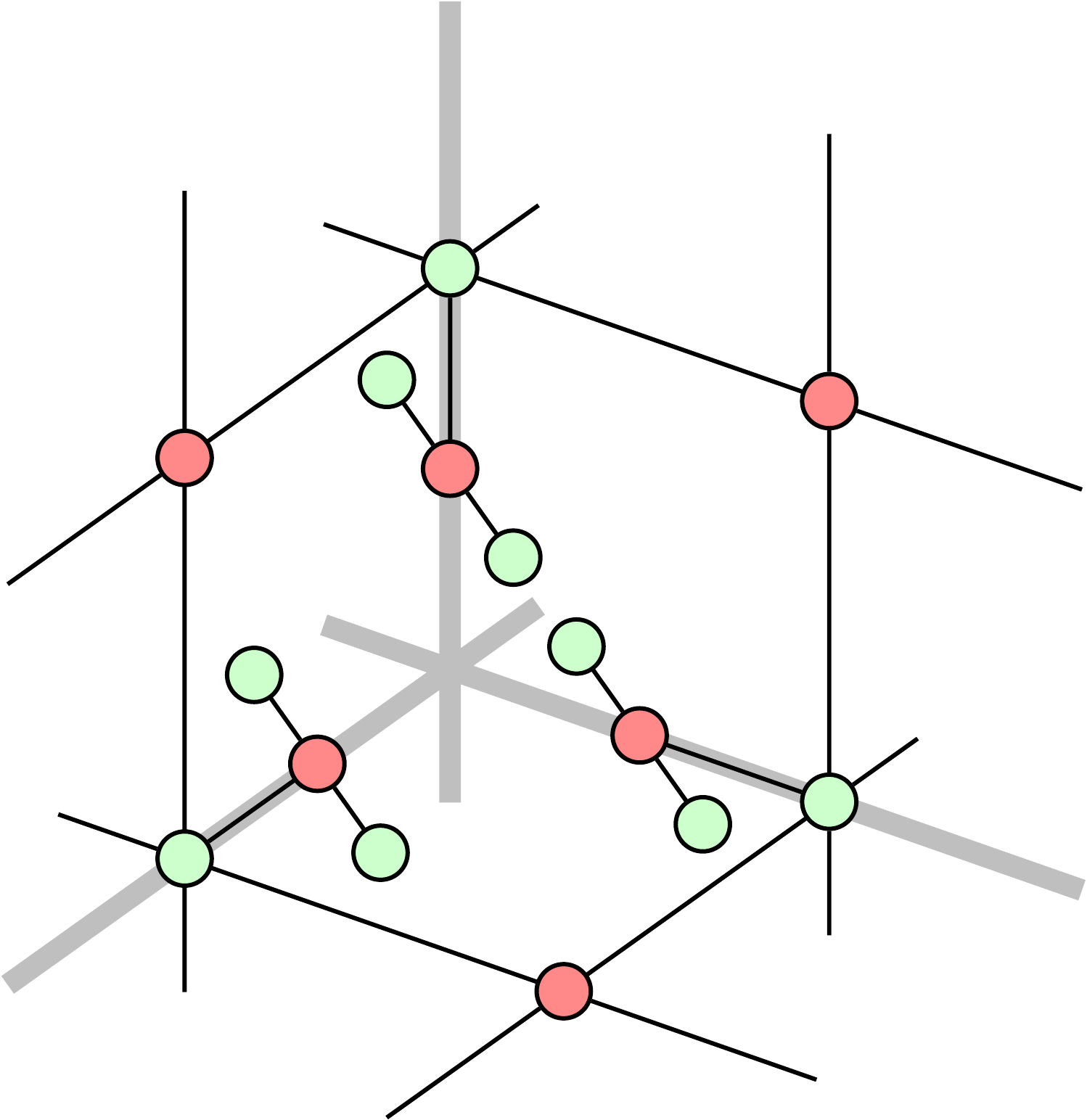}}
\overset{\ref{sc},\ref{sf},\ref{s}}{=} \underbrace{\frac{2^{V/2}\cdot 2^{3V}}{2^{3V/2}}}_{2^{2V}}~\raisebox{-0.5\height}{\includegraphics[scale=0.17]{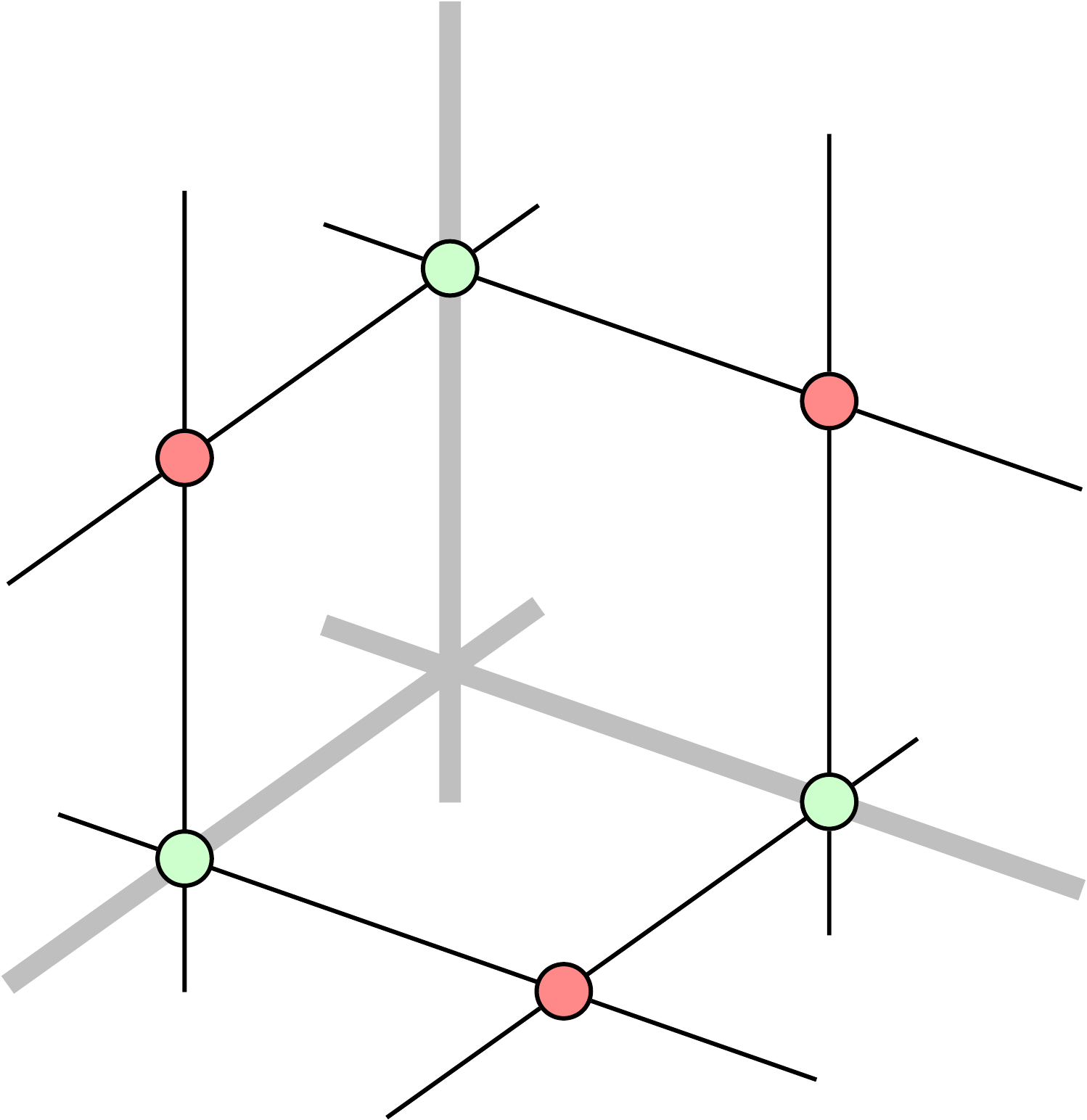}}~.
\fe
Multiplying $T_{\text{l}\leftarrow\text{p}}$ on the left on both sides of \eqref{DC=<C>D} gives $\mathsf D \mathsf C = \<{+}\cdots{+}| \mathsf C |{+}\cdots{+}\> \mathsf D$.\footnote{The matrix element $\<{+}\cdots{+}| \mathsf C |{+}\cdots{+}\>$ can be interpreted as the partition function of the 3d (classical) topological $\mathbb Z_2$ gauge theory on $T^3$. This is the lattice counterpart of the continuum result in the second line of \eqref{contDD2}, where the fusion coefficient $(\mathcal Z_2)_0$ is the partition function of the 3d $\mathbb Z_2$ gauge theory.} 
The matrix element here can be computed using \eqref{C-Pauli-def}:
\ie
\<{+}\cdots{+}| \mathsf C |{+}\cdots{+}\> = \frac12\cdot2^3 = 4~,
\fe
from which, we get $\mathsf D \mathsf C = 4 \mathsf D$. Similarly, one finds $\mathsf C\mathsf D =4\mathsf D$.

\subsection{Action of $\mathsf D$ on the states}\label{app:D-action-states}

Below we derive the action of $\mathsf D$ in \eqref{Daction} using ZX-calculus. 
First, using \eqref{D-eta,D-G}, we have
\ie
\mathsf D |\xi\> = 2^{(V-1)/2} \mathsf D |0\cdots0\>~.
\fe
Now, we have
\ie
\mathsf D_{\text{p}\leftarrow\text{l}} |0\cdots0\> = \underbrace{\frac{2^{4V}}{2^{9V/2}}}_{1/2^{V/2}}~\raisebox{-0.5\height}{\includegraphics[scale=0.17]{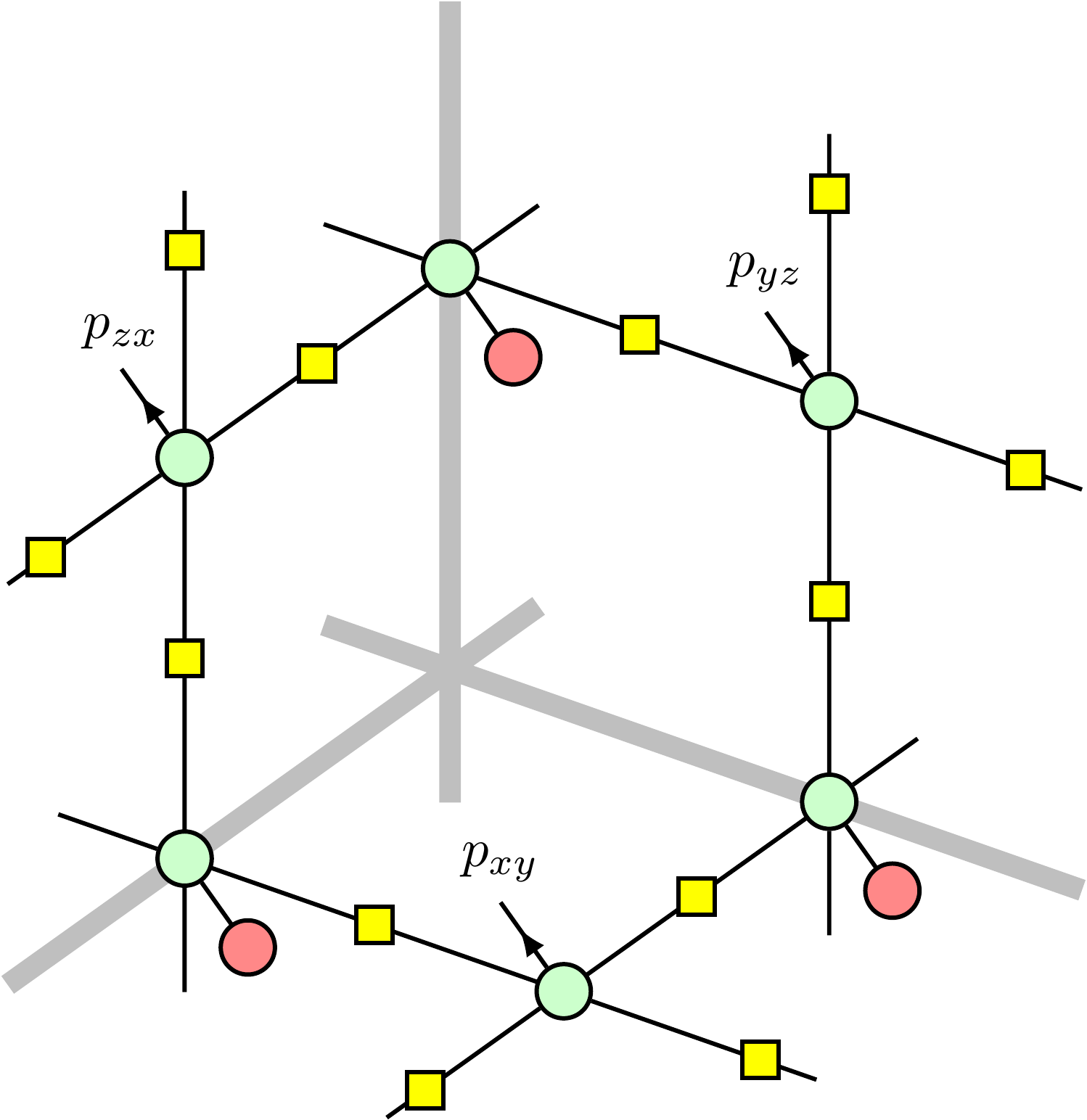}}~.
\fe
Applying state copy \ref{sc} through green dots on the links, color change \ref{cc'}, and spider fusion \ref{sf} into the green dots on the plaquettes, we get
\ie
\mathsf D_{\text{p}\leftarrow\text{l}} |0\cdots0\> &= \frac{1}{2^{V/2}\cdot 2^{9V/2}}~\raisebox{-0.5\height}{\includegraphics[scale=0.17]{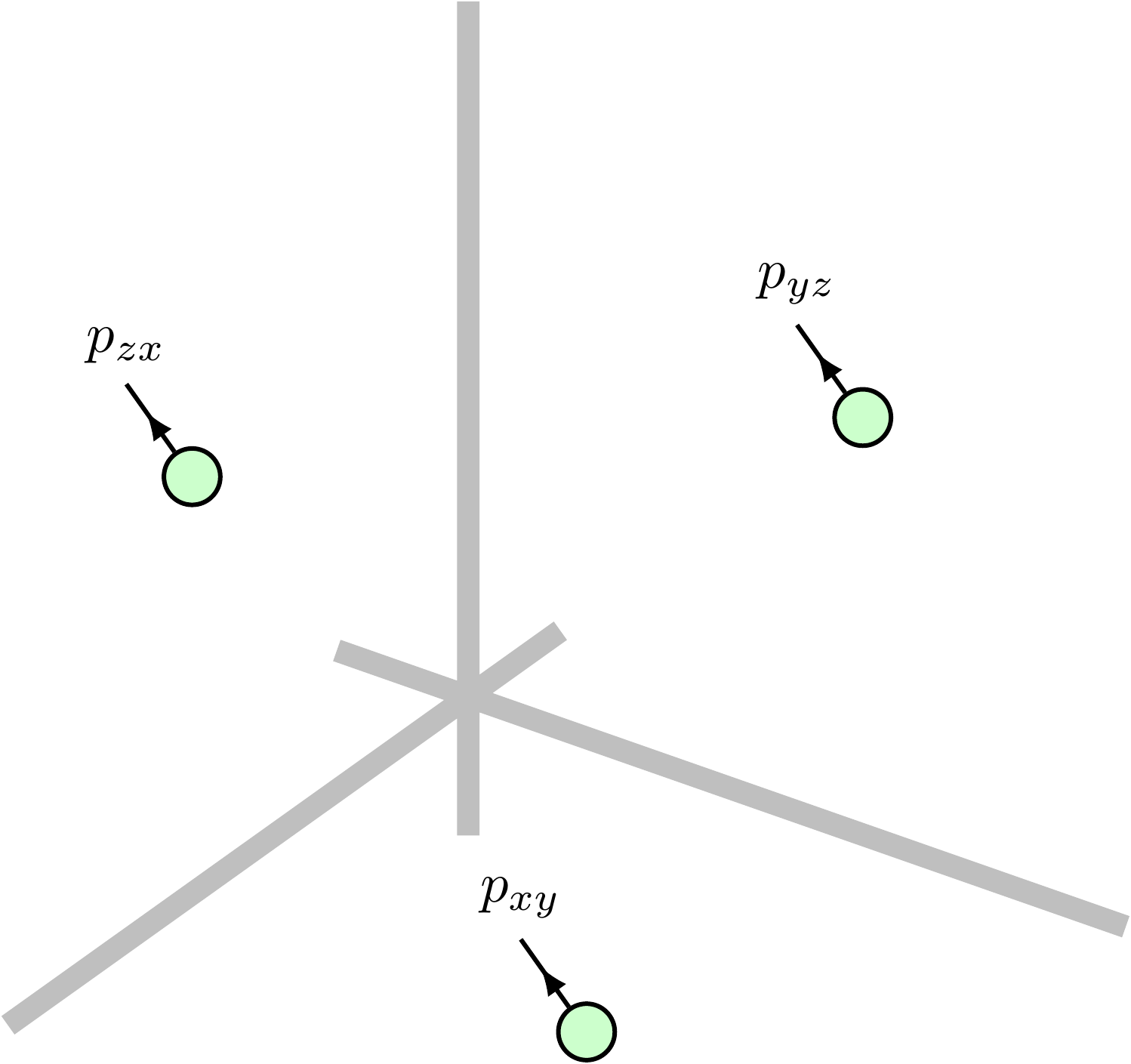}}~.
\fe
Multiplying by $T_{\text{l}\leftarrow\text{p}}$ on the left on both sides, we get
\ie
\mathsf D |0\cdots0\> = \frac{1}{2^{V/2}} |{+}\cdots{+}\> \implies \mathsf D|\xi\> = \frac1{\sqrt2} |{+}\cdots{+}\>~.
\fe
Using \eqref{D-fusion} and \eqref{sumofeta}, we further have
\ie
\mathsf D|{+}\cdots{+}\> = \sqrt2\,\mathsf D^2|\xi\> = \frac1{\sqrt2} \prod_{i<j} (1+\eta_{ij}) \prod_s \left( \frac{1+G_s}{2} \right) |\xi\> = \frac1{\sqrt2} \sum_{\xi\in\{0,1\}^3} |\xi\>.
\fe
Hence, we have derived \eqref{Daction}.

\subsection{Ground states of 3+1d toric code}\label{app:TC}
Here, we relate the loop-gas and membrane-gas representations of the ground states of the 3+1d toric code.

Recall that the ground states of the 3+1d toric code are given by \eqref{TC},
\ie
|\xi\> := 2^{(V-1)/2}\prod_{i<j} \eta_{ij}^{\xi_{ij}} \prod_{s} \left(\frac{1+G_s}{2}\right) |0\cdots0\>~,
\fe
where $\xi = (\xi_{xy},\xi_{yz},\xi_{zx}) \in \{0,1\}^3$ with $\xi_{ij}$ defined modulo 2. An alternative basis for these states is
\ie
|\zeta\> &:= \frac1{2\sqrt2} \sum_{\xi \in \{0,1\}^3} (-1)^{\sum_{i<j}\zeta_{ij}\xi_{ij}} |\xi\>~,
\fe
where $\zeta = (\zeta_{xy},\zeta_{yz},\zeta_{zx}) \in \{0,1\}^3$ with $\zeta_{ij}$ defined modulo 2.

It is instructive to write the state $|\zeta\>$ as
\ie\label{zeta-gas-rep}
|\zeta\>&= 2\cdot 2^{V/2} \prod_{i<j} \left( \frac{1+(-1)^{\zeta_{ij}}\eta_{ij}}{2} \right) \prod_{s} \left(\frac{1+G_s}{2}\right) |0\cdots0\>
\\
&= 2^{V-1} \prod_{i,j,k\atop\text{cyclic}} W_k^{\zeta_{ij}} \prod_p \left(\frac{1+\prod_{\ell\in p}Z_\ell}{2}\right) |{+}\cdots{+}\>~,
\fe
where $W_k := W(\gamma_k)$ is the Wilson line operator \eqref{Wilson-op} along the curve $\gamma_k$, which is a non-contractible cycle in the $k$th spatial direction.\footnote{Since $W_k$ is multiplied by $\prod_p \left(\frac{1+\prod_{\ell\in p}Z_\ell}{2}\right)$ in \eqref{zeta-gas-rep}, the choice of $\gamma_k$ does not matter.} One can recognize the first (second) line of \eqref{zeta-gas-rep} as the membrane-gas (loop-gas) representation of the ground states of the 3+1d toric code. 

We can use ZX-calculus to quickly see the equality of these two representations. Consider the state $|\zeta=0\>$ for simplicity (the other states are obtained by multiplying $\prod_{i,j,k\atop\text{cyclic}} W_k^{\zeta_{ij}}$). We have
\ie
|\zeta=0\>&=2\cdot 2^{V/2} \prod_{i<j} \left( \frac{1+\eta_{ij}}{2} \right) \prod_{s} \left(\frac{1+G_s}{2}\right) |0\cdots0\>
\\
&=\frac{2^{V/2}}{2} \mathsf C |0\cdots 0\> = \underbrace{\frac{2^{V/2} \cdot 2^{7V/2}}{2\cdot 2^{3V/2}}}_{\frac{2^{5V/2}}{2}}~\raisebox{-0.5\height}{\includegraphics[scale=0.17]{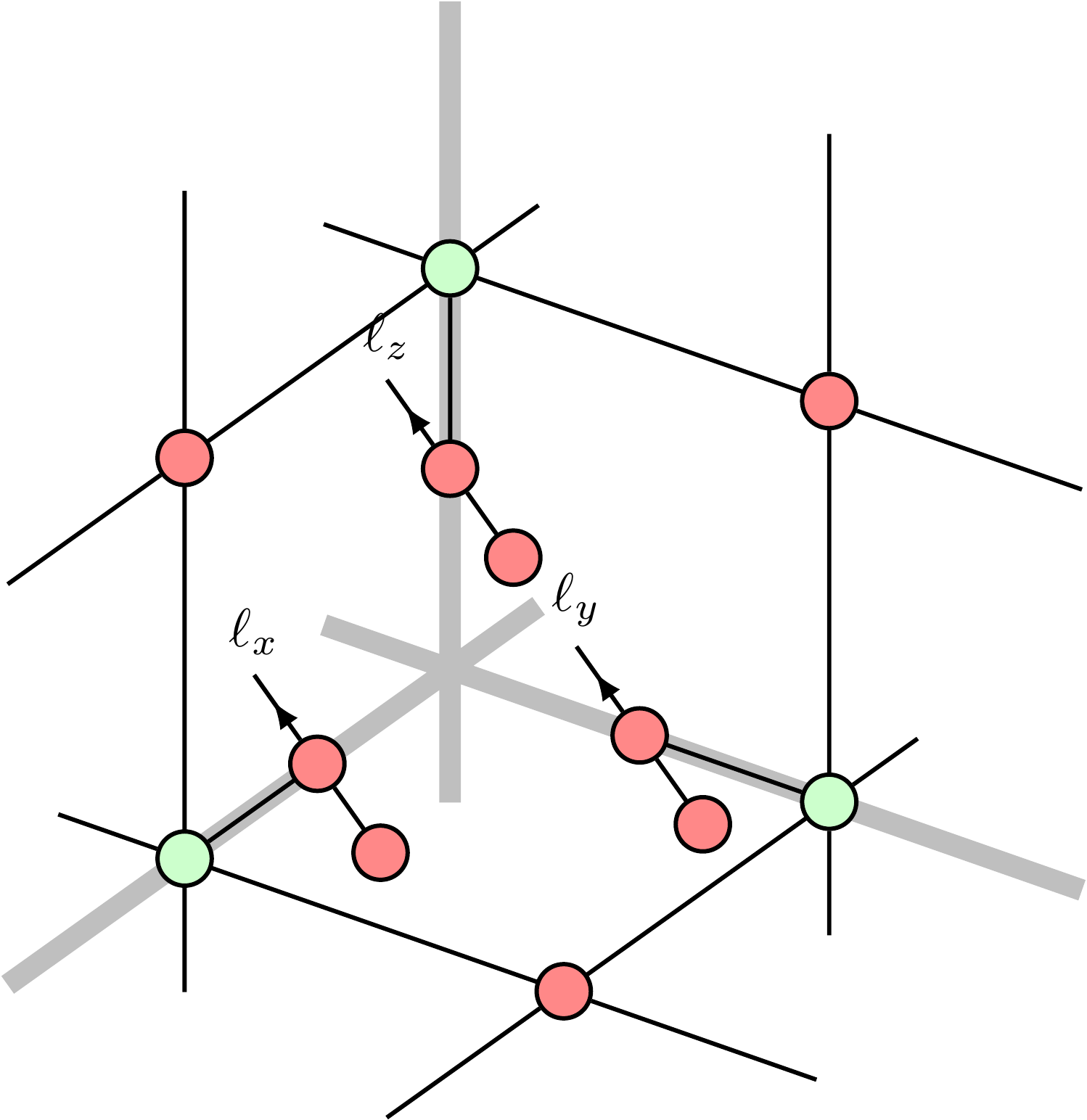}}~.
\fe
Applying spider fusion \ref{sf}, identity removal \ref{i}, and spider fusion \ref{sf} again, we get
\ie\label{zeta=0-zx}
|\zeta=0\> = \frac{2^{5V/2}}{2}~\raisebox{-0.5\height}{\includegraphics[scale=0.17]{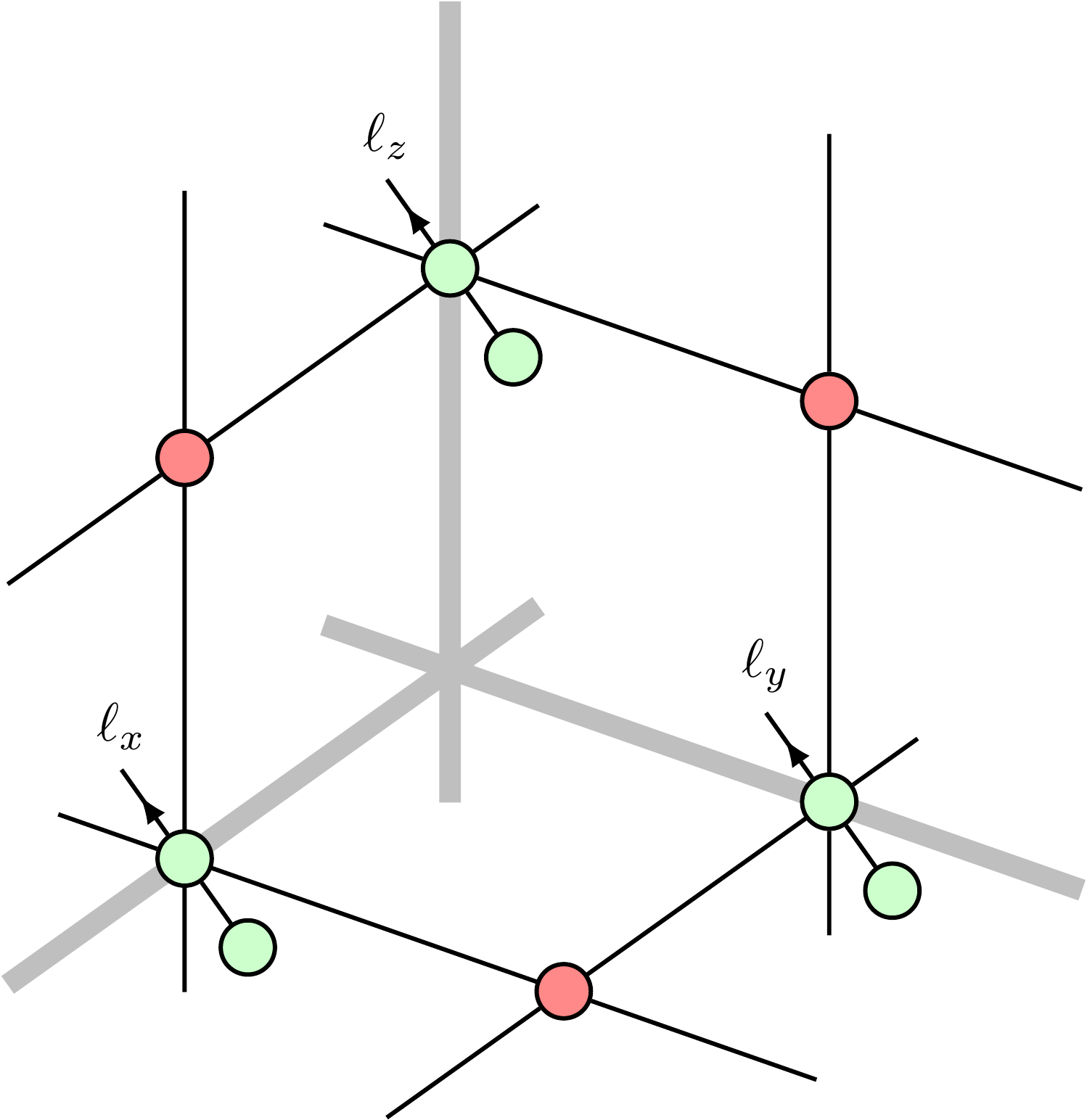}}~.
\fe
This diagram is called the 2-form representation of the state $|\zeta=0\>$ in \cite{Williamson:2020hxw}. (See also \cite{Delcamp:2020rds} for related discussion.) Using \eqref{projection-identity}, with Z/X spiders exchanged, on all the red dots, we get 
\ie
|\zeta=0\>
&= 2^{V-1} \prod_p \left(\frac{1+\prod_{\ell\in p}Z_\ell}{2}\right) |{+}\cdots{+}\>~,
\fe
which proves the equality of the first and second lines of \eqref{zeta-gas-rep}. As a consequence, we also have
\ie\label{xi=0}
|\xi\> &= 2^{(V-1)/2}\prod_{i<j} \eta_{ij}^{\xi_{ij}} \prod_{s} \left(\frac{1+G_s}{2}\right) |0\cdots0\>
\\
&= \sqrt2 \cdot 2^V \prod_{i,j,k\atop\text{cyclic}} \left( \frac{1+(-1)^{\xi_{ij}}W_k}{2} \right) \prod_p \left(\frac{1+\prod_{\ell\in p}Z_\ell}{2}\right) |{+}\cdots{+}\>~.
\fe
For the sake of completeness, we note that the state $|\xi=0\>$ can be represented as
\ie\label{xi=0-zx}
|\xi = 0\> &= \frac{2^{V}}{\sqrt2}~\raisebox{-0.5\height}{\includegraphics[scale=0.17]{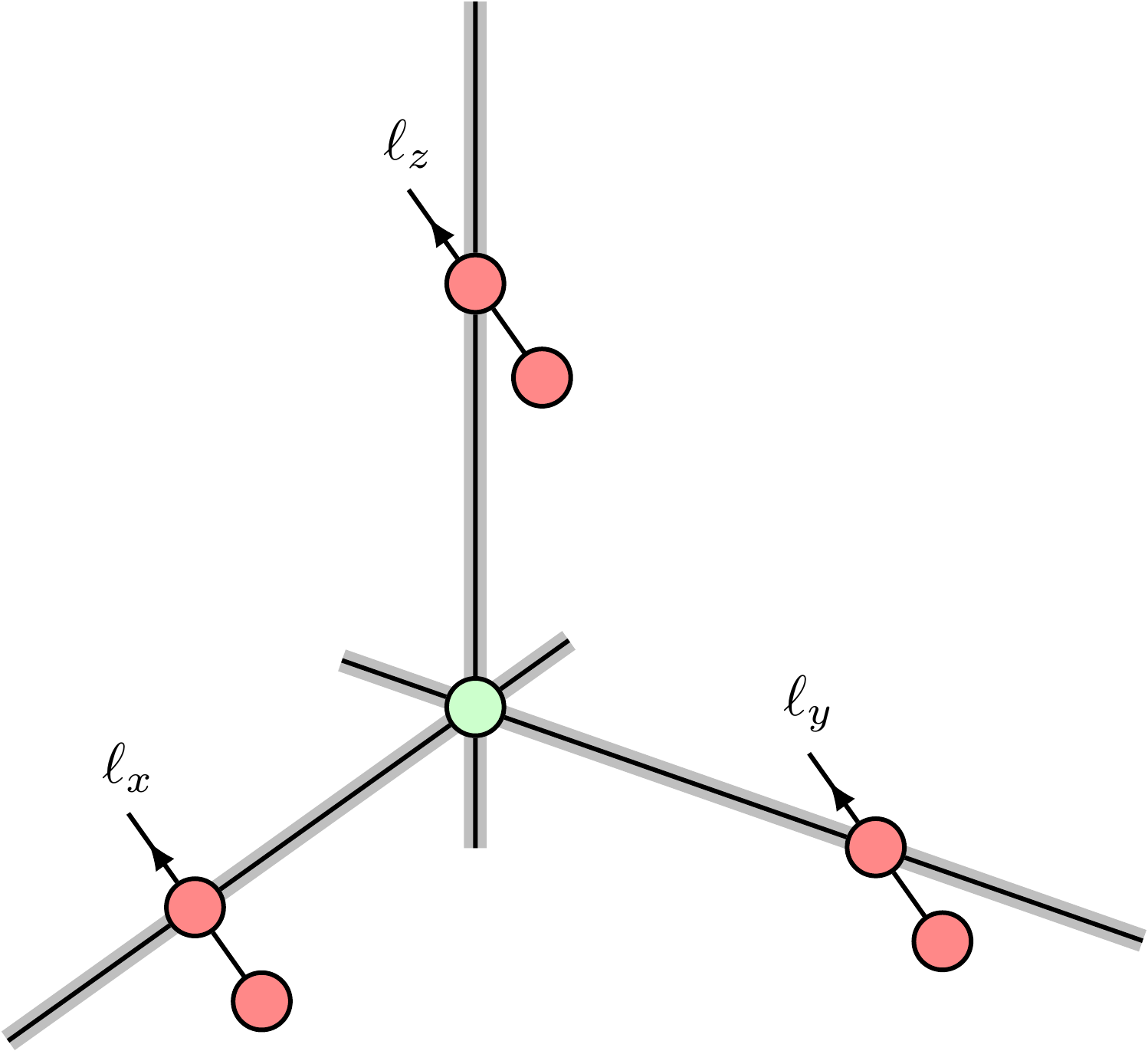}} \overset{\ref{sf},\ref{i}}{=} \frac{2^V}{\sqrt2}~\raisebox{-0.5\height}{\includegraphics[scale=0.17]{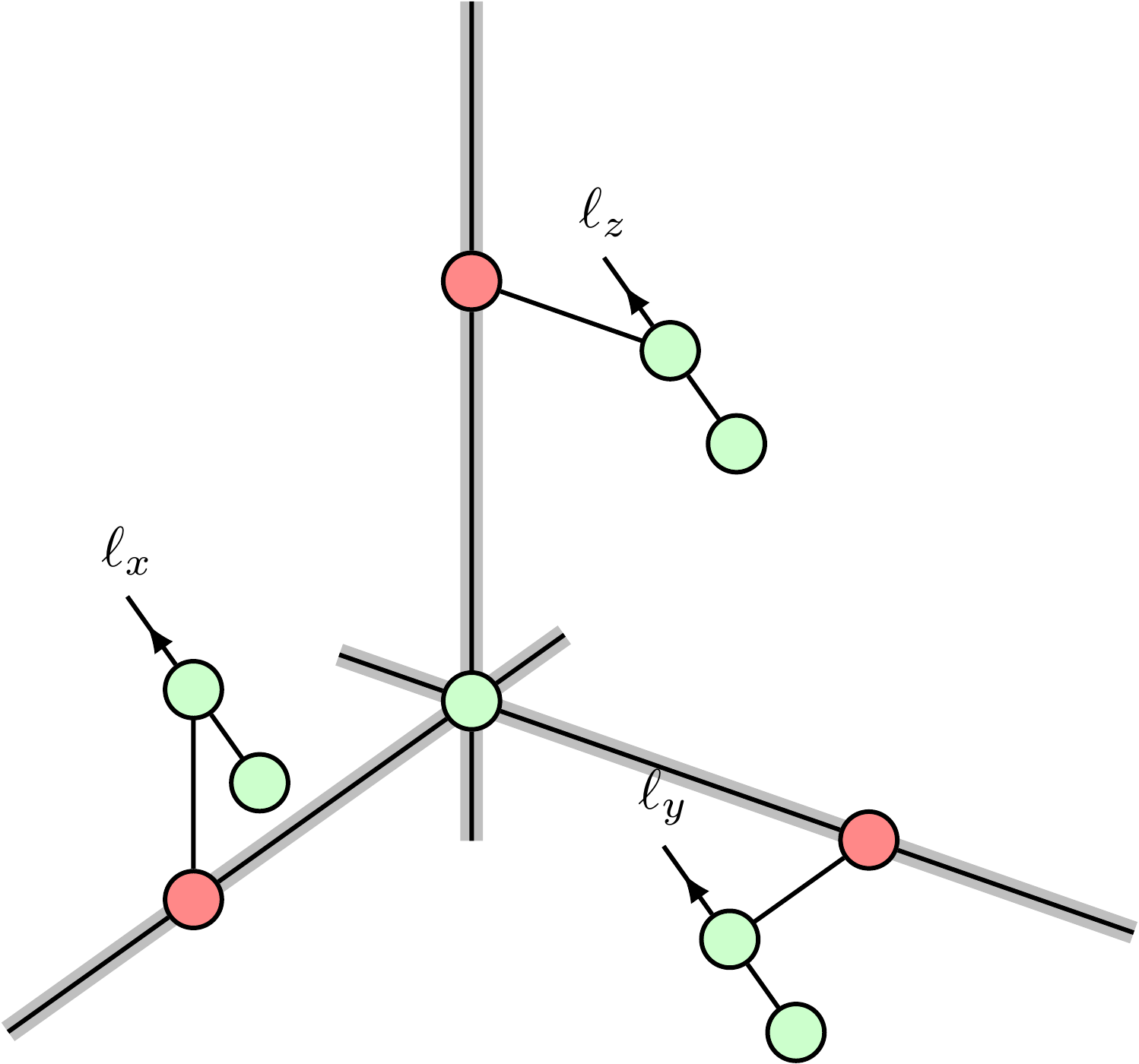}}~.
\fe
These diagrams correspond to the first and second lines of \eqref{xi=0}, respectively. The first diagram is called the 1-form representation of the state $|\xi=0\>$ in \cite{Williamson:2020hxw}. (See also \cite{Delcamp:2020rds} for related discussion.) On the other hand, the second diagram can be thought of as the action of the condensation operator $\mathsf C^{(2)}$ for the 2-form symmetry of the 3+1d toric code on the product state $|{+}\cdots{+}\>$, i.e.,
\ie
|\xi=0\> = \frac{2^{V}}{\sqrt2} \mathsf C^{(2)} |{+}\cdots{+}\>~,
\fe
where
\ie
\mathsf C^{(2)} &:= 2^{3V/2}~\raisebox{-0.5\height}{\includegraphics[scale=0.17]{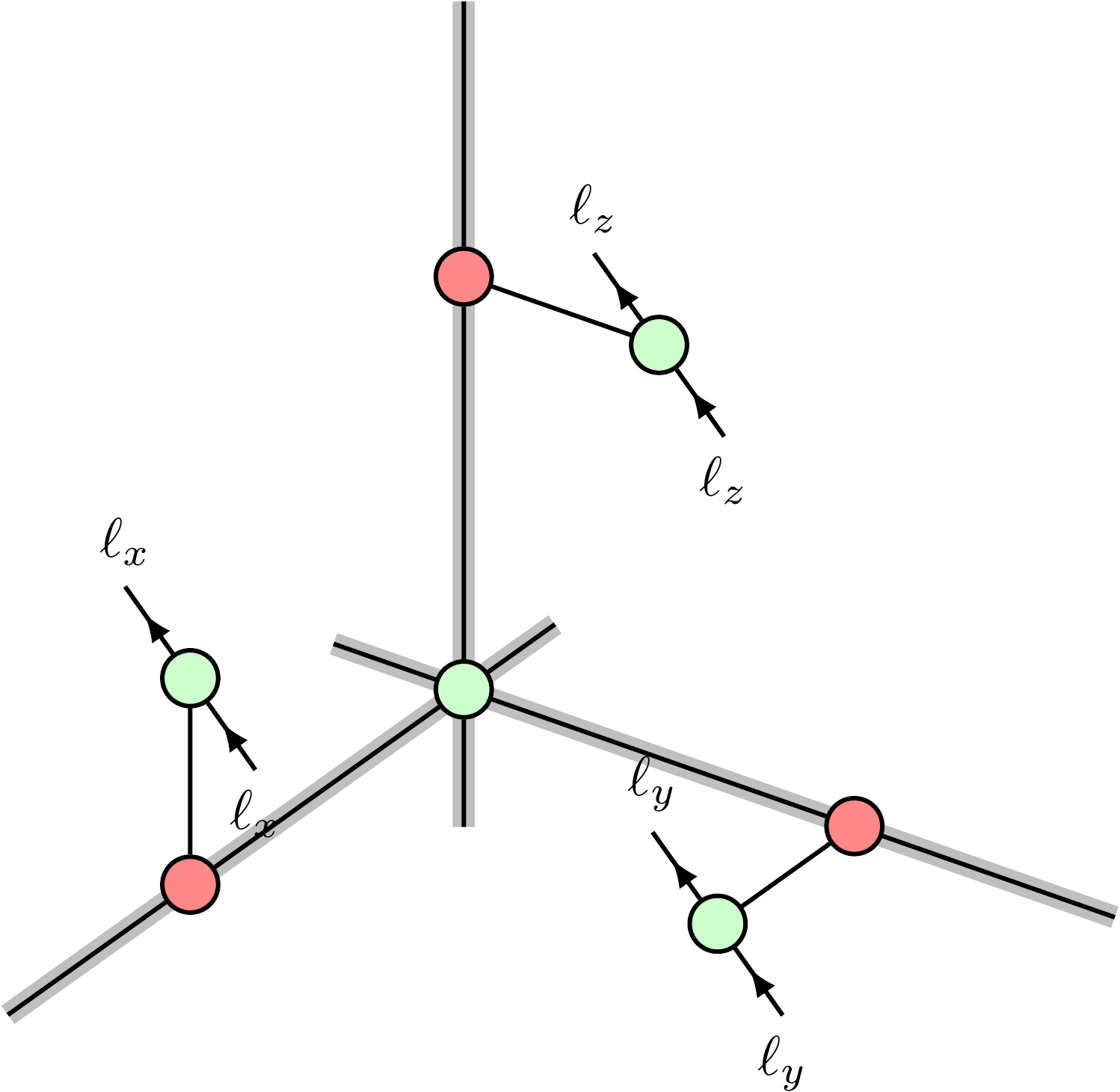}}
\\
&= \frac1{2^{2V}} \sum_\gamma W(\gamma) = \frac14 \prod_k (1+W_k) \prod_p \left( \frac{1+\prod_{\ell \in p} Z_\ell}{2} \right)~.
\fe

\section{Higher quantum symmetry}\label{app:higher-quan-sym}

It is well-known that gauging a finite global symmetry leads to a quantum (or dual) finite symmetry \cite{Vafa:1989ih}. 
For example, gauging a $\mathbb Z_N$ $p$-form symmetry in $d$-dimensional spacetime gives a dual $\mathbb Z_N$ $(d-p-2)$-form symmetry \cite{Gaiotto:2014kfa,Tachikawa:2017gyf}. Similarly, ``higher gauging,'' where the symmetry is gauged on a submanifold of positive codimension, results in a dual symmetry referred to as ``higher quantum symmetry'' \cite{Roumpedakis:2022aik}. One way to construct a higher quantum symmetry is by pushing a charged object from the bulk across a condensation defect. In this appendix, we use this to construct the higher quantum symmetry operators in the 3+1d lattice $\mathbb Z_2$ gauge theory.

The ZX-diagram for the Wilson line operator is
\ie\label{Wilson-op-ZX}
W(\gamma) = \prod_{\ell \in \gamma} Z_\ell = \raisebox{-0.5\height}{\includegraphics[scale=0.17]{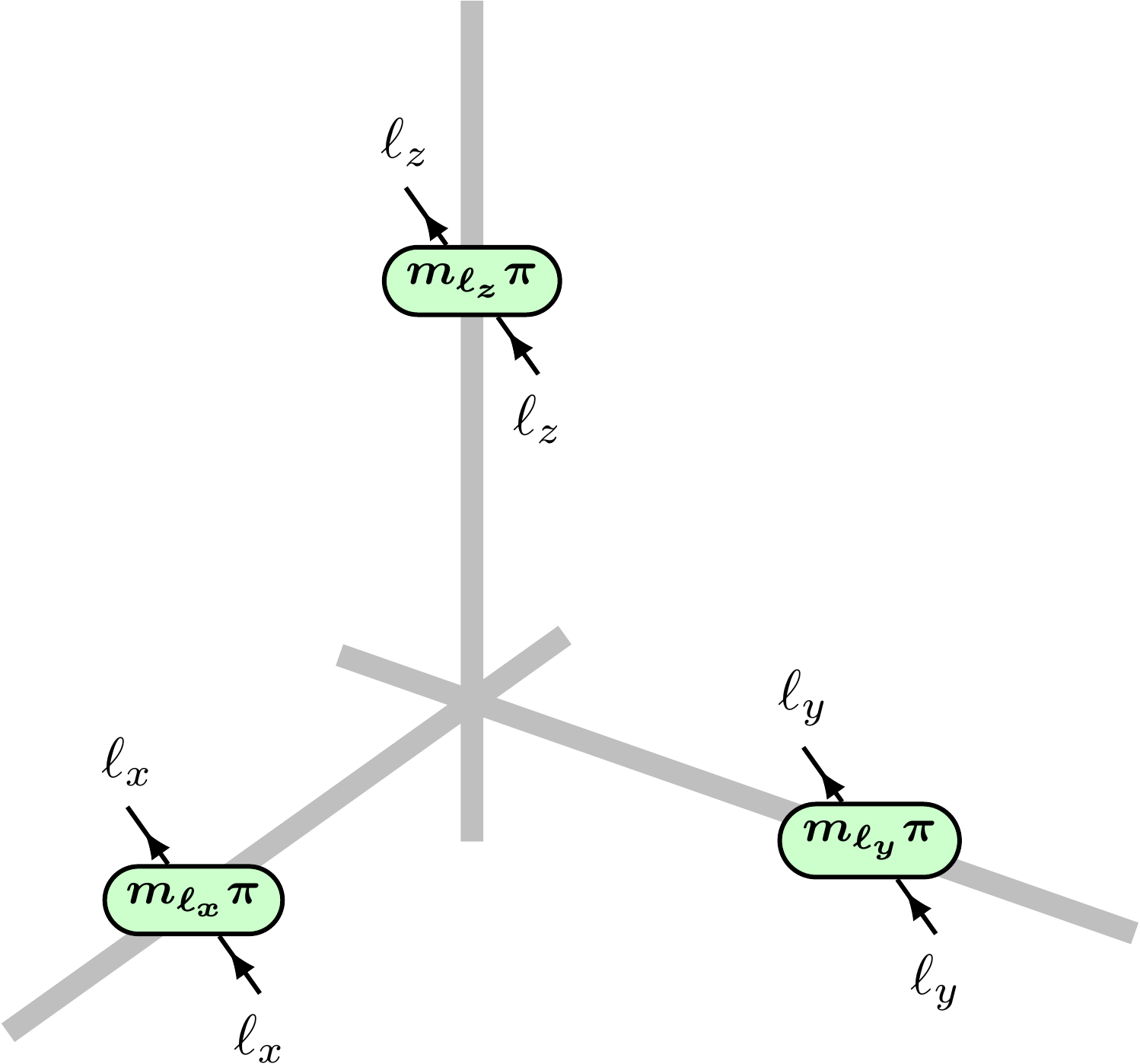}}~,\qquad m_\ell = \delta_{\gamma}(\ell)~,
\fe
where $\gamma$ is any (not necessarily connected) curve along the links of the lattice and $\delta_{\gamma}(\ell) = 1$ when $\ell \in \gamma$ and $0$ otherwise, i.e., $\delta_{\gamma}$ is the indicator function for $\gamma$ on the links. Recall that, when the Gauss laws are imposed exactly, $W(\gamma)$ is gauge invariant only when $\gamma$ is closed. On the other hand, when the Gauss laws are imposed energetically, $\gamma$ is allowed to be open and the end points of $\gamma$ create Gauss law violations.

Consider the commutation of condensation operator $\mathsf C$ with Wilson line operator $W(\gamma)$:
\ie
\mathsf C W(\gamma) = 2^{7V/2}~\raisebox{-0.5\height}{\includegraphics[scale=0.17]{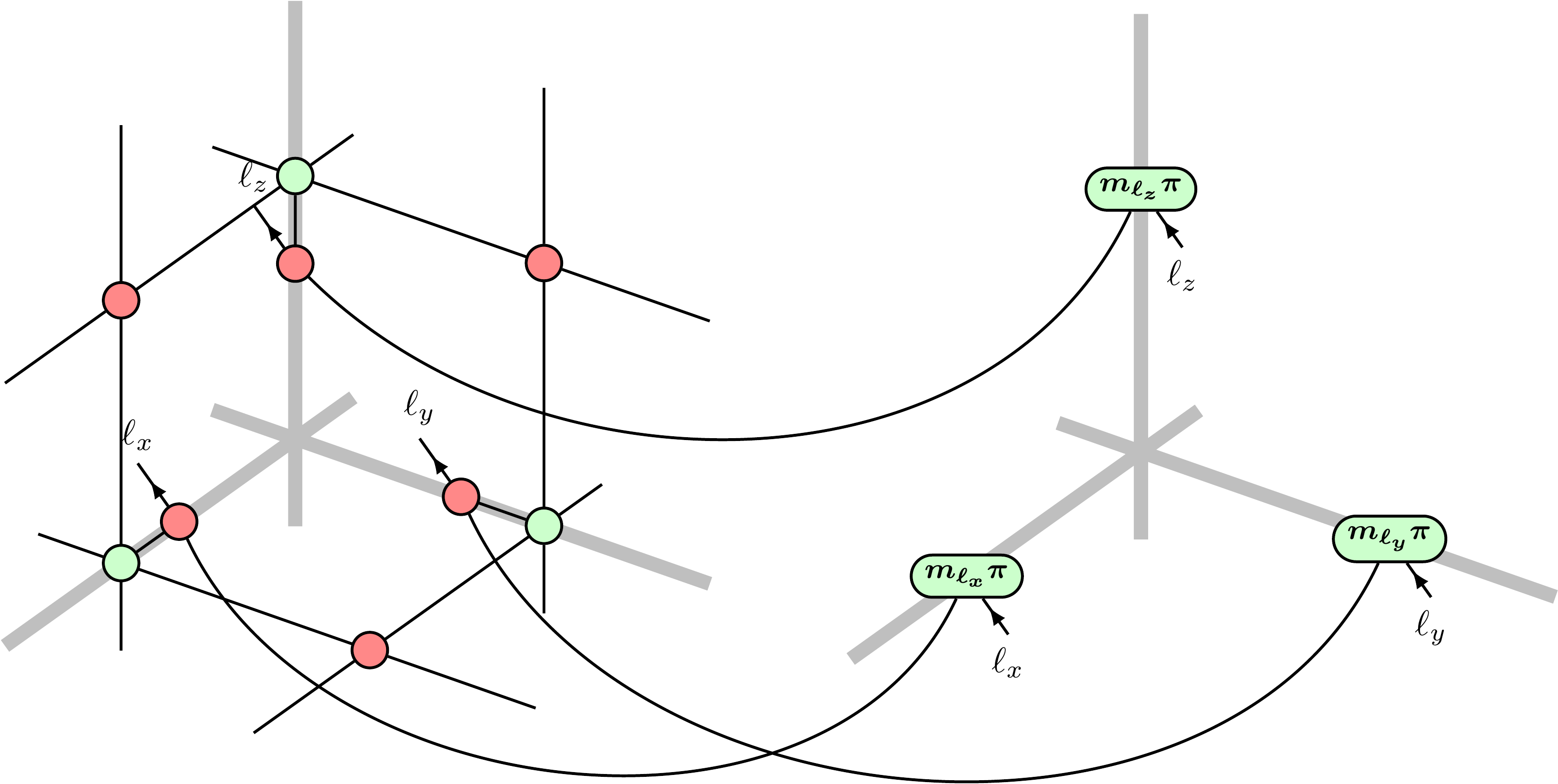}}~,\qquad m_\ell = \delta_{\gamma}(\ell)~.
\fe
Using $\pi$ commutation \ref{pi} or identity removal \ref{i} followed by spider fusion \ref{sf}, we get\footnote{
\tikzexternalenable
The 1+1d counterpart of this equation is
\tikzsetfigurename{C-Z}
\ie
(1+\eta) Z_i = 2^{L/2}~\begin{ZX}[]
& \zxZ[xshift=8pt,yshift=-4pt]{} \ar[dr] \ar[ddr] \ar[dddr] & \zxElt{\vdots} & \zxN{}\\[-4pt]
\rar & \zxN{} \rar & \zxX{} \rar & \zxN{}\\
\zxN-{i} \rar & \zxZ{\pi} \rar & \zxX{} \rar & \zxN-{i}\\
\rar & \zxN{} \rar & \zxX{} \rar & \zxN{}\\[-4pt]
& & \zxElt{\smash\vdots} & \zxN{}
\end{ZX}
\overset{\ref{pi},\ref{sf}}{=} 2^{L/2}~\begin{ZX}[]
\zxZ[xshift=8pt,yshift=-4pt]{\pi} \ar[dr] \ar[ddr] \ar[dddr] & \zxElt{\vdots} & \zxN{} &\\[-4pt]
\rar & \zxX{} \rar & \zxN{} \rar &\\
\zxN-{i} \rar & \zxX{} \rar & \zxZ{\pi} \rar & \zxN-{i}\\
\rar & \zxX{} \rar & \zxN{} \rar &\\[-4pt]
& \zxElt{\smash\vdots} & \zxN{}
\end{ZX}
= Z_i (1-\eta)~,
\fe
where  $\mathsf C_0 = 1+\eta$ can be viewed as the condensation operator for an ordinary $\bZ_2$ symmetry. The higher quantum symmetry is $\mathsf C_1= 1-\eta$.
\tikzexternaldisable}
\ie
\mathsf C W(\gamma) &= 2^{7V/2}~\raisebox{-0.5\height}{\includegraphics[scale=0.17]{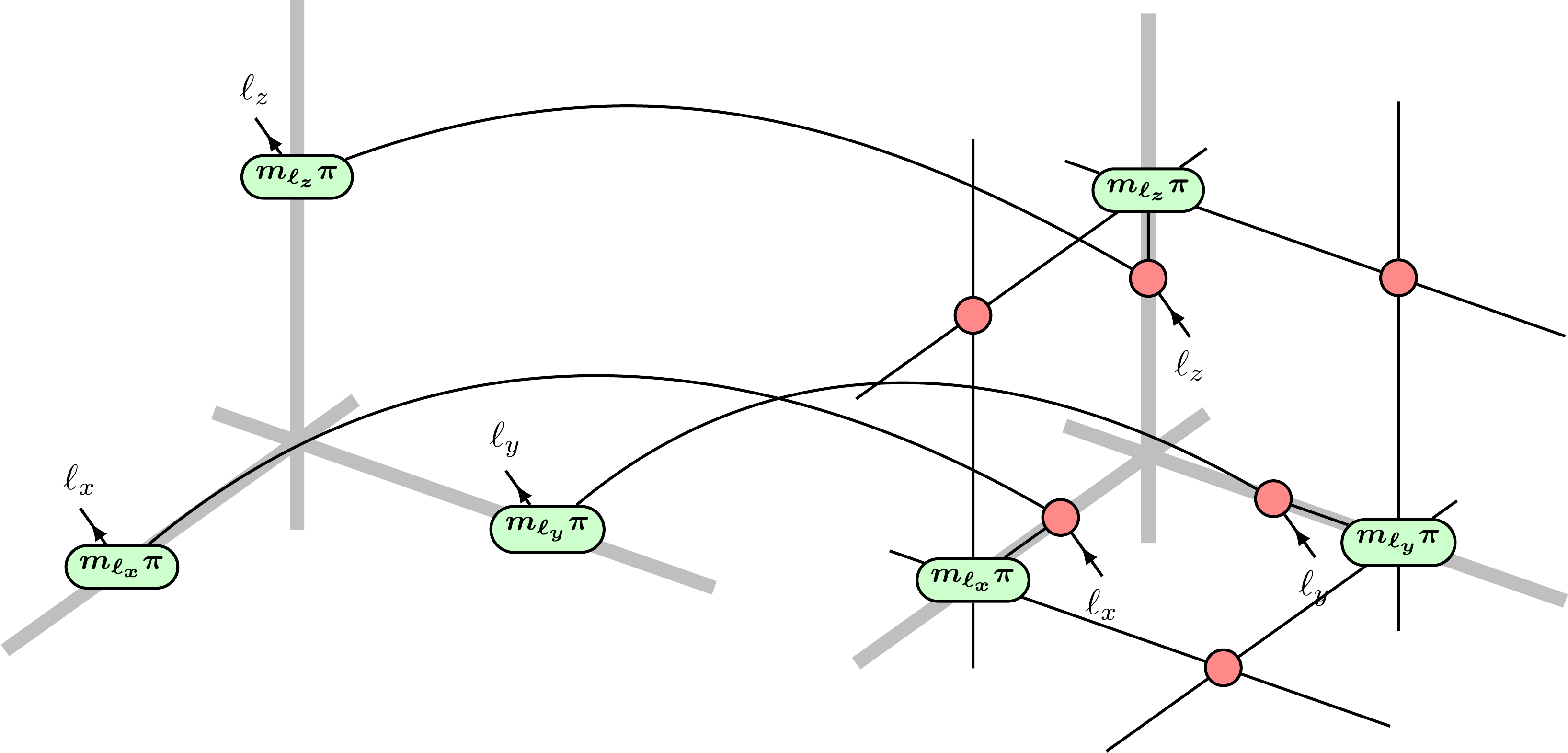}}~,\qquad m_\ell = \delta_{\gamma}(\ell)
\\
&= W(\gamma) \mathsf C(\gamma)~,
\fe
where we defined
\ie
\mathsf C(\gamma) := 2^{7V/2}~\raisebox{-0.5\height}{\includegraphics[scale=0.17]{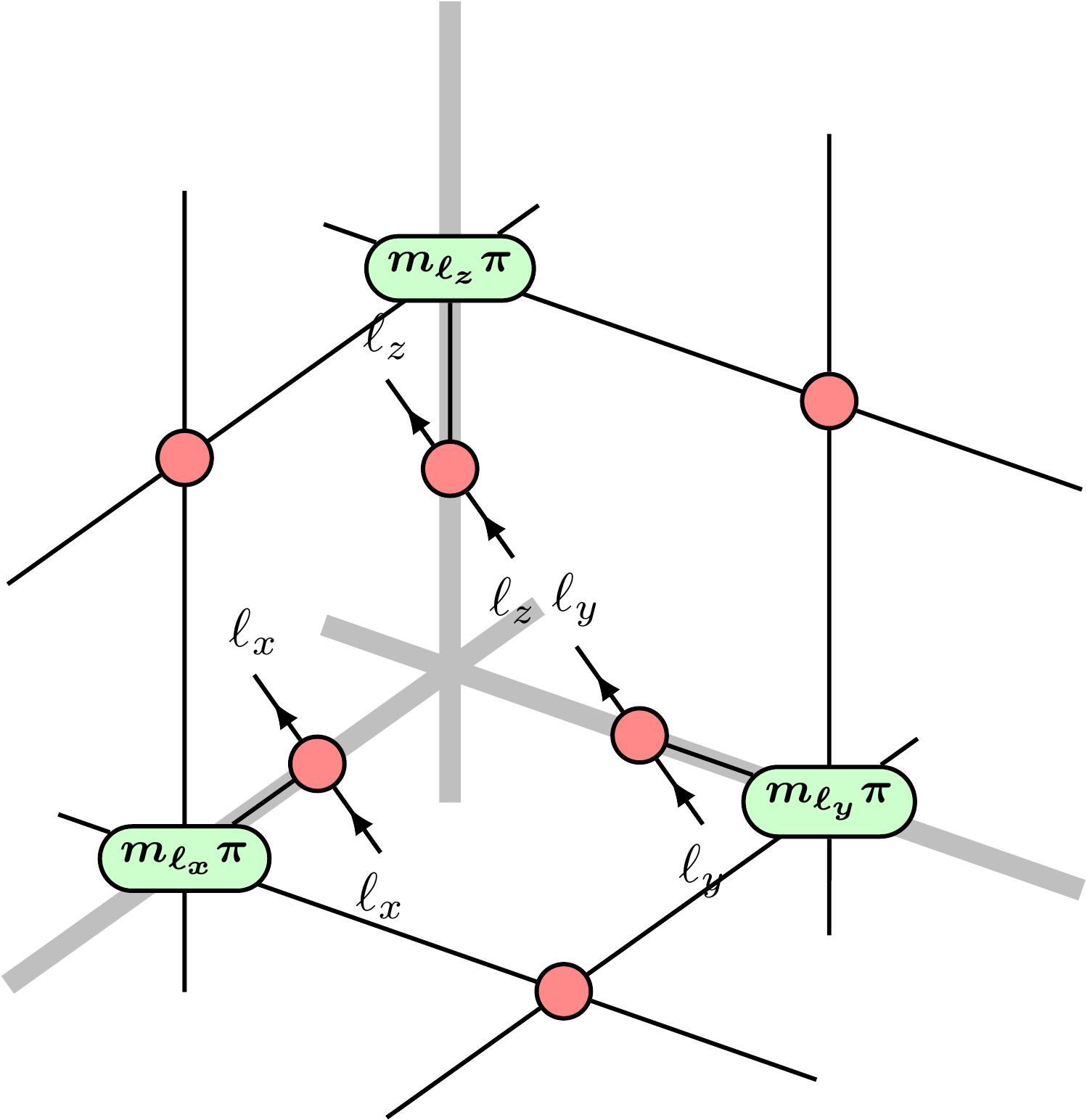}}~,\qquad
m_\ell = \delta_{\gamma}(\ell)~.
\fe
Note that a similar derivation as in Appendix \ref{app:condensation} gives
\ie
\mathsf C(\gamma) = \frac12 \prod_{i<j}[1+ (-1)^{\<\gamma,\widehat \Sigma_{ij}\>} \eta_{ij}]\prod_s \left(\frac{1+(-1)^{\delta_{\partial \gamma}(s)}G_s}{2}\right)~,
\fe
where $\<\gamma,\widehat \Sigma_{ij}\>$ is the intersection number of $\gamma$ and $\widehat \Sigma_{ij}$, and $\delta_{\partial \gamma}(s) = 1$ when $s\in \partial \gamma$ and $0$ otherwise, i.e., $\delta_{\partial \gamma}$ is the indicator function for $\partial \gamma$, the set of end points of $\gamma$.

Recall that higher gauging a $\mathbb Z_2$ 1-form symmetry on a 3-dimensional submanifold gives a $\mathbb Z_2$ 1-form higher quantum (or dual) symmetry on that submanifold. The operator $\mathsf C(\gamma)$ is the higher quantum symmetry operator at a fixed time slice.

The higher quantum symmetry operator $\mathsf C(\gamma)$ depends only on the topology of $\gamma$. Indeed, using spider fusion \ref{sf} and $\pi$ commutation \ref{pi}, one can show that $\mathsf C(\gamma) = \mathsf C$ whenever $\gamma$ is a closed contractible curve. For example, when $\gamma$ is the boundary of a plaquette $p$, i.e., $\gamma = \partial p$, we have
\ie
\mathsf C(\partial p) &= 2^{7V/2}~\raisebox{-0.5\height}{\includegraphics[scale=0.17]{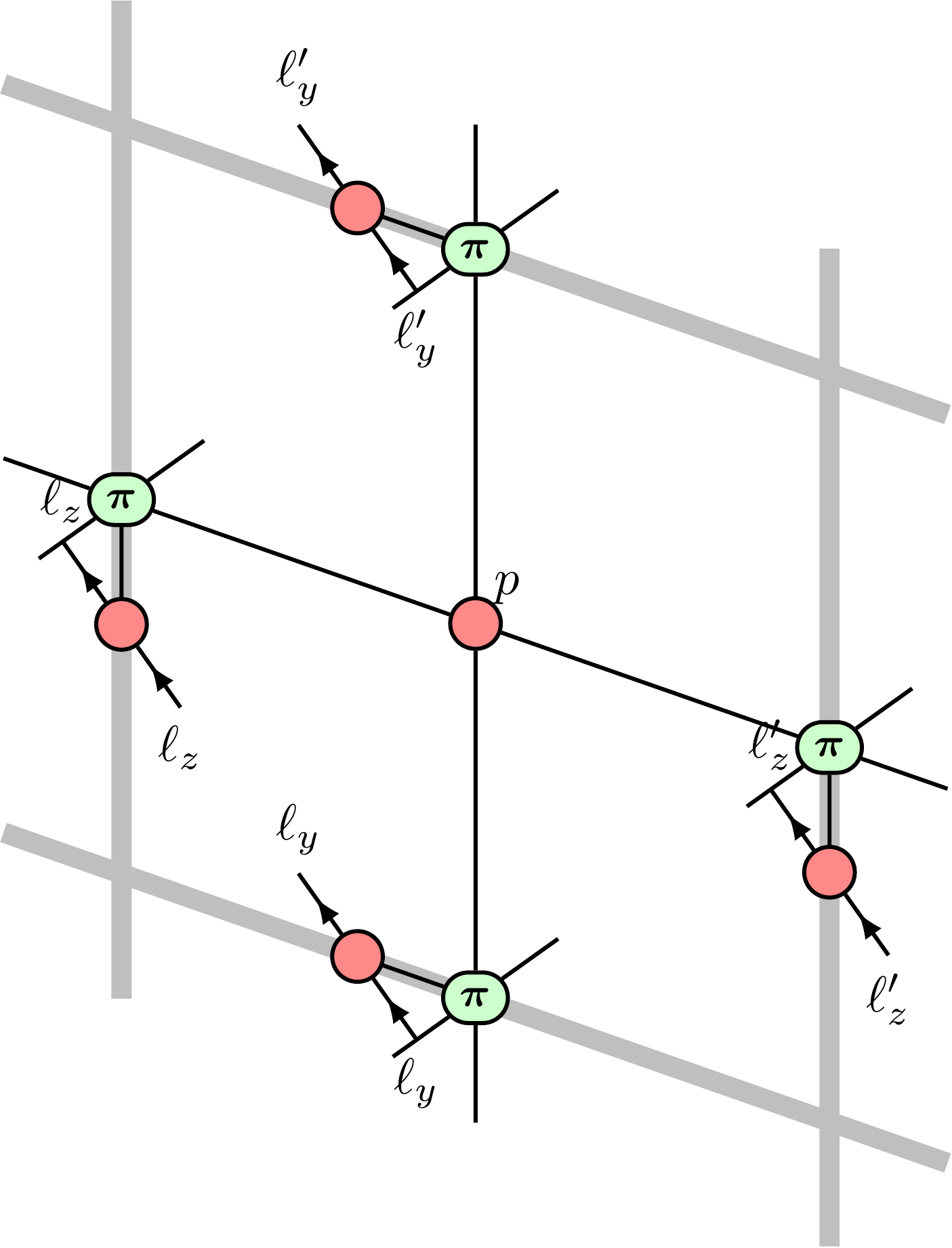}}
\overset{\ref{sf}}{=} 2^{7V/2}~\raisebox{-0.5\height}{\includegraphics[scale=0.17]{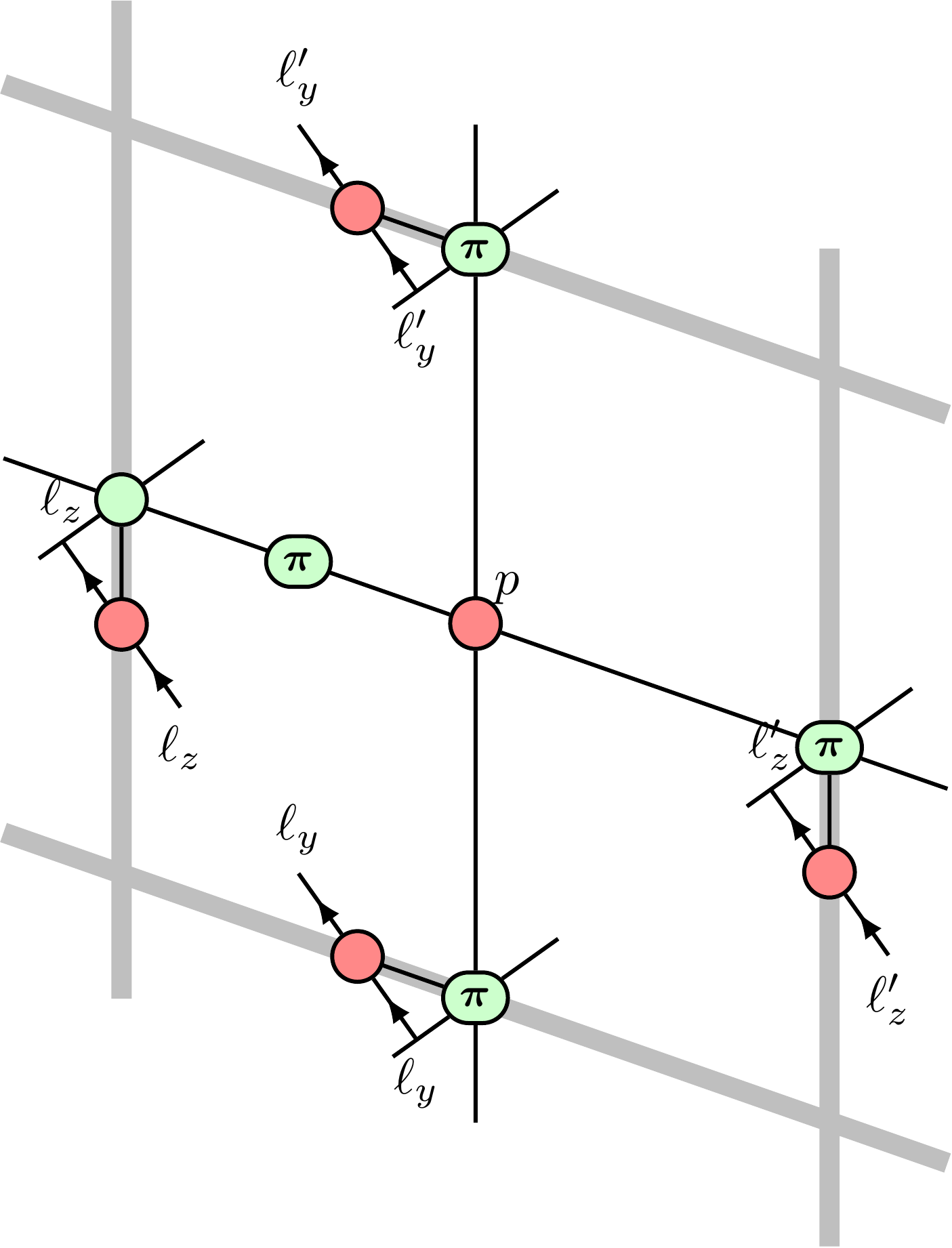}}
\\
&\overset{\ref{pi}}{=} 2^{7V/2}~\raisebox{-0.5\height}{\includegraphics[scale=0.17]{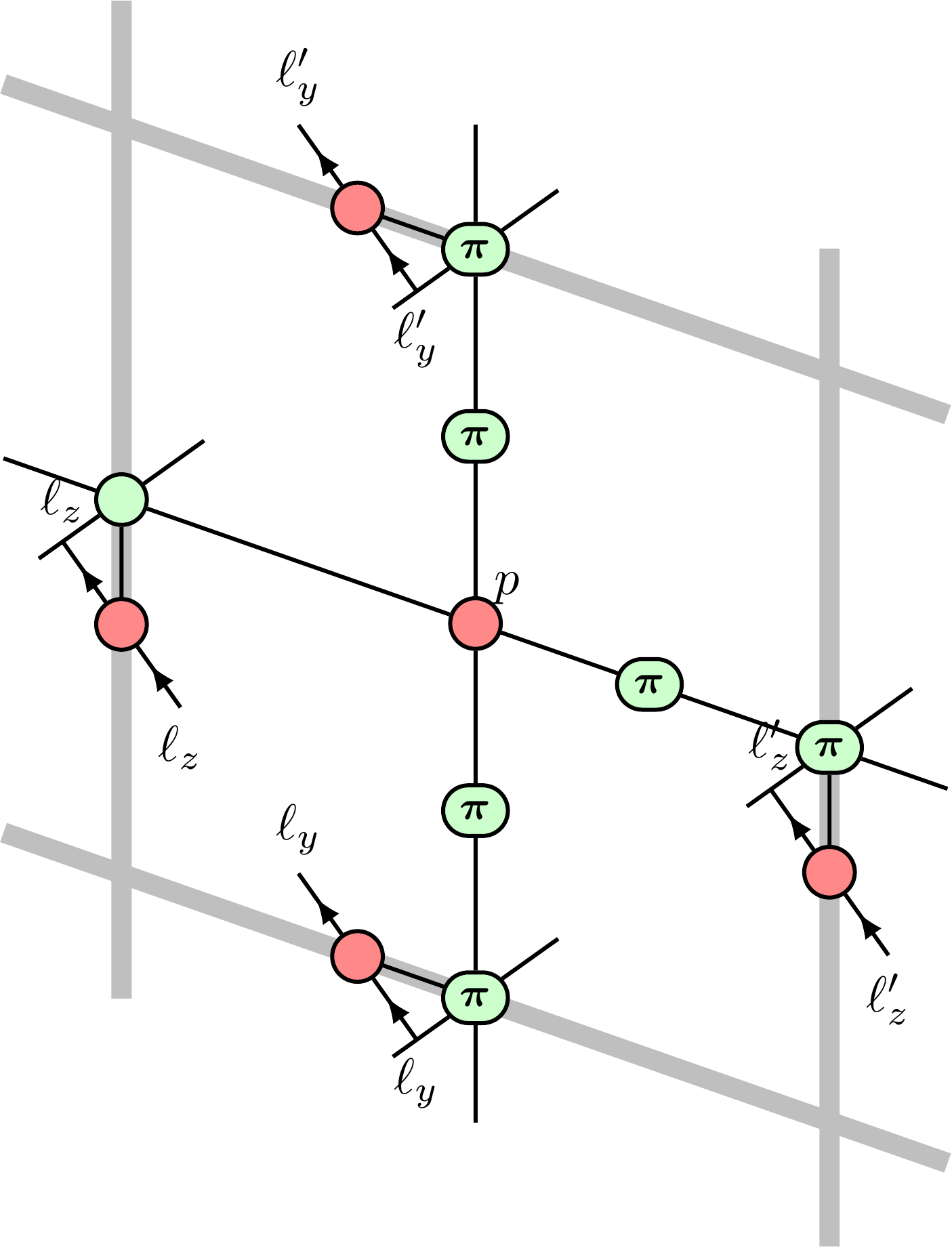}}
\overset{\ref{sf}}{=} 2^{7V/2}~\raisebox{-0.5\height}{\includegraphics[scale=0.17]{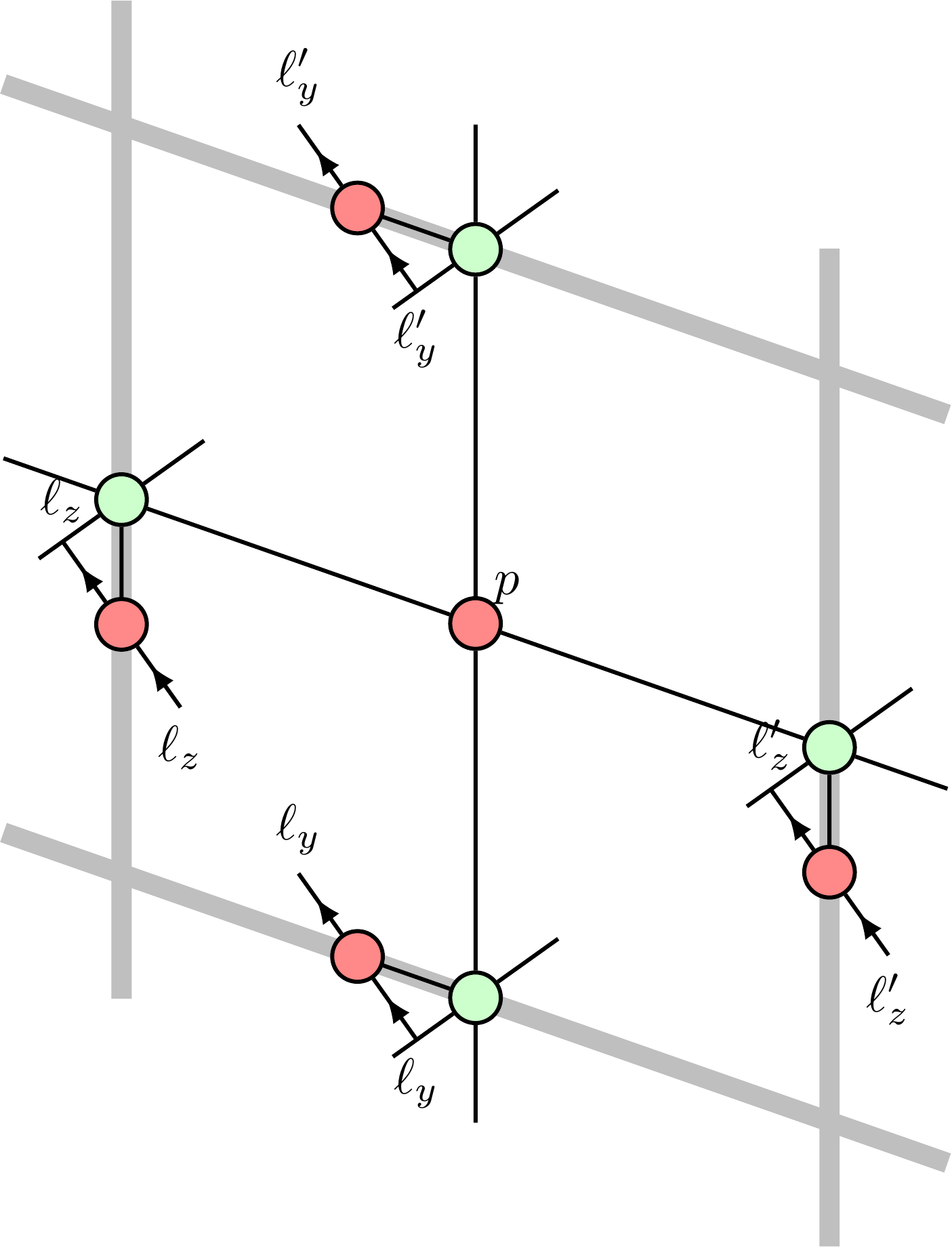}}
= \mathsf C~,
\fe
where we show only a part of the ZX-diagram around the plaquette $p$. It follows that $\mathsf C(\gamma) = \mathsf C(\gamma')$ whenever $\gamma$ and $\gamma'$ have the same end points and $[\gamma - \gamma'] = 0 \in H_1(M,\mathbb Z_2)$, where $M$ is the original lattice.

\bibliographystyle{JHEP}
\bibliography{Duality_draft}

\end{document}